\documentclass[10pt,journal,comsoc]{IEEEtran}
\usepackage{amsmath}
\usepackage{amsfonts}
\usepackage{multirow}

\usepackage{amssymb}
\usepackage{bm}
\usepackage{algorithm}
\usepackage{algpseudocode}
\usepackage{xspace}
\usepackage{caption}
\usepackage{subcaption}
\usepackage{graphicx}
\usepackage{url}
\usepackage{caption}
\usepackage[noadjust]{cite}
\usepackage{graphicx}
\usepackage{multirow}
\usepackage{comment}
\usepackage{makecell}
\usepackage{ragged2e}
\usepackage{rotating}
\usepackage{lscape}
\usepackage[table]{xcolor} 
\usepackage{nomencl}
\usepackage{stfloats}
\makenomenclature
\usepackage{wrapfig}
\usepackage{makecell}
\usepackage{balance}
\usepackage[utf8]{inputenc}
\usepackage{ragged2e}
\usepackage{booktabs}% http://ctan.org/pkg/booktabs

\usepackage{pifont}
\usepackage{tabulary}
\usepackage{comment}
\usepackage{lipsum}
\usepackage{array, tabularx, multirow, booktabs, diagbox}
\usepackage{adjustbox}

\newcommand{\cmark}{\ding{51}}%
\newcommand{\xmark}{\ding{55}}%

\begin{document}
	\title{Federated Learning for Internet of Things: Recent Advances, Taxonomy, and Open Challenges}
	\author{Latif~U.~Khan,~Walid~Saad,~\IEEEmembership{Fellow,~IEEE},~Zhu~Han,~\IEEEmembership{Fellow,~IEEE},~Ekram~Hossain,~\IEEEmembership{Fellow,~IEEE},~and~Choong~Seon~Hong,~\IEEEmembership{Senior~Member,~IEEE}
		
		% ~Nguyen~H.~Tran,\IEEEmembership{~Senior~Member,~IEEE,~}

		\IEEEcompsocitemizethanks{
			%\IEEEcompsocthanksitem This work was supported by the National Research Foundation of Korea(NRF) grant funded by the Korea government(MSIT) (No. No. 2020R1A4A1018607). *Dr. CS Hong is the corresponding author.
			
			\IEEEcompsocthanksitem L.~U.~Khan~and~C.~S.~Hong are with the Department of Computer Science \& Engineering, Kyung Hee University, Yongin-si 17104, South Korea.
			\IEEEcompsocthanksitem Walid Saad is with the  Wireless@VT, Bradley Department of Electrical and Computer Engineering, Virginia Tech, Blacksburg, VA 24061 USA.
			\IEEEcompsocthanksitem Zhu Han is with the Electrical and Computer Engineering Department, University of Houston, Houston, TX 77004 USA, and also with the Computer Science Department, University of Houston, Houston, TX 77004 USA, and the Department of Computer Science and Engineering, Kyung Hee University, South Korea.
			\IEEEcompsocthanksitem E. Hossain is with Department of Electrical and Computer Engineering at University of Manitoba, Winnipeg, Canada.

		}

		\thanks{
	}}

	%The paper headers
	\markboth{IEEE Communications Surveys and Tutorials}{}%
	
	% Traditional machine learning is based on the migration of data from massively distributed IoT devices to a centralized cloud for training and thus, poses serious security and privacy concerns.
	
	\IEEEcompsoctitleabstractindextext{%
		\justify
		\begin{abstract} 
			The Internet of Things (IoT) will be ripe for the deployment of novel machine learning algorithm for both network and application management. However, given the presence of massively distributed and private datasets, it is challenging to use classical centralized learning algorithms in the IoT. To overcome this challenge, federated learning can be a promising solution that enables on-device machine learning without the need to migrate the private end-user data to a central cloud. In federated learning, only learning model updates are transferred between end-devices and the aggregation server. Although federated learning can offer better privacy preservation than centralized machine learning, it has still privacy concerns. In this paper, first, we present the recent advances of federated learning towards enabling federated learning-powered IoT applications. A set of metrics such as sparsification, robustness, quantization, scalability, security, and privacy, is delineated in order to rigorously evaluate the recent advances. Second, we devise a taxonomy for federated learning over IoT networks. Finally, we present several open research challenges with their possible solutions.                 
		\end{abstract}

		\begin{IEEEkeywords}
			Federated learning, Internet of Things, wireless networks.
	\end{IEEEkeywords}}
	\maketitle
	\IEEEdisplaynotcompsoctitleabstractindextext
	\IEEEpeerreviewmaketitle
	
	%Numerous devices involved in the training of the shared global learning model first perform local training, which is followed by transmission of local model parameters only to a remote cloud or edge server. At the remote cloud, aggregation of the local models parameters is performed and the resulting global model parameters are sent back to all the involved devices. This process continues in an iterative fashion until convergence of the global model to some accuracy level.
	
	% These developments include smart transportation, smart healthcare, smart grid, smart buildings, smart banking, smart waste management, smart agriculture, and smart surveillance, to name a few.
	
	% Federated learning models based on context-awareness can be divided into two categories, such as global context federated learning and local context federated learning. Local context federated learning involves a set of devices within close vicinity and computation of global learning model at edge server, whereas global context federated learning model involves computation of global learning model at a cloud for a massively geographically distributed devices
	
	% \setlength{\parindent}{0.25ex}
	
	%According to statistics, the number of IoT devices will reach to 500 billion in 2030\protect\footnote{Access Date: March 12, 2019. https://www.cisco.com/c/dam/en/us/products/collateral/se/internet-of-things/at-a-glance-c45-731471.pdf}.

	\section{Introduction}
	\setlength{\parindent}{0.7 cm}Internet of Things (IoT) applications can lead to the true realization of smart cities \cite{6Gvisionlatif}. Smart cities can offer several critical applications such as intelligent transportation, smart industries, healthcare, and smart surveillance, among others \cite{khan2019edge, sanchez2019smart, perera2017fog,khan2021digital}. To successfully deploy these smart services, a massive number of IoT devices are required \cite{saad2019vision, al2015internet, perera2013context}. According to statistics, the number of IoT devices will reach $125$ billion by $2030$ \cite{IOTdevices_1}. These IoT devices will generate enormous amount of data. A recent report in \cite{IOTdevices_2} revealed that the data generated by IoT devices will reach $79.4$ zettabytes (ZB) by $2025$. Such a remarkable increase in both IoT network size and accompanying data volume open up attractive opportunities for artificial intelligence \cite{qi2020popularity}. To this end, one can use centralized machine learning schemes to enable these smart IoT applications. However, centralized machine learning techniques suffer from the inherent issue of users' privacy leakage because of the need to transfer the end-devices data to a centralized third party server for training. Furthermore, centralized machine learning might not be feasible when the data size is very large (e.g., astronomical data \cite{raicu2006astroportal}) and located at multiple locations \cite{bernstein2009principles, verbraeken2019survey}. \par
	
	To cope with the aforementioned challenge of large, distributed datasets, we can use distributed machine learning to distribute the machine learning workload to multiple machines \cite{Distributed_ML1, 8038464, qiu2016survey, peteiro2013survey}. These multiple machines can then train several models at distributed locations in parallel to create a machine learning model. Distributed learning enables high-performance computing via parallel computation of a machine learning model. Two types of approaches such as data-parallel and model-parallel approaches can be used for distributed machine learning. In a data-parallel approach, the whole data is divided among a set of distributed servers, while each server using the same model for training. The model-parallel approach uses exactly the same data for all servers to train distinct portions of a machine learning model \cite{verbraeken2019survey}. The model-parallel approach might not be suitable for some applications because all the machine learning model can not be divided into parts.\par
	
	\begin{comment}
	
	\begin{table}[]
	\caption {List of abbreviations.} \label{tab:performance_parameters} 
	\begin{center}
	\begin{tabular}{p{1.5cm}p{6cm}}
	\toprule
	\hline
	\textbf{Abbreviation}  & \textbf{Word} \\ \hline
	IoT & Internet of Things & BS & Base station \\ 
	
	CAGR & Compound annual growth rate \\ FNN & Feed-forward neural network\\
	
	CNN & Convolutional neural neural network \\ SVM & Support vector machine \\
	
	LSTM &  Long-short term memory \\ SGD & Stochastic gradient descent   \\
	
	QSGD & SGD via gradient quantization and encoding \\ URLLC & Ultra-Reliable
	Low Latency Communication \\
	
	BCH & Bose, Chaudhuri, and Hocquenghem \\ IID & Independent and identically distributed\\
	
	Non-IID & Non independent and identically distributed \\ MPC & MultiParty computation \\
	
	TFF & TensorFlow Federated \\ FL-Block & Blockchain-enabled federated learning \\
	
	UAV & unmanned aerial vehicles \\ PoW &  Proof of work \\
	
	FedAvg & Federated averaging \\  FML  &  Federated multi-task learning   \\
	
	RSU & Road side units \\ IoE & Internet of Everythings \\
	
	D2D & Device-to-device \\  LDPC  & Low density parity-check  \\
	
	ReLU & Rectified linear unit \\  FedCS  & Federated client selection   \\

	\hline
	\bottomrule
	\end{tabular}
	\end{center}
	\end{table}
	
	\end{comment}

	\begin{figure*}[!t]
		\centering
		\captionsetup{justification=centering}
		\includegraphics[width=18cm, height=13.5cm]{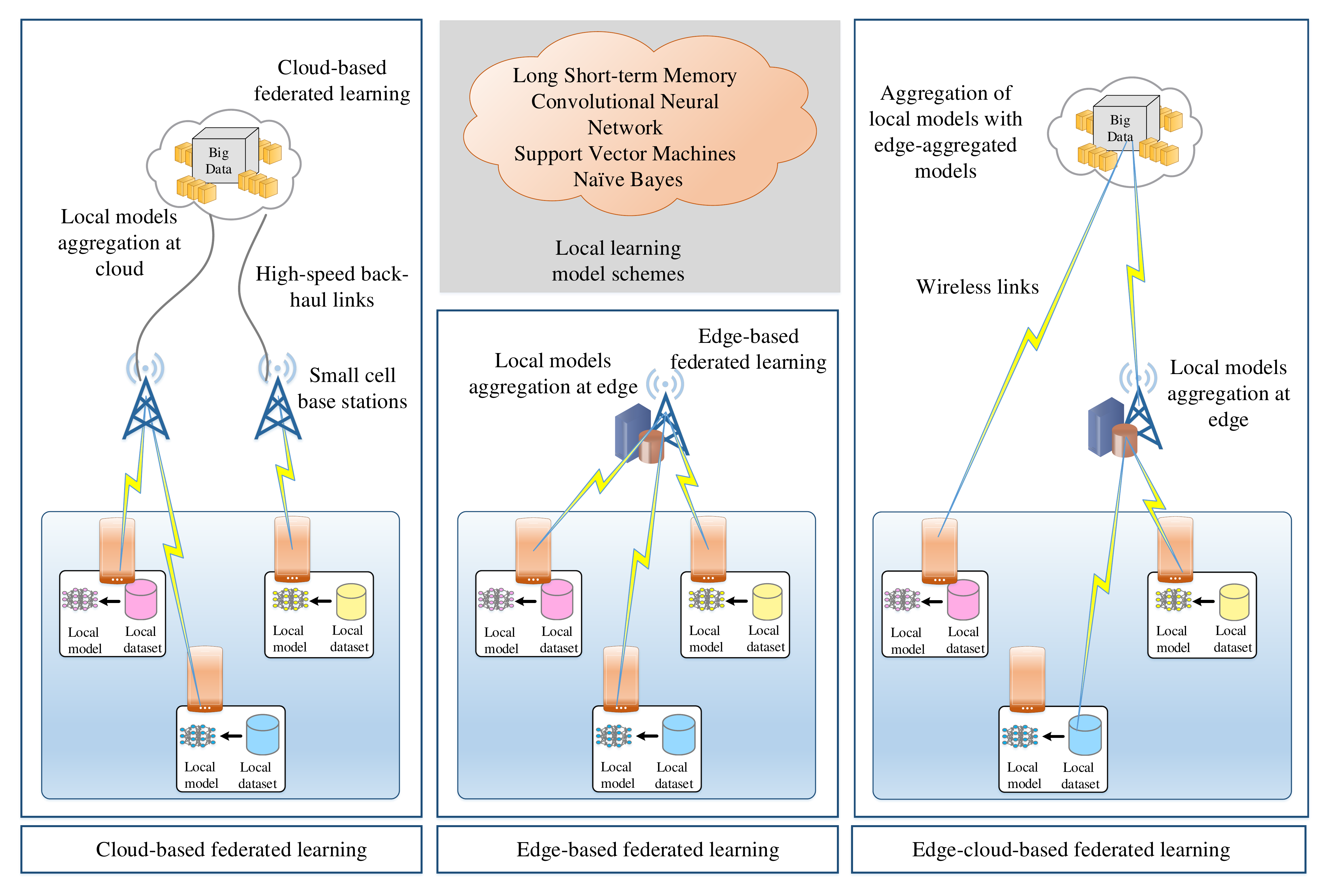}
		\caption{An overview of federated learning for IoT networks.}
		\label{fig:federatedoverview}
	\end{figure*}
	
	Although distributed machine learning offers parallel computation of models at geographically distributed locations, it does not explicitly addressed some more practical challenges of data and system heterogeneity \cite{mcmahan2017communication}. Here, system heterogeneity refers to variations in computational resources (CPU-cycles/sec), communication resources, and storage capacity of end-devices, whereas data heterogeneity deals with non-independent and identical distribution (non-IID) of data \cite{yang2020heterogeneity, chai2019towards}. Additionally, traditional distributed machine learning algorithms that rely on the model-parallel or the data-parallel approach do not truly preserve the privacy of users. These distributed machine learning schemes use training of machine learning models at distributed servers that rely on data from multiple end-users, and thus suffers from privacy leakage issue. To ensure privacy preservation while considering system and data heterogeneity in machine learning, federated learning\footnote{Federated learning itself has privacy concerns and does not completely guarantee privacy.} has been introduced \cite{mcmahan2017communication}. In contrast to traditional machine learning schemes, federated learning does not require migration of data from end-devices to a centralized edge/cloud server for training. The process of federated learning involves local learning model computation at end-devices, which is followed by sending the local learning model parameters to the edge/ cloud server for global model aggregation. Finally, the global model parameters are sent back to the end-devices \cite{konevcny2016federated}. To implement federated learning for IoT networks, we will face some challenges. Computing local learning model for federated learning requires significant computational power and backup energy. Additionally, implementing federated learning for IoT applications where the IoT sensors generate fewer data seems challenging \cite{akyildiz2002wireless, jiang2020federated, nguyen2020efficient}. To train a local learning model, a sufficient amount of data is required for better learning accuracy \cite{Dusit_FL_survey, khan2019federated, kairouz2019advances}. Therefore, training a local learning model with better accuracy for computing resource-constrained devices with less data is challenging. To address this challenge one can migrate the data from resource-constrained IoT sensors (e.g., temperature monitoring sensors) to the nearby trustworthy edge for local model training \cite{edgeAI}. On the other hand, autonomous driving cars generate a significant amount of data (i.e., $4,000$ gigaoctet) everyday \cite{FL_industry_1}. Therefore, sending everyday autonomous driving cars will consume more communication resources. One can only send a local learning model to the aggregation server that will consume fewer communication resources than the whole data. Additionally, autonomous cars have more computational power with sufficient energy. Therefore, federated learning can be easily implemented for autonomous driving cars from the perspective of resource usage.  \par

	An overview of federated learning for IoT networks is presented in Fig.~\ref{fig:federatedoverview}. First of all, end-devices iteratively compute the local learning models using their local datasets. After local model computation, the local learning model updates are sent by end-devices to the aggregation server for the global aggregation. Generally, aggregation can be performed in federated learning either at the edge or cloud. Therefore, we can say that federated learning can be either edge-based or cloud-based depending on the location of the global model aggregation \cite{khan2019federated}. Edge-based federated learning performs global model aggregation at the edge server, whereas cloud-based federated learning performs global federated learning model aggregation at the remote cloud. The edge server serves a small geographical area with a limited number of devices. Therefore, edge-based federated learning is preferable for use in applications that have local context-awareness and require specialized machine learning models for a set of users located in close vicinity to each other \cite{Dusit_FL_survey}. For instance, consider the training of keyboard keyword suggestion model for a language of a particular region using federated learning. There can two ways to train such type of model: a specialized model for the regional language via edge-based federated learning and a generalized model for languages of different regions via cloud-based federated learning. A model using edge-based federated can offer more promising results in such a scenario than cloud-based federated learning \cite{khan2019federated}. On the other hand, the cloud serves a large number of users located over a large distributed area. Therefore, federated learning that uses aggregation at the cloud (i.e., cloud-based federated learning) is more suitable for training more generalized models than edge-based federated learning by considering a large number of geographically distributed users. It must be noted that we derive the use of terms \emph{edge-based federated learning} and \emph{cloud-based federated learning} from \cite{khan2019federated}. Other than edge-based federated learning and cloud-based federated learning, we can jointly use both edge and cloud servers for the aggregation of learning models. We refer to federated learning that uses aggregation at both cloud and edge as edge-cloud-based federated learning \cite{liu2019client}. Communication with an edge is less expansive in terms of communication resources consumption and latency than the remote cloud \cite{khan2019edge}. Therefore, end-devices can easily communicate with the edge server that will perform aggregation of local learning models to yield an edge-aggregated learning model. On the other hand, edge server has generally small coverage area and has a limited capacity to serve end-devices. Therefore, few of the end-devices located in a different geographical area as that of the edge server will directly communicate with the cloud. The edge-cloud-based federated learning will perform aggregation of local learning models of these devices with the edge-aggregated learning models. The main advantage of edge-cloud-based federated learning lies in enabling the participation of more devices in the learning process, and thus improves federated learning performance \cite{liu2019client, mcmahan2017communication, khan2019federated}. \par

	Although federated learning enables on-device machine learning, it suffers from security, robustness, and resource (both computational and communication) optimization challenges. To truly benefit from the use of federated learning in IoT networks, there is a need to resolve these challenges. To cope with resource constraints issues, one can develop federated learning protocols based on device-to-device (D2D) communication. An IoT device might not be able to communicate with a central base station (BS) due to communication resource constraints. To enable participation of such IoT devices in federated learning, one can use collaborative federated learning \cite{chen2020wireless}. Collaborative federated learning allows resource-constrained IoT devices to send their local learning models to nearby devices rather than the BS, where local aggregation takes place between the receiving device local model and the received local model. Then, the locally aggregated model is sent to the centralized edge/cloud server for global aggregation. Similar to traditional federated learning, collaborative federated learning can be either edge-based or cloud-based depending on the context of the machine learning model. On the other hand, a malicious aggregation server can infer end-devices information from their learning model parameters, and thus might cause privacy leakage \cite{nasr2019comprehensive, Dusit_FL_survey}. To cope with this issue, one can use differential privacy-aware federated learning \cite {geyer2017differentially}. Next, we discuss market statistics and research trends of federated learning and IoT-networks. \par    
	\begin{figure}[!t]
		\centering
		\captionsetup{justification=centering}
		\includegraphics[width=8cm, height=6cm]{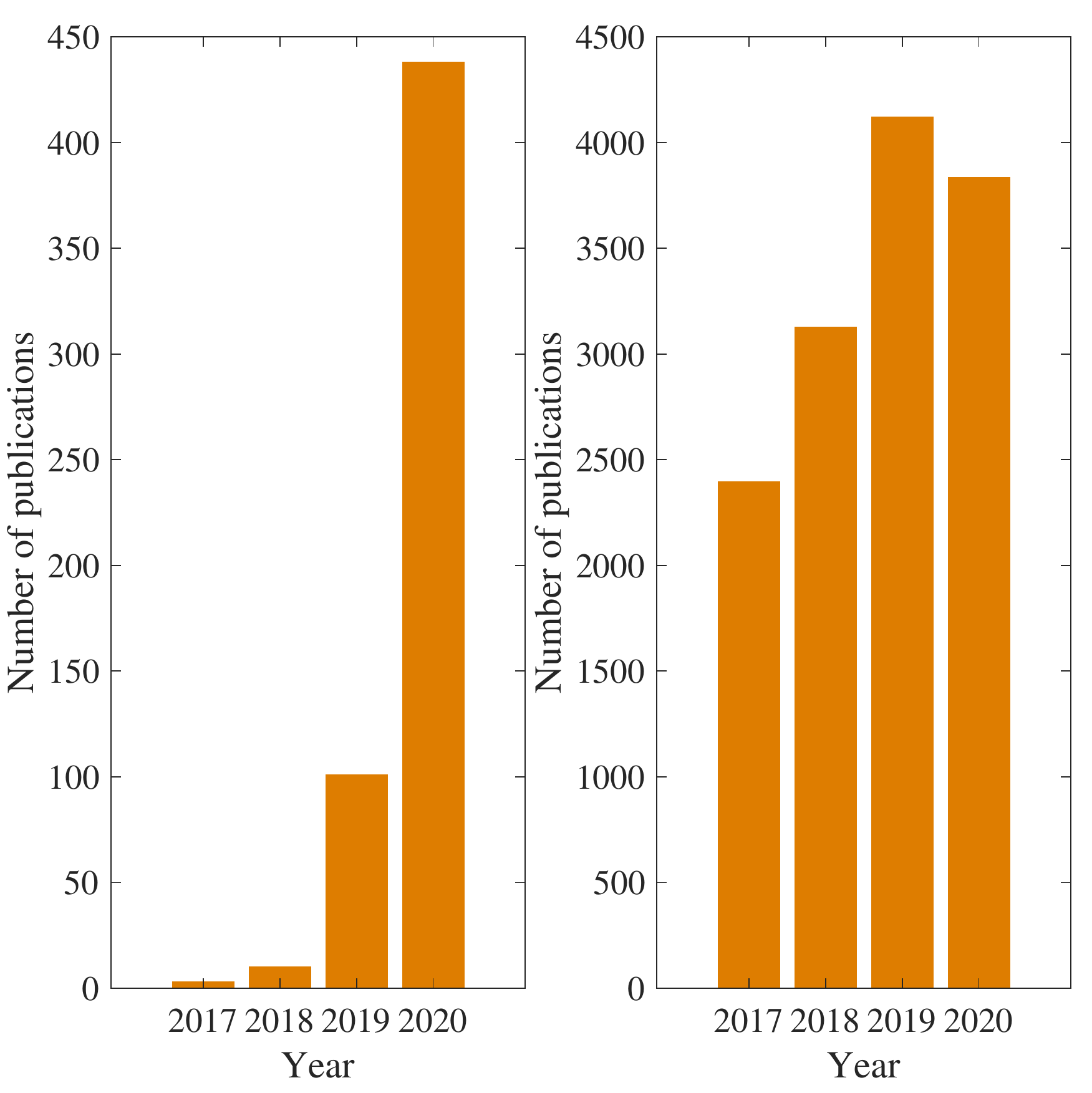}
		\caption{Year wise publications of (a) federated learning, and (b) Internet of Things [Scopus data, Access month: February 2021].}
		\label{fig:count}
	\end{figure}
	
	%A set of users located in a close vicinity can can communicate with each other. Then, a cluster-head is selected based on its highest social relationship with other devices compared to members of its social group. Initially, a sub-global federated learning model is trained in an iterative manner within multiple social groups as shown in figure~\ref{}. Then, the sub-global model is sent to the base station for global model aggregation. Finally, the global model parameters are sent to the cluster heads which disseminate the global model parameters to the end devices. This approach has the advantage of using already occupied bandwidth by other users for D2D communication within clusters while protecting the other users by keeping the interference level below the maximum allowed limit. \par

	%In \cite{kim2019blockchained}, the authors presented blockchain-based federated learning for providing robustness and security. However, they obtain robustness and security at the cost of extra communication overhead and high computational complexity compared to traditional federated learning. In another study, Tran \textit{et al.} studied resource optimization in federated learning for wireless networks \cite{tran2019federated}. Other than resources optimization, there must be some incentive mechanism for attracting the users to participate in federated learning process. In \cite{pandey2019crowdsourcing}, Pandey \textit{et al.} proposed a Stackelberg-based incentive mechanism for federated learning for edge network.\par 

	\subsection{Market Statistics and Research Trends}
	According to statistics, the market of artificial intelligence is expected to grow at a Compound Annual Growth Rate (CAGR) of 33.2\% from 27.23 billion US Dollar in 2019 to 266.93 billion US Dollar in 2027 \cite{ML_Market}. The key driver of the increase in the artificial intelligence market is the proliferation of technological advancement and data generation. Banking, financial and insurance (BFSI) vertical, IT and telecom, retail, healthcare, automotive, advertising and media, among others are various industries that are mainly contributing to the artificial intelligence market. On the other hand, North America has shown significant growth in the artificial intelligence market in 2019 with the US as the highest contributor. The Asia Pacific is also expected to have significant growth in the artificial intelligence market. China will add a major portion to the artificial intelligence market in the Asia Pacific region. \par        
	The IoT market is expected to reach 1463.19 billion US Dollar from 250.72 billion US Dollar in 2027 compared to 2019 at a CAGR of 24.9\% \cite{IoT_Market}. The key drivers of IoT markets are telecommunication, transportation, manufacturing, health-care, government, retail, BFSI, among others. The highest share is hold by BFSI sector. The highest revenue generated by the Asia Pacific region in 2018 is 98.86 billion US Dollar, which is further expected to lead the IoT market in future. Within Asia Pacific region, China has the highest share. Other than market share of machine learning and IoT, the number of research publications versus year are shown in Fig.~\ref{fig:count}. It is clear from the Fig.~\ref{fig:count} that there is significant increase in number of publications every year for both federated learning and IoT. Based on the above-discussed facts, machine learning and IoT are expected to opens up novel opportunities for research and development in the foreseeable future. \par

	\begin{table*}[]
		\rowcolors{2}{gray!25}{white}
		\caption {Summary of existing surveys and tutorials with their primary focus.} \label{tab:surveyssummaries} 
		\begin{center}
			\begin{tabular}{p{1.8 cm}p{1.8cm}p{1.8cm}p{1.8 cm}p{2.5 cm}p{1.8 cm}p{3.5 cm}}
				\toprule 
				\textbf{Reference}   & \textbf{Internet of Things}& \textbf{Resource optimization for federated learning} & \textbf{Incentive mechanism for federated learning} & \textbf{Learning algorithm design for federated learning}  &\textbf{Taxonomy}&\textbf{Remark} \\ \midrule
				Lim \textit{et al.}, \cite{Dusit_FL_survey} &\xmark & \cmark & \cmark & \cmark &  \xmark&Federated learning at network edge is considered.\\ \midrule
				Li \textit{et al.}, \cite{li2019federated} & \xmark & \xmark & \xmark & \cmark & \xmark & N.A\\ \midrule
				Li \textit{et al.}, \cite{li2019federated2} & \xmark & \xmark & \xmark & \cmark &\xmark& N.A\\ \midrule
				Wang \textit{et al.}, \cite{edgeAI} & \xmark & \xmark & \xmark &\xmark & \xmark& Mainly considered federated learning for caching and computational offloading.\\ \midrule
				Lyu \textit{et al.}, \cite{lyu2020threats} & \xmark & \xmark & \xmark & \xmark   & \xmark& Surveyed security in federated learning.\\ \midrule
				Kairouz \textit{et al.}, \cite{kairouz2019advances} & \xmark & \xmark & \xmark & \cmark   & \xmark& N.A\\ \midrule
				Khan \textit{et al.}, \cite{khan2020dispersed}  & \xmark & \xmark & \xmark & \xmark  & \xmark& The work introduced dispersed federated learning, devise its taxonomy, and presented future directions.\\ \midrule
				
				Rahman \textit{et al.},~\cite{FL_Survey_IOT_jour} & \xmark  & \cmark   &\xmark  &\cmark  & \xmark & Surveyed federated learning design aspects (i.e., learning algorithm design and resource optimization) and open challenges. \\ \midrule
				Aledhari \textit{et al.},\cite{aledhari2020federated} & \xmark  & \xmark   & \xmark  &\cmark  & \xmark &  Focused mainly on federated learning architecture and enablers.\\ \midrule
				Our Tutorial & \cmark  & \cmark   &\cmark   &  \cmark& \cmark& N.A\\ 
				\bottomrule 
			\end{tabular}
		\end{center}
	\end{table*} 
	
	\subsection{Existing Surveys and Tutorials}
	Numerous surveys and tutorials primarily reviewed federated learning \cite{Dusit_FL_survey, li2019federated, li2019federated2, edgeAI,lyu2020threats, kairouz2019advances, FL_Survey_IOT_jour},and\cite{ aledhari2020federated}. Lim \textit{et al.} presented a comprehensive survey of federated learning for mobile edge networks \cite{Dusit_FL_survey}. First, the authors provided an overview of federated learning. Second, the key implementation challenges with existing solutions and several applications of federated learning in mobile edge networks are discussed. Finally, several open research challenges are presented. In \cite{li2019federated}, Li \textit{et el.} surveyed implementation challenges of federated learning, existing approaches, and future research directions. The work in \cite{li2019federated2} presented an overview of federated learning, devised taxonomy, and discussed the existing solutions with their limitations in enabling federated learning. In \cite{edgeAI}, Wang \textit{et al.} proposed an “In-Edge AI” framework for edge networks to enable efficient collaboration between end-devices and edge servers for learning model updates exchange. They considered two use cases such as edge caching and computation offloading. To efficiently enable these use cases, a double deep Q-learning agent was trained using federated learning. Finally, several open research challenges were presented. In \cite{lyu2020threats}, Lyu \textit{et al.} presented a survey of threats to federated learning. In \cite{kairouz2019advances}, Kairouz \textit{et al.} comprehensively surveyed federated learning, and they discussed basics, privacy preservation, robustness, and ensuring fairness. The authors in \cite{khan2020dispersed} presented a vision of dispersed federated learning that is based on true decentralization. Furthermore, a taxonomy and future directions are presented for dispersed federated learning. The authors in \cite{FL_Survey_IOT_jour} primarily surveyed federated learning design aspects (i.e., learning algorithm design and resource optimization) and presented open challenges. Aledhari \textit{et al.} in \cite{aledhari2020federated} presented enabling technologies, protocols, and few applications of federated along with open challenges. On the other hand, there are few surveys \cite{hussain2020machine, waheed2020security } that have briefly discussed federated learning as a solution to some existing problems (e.g., caching at the network edge). However, their primary perspective was to use machine learning for enabling IoT. In contrast, our tutorial primarily focus on federated learning for IoT.

	\begin{figure}[!t]
		\centering
		\captionsetup{justification=centering}
		\includegraphics[width=8cm, height=17.5cm]{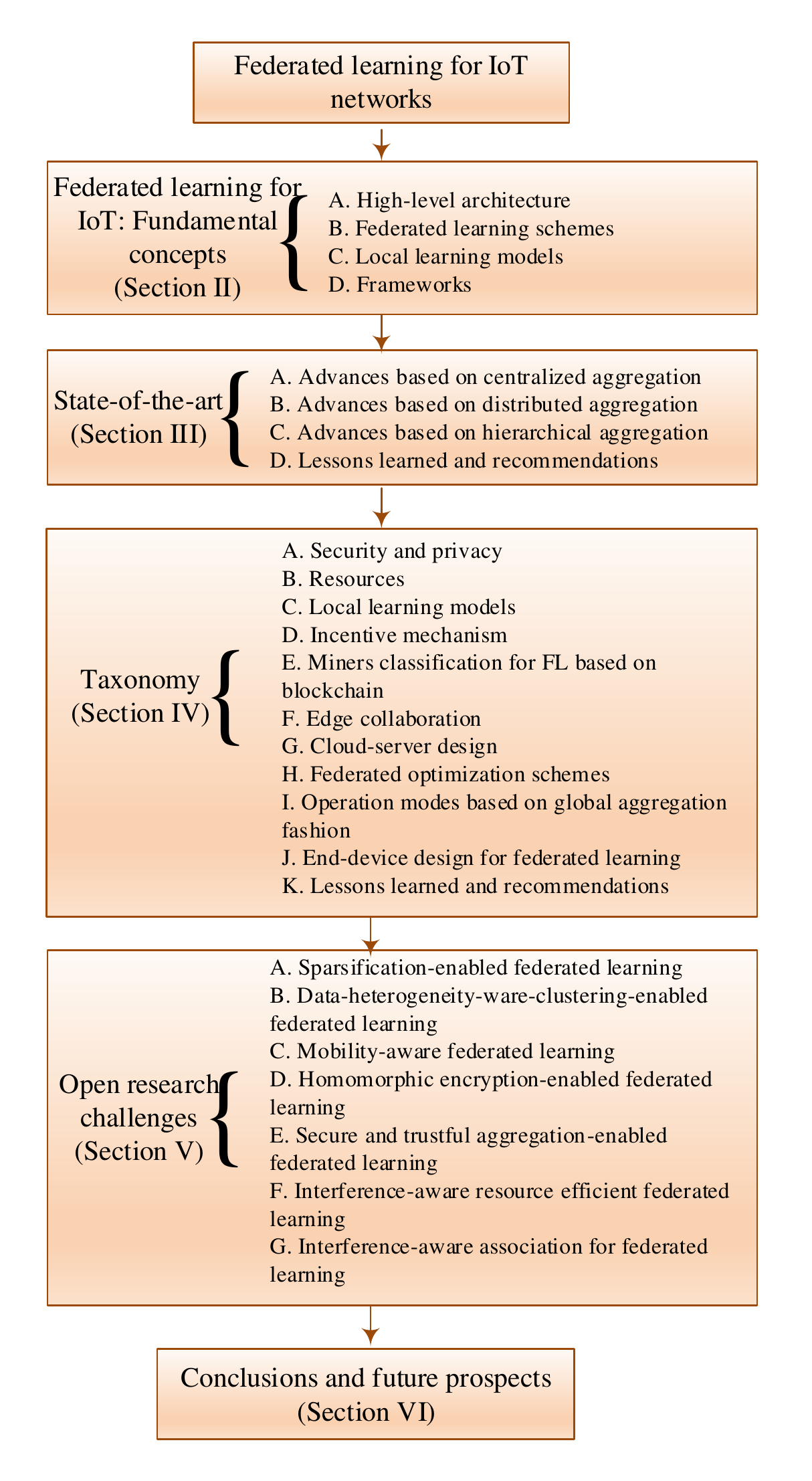}
		\caption{Overview of our tutorial.}
		\label{fig:roadmap}
	\end{figure}
	
	\subsection{Our Tutorial}
	Mostly, the existing surveys and tutorials \cite{Dusit_FL_survey, li2019federated, li2019federated2, edgeAI,lyu2020threats, kairouz2019advances, khan2020dispersed, FL_Survey_IOT_jour, aledhari2020federated} considered resource optimization, incentive mechanism, and learning algorithm design for federated learning. In contrast (as given in Table~\ref{tab:surveyssummaries}) to work in \cite{Dusit_FL_survey, li2019federated, li2019federated2, edgeAI,lyu2020threats, kairouz2019advances, khan2020dispersed, FL_Survey_IOT_jour, aledhari2020federated}, we present state-of-the-art advances of federated learning towards enabling IoT applications and devise a taxonomy. Furthermore, the open research challenges in our paper are significantly different than those in \cite{Dusit_FL_survey, li2019federated, li2019federated2, edgeAI,lyu2020threats, kairouz2019advances, khan2020dispersed}. \par 
	The contributions of this tutorial (whose overview is given in Fig.~\ref{fig:roadmap}) are as follows:
	\begin{itemize}
		%\item First, we present the background and motivation of machine learning and IoT-based smart environments. Moreover, we discuss the advantages of these machine learning enabled-smart environments.
		% \item First, we present the key design aspects of federated learning. These key design aspects include resources optimization, incentive mechanism design, and learning algorithm design. 
		\item First, we discuss and critically evaluate the recent federated learning literature, with a focus on the works that are pertinent to IoT applications. To assess these recent works, metrics that serve as key evaluation criteria: Scalability, quantization, robustness, sparsification, security and privacy, are considered. Furthermore, a summary of key contributions and evaluation metrics are presented.     
		\item Second, we derive a taxonomy for federated learning based on different parameters. These parameters include federated optimization schemes, incentive mechanism, security and privacy, operation modes depending on global model aggregation fashion, end-device design, local learning models, resources, miners classification, edge collaboration, and cloud server design.  
		%\item Third, we present two practical use cases of dispersed federated learning that is based on true decentralization to offer robustness and efficient communication resources utilization. 
		\item Third, we present several key open research challenges with their possible solutions. 
		
	\end{itemize}\par
	The rest of the tutorial is organized as follows: Section~\ref{sec:fundamentals} outlines fundamental concepts of federated learning. Recent works of federated learning leading towards IoT applications are presented in Section~\ref{sec:advances}. A taxonomy is devised in Section~\ref{Taxonomy} using various parameters. Finally, open research challenges with possible solutions are presented in Section~\ref{sec:challenges} and paper is concluded in Section~\ref{conclusions}.

	\begin{figure*}[!t]
		\centering
		\captionsetup{justification=centering}
		\includegraphics[width=18cm, height=16cm]{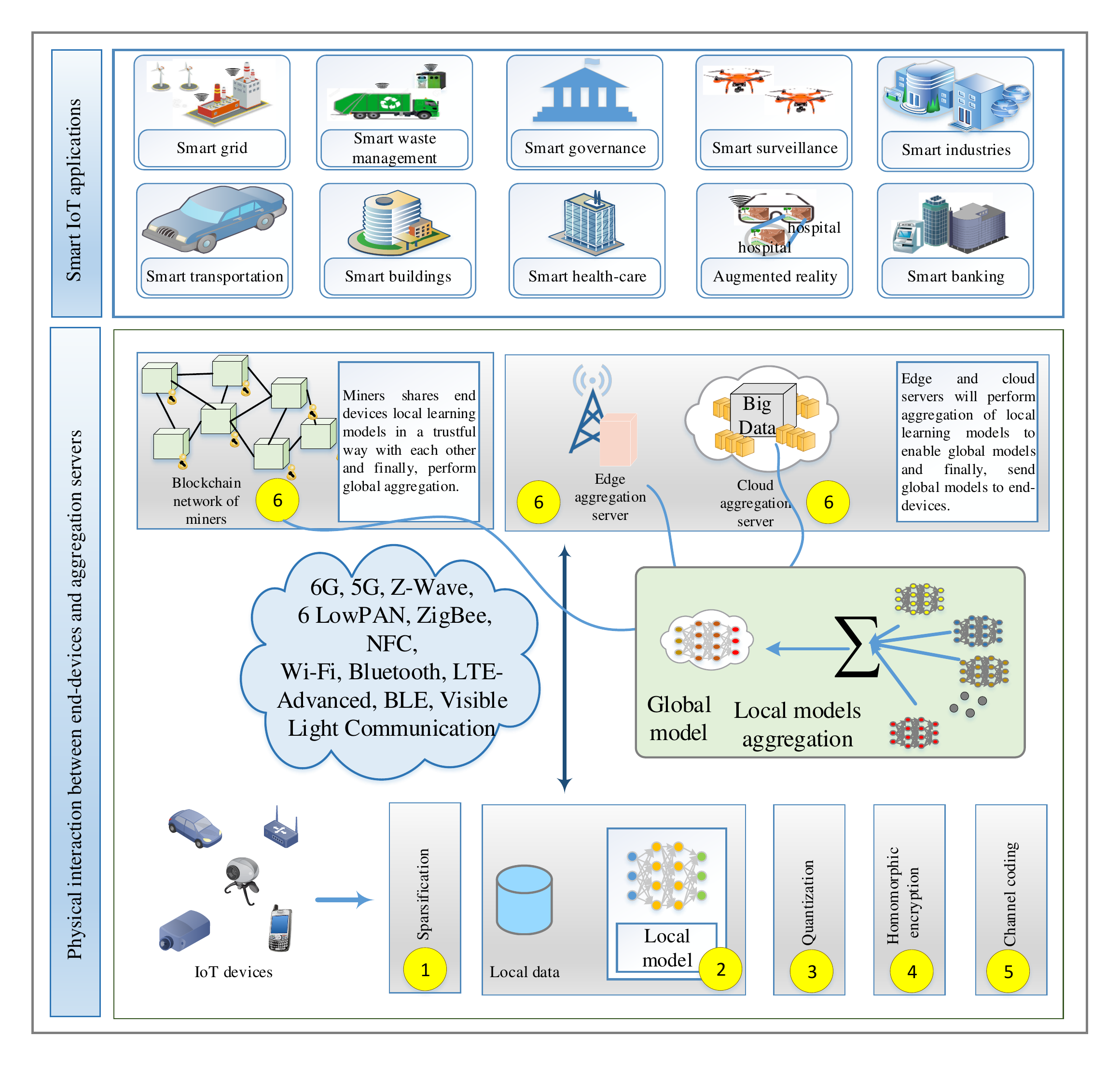}
		\caption{Architectural overview of federated learning for IoT networks.}
		\label{fig:architecture}
	\end{figure*}
	
	\section{Federated Learning for IoT: Fundamental Concepts}
	\label{sec:fundamentals}
	This section presents a high-level architecture, federated optimization schemes, and various frameworks for federated learning-enabled IoT networks. Next, we briefly discuss various aspects of federated learning-enabled IoT networks. \par
	
	\subsection{High-Level Architecture}
	A high-level architecture of federated learning-enabled IoT networks is shown in Fig.~\ref{fig:architecture}. The process starts with selection of a set of end-devices according to a suitable criterion \cite{nishio2019client}. Generally, only a set of devices can participate due to communication resources constraints \cite{khan2020self, industriallatif, khan2020federatedc}. Therefore, we must select devices according to an effective criterion using effective sparsification scheme (step 1). For instance, the criteria can be end-devices with clean datasets, more computational resources, and high-backup power, to enable accurate computation of local learning model within less time \cite{khan2019edge}. To compute the local learning models using local datasets, we can use various schemes at end-devices, such as feed-forward neural networks (FNN), convolutional neural neural networks (CNN), support vector machines (SVM), and long-short term memory (LSTM), among others (step 2). After computation of the local learning model, we can employ effective quantization scheme to reduce the number of bits used for representation of the local learning model updates (step 3). Various quantization schemes for federated learning were introduced in \cite{amiri2020federated,elkordy2020secure,9054168}. In \cite{amiri2020federated}, a lossy federated learning (LFL) scheme based on quantization was proposed. In LFL, both local learning updates and global model updates are quantized by slightly modifying Communication-Efficient stochastic gradient descent (SGD) via Gradient Quantization and Encoding (QSGD) \cite{alistarh2017qsgd} prior to transmission. LFL has shown performance near to federated learning without quantization while reducing communication resources consumption. In another work \cite{elkordy2020secure}, the authors proposed a secure aggregation scheme using heterogeneous quantization for federated learning. \cite{9054168} presented a subtractive dithered lattice quantization-based federated learning scheme. Moreover, dithered hexgonial lattice quantization has shown less compression error compared to dithered scalar quantization, uniform quantization with random unitary rotation, and subsampling with $3$ bits quantizers. Although various lossy quantization schemes of \cite{amiri2020federated,elkordy2020secure,9054168} can offer a tradeoff between accuracy and communication cost for federated learning, loss in accuracy due to a quantization must be further reduced. \par
	Next to quantization, we can use some effective encryption algorithm (step 4). A malicious end-device and aggregation server can infer end-devices sensitive information using their local learning model updates \cite{Dusit_FL_survey,khan2020dispersed}. To avoid this kind of privacy leakage, we can use additive homomorphic encryption in federated learning \cite{xu2019hybridalpha, yang2020fpga}. Although homomorphic encryption will allow aggregation server to perform global model update from encrypted local learning models without the need for decryption, it will add computation and communication cost. The decryption and encryption in case of homomorphic encryption require multiple modular exponentiation and multiplication operations which are complex operations \cite{tran2020implementing}. Moreover, larger ciphertexts are used by homomorphic encryption that will require more communication resources for transmission. To address these issues, the authors in \cite{zhang2020batchcrypt} proposed a BatchCrypt scheme that is based on encrypting a batch of quantized gradients. The batch of quantized gradients will be then encrypted and gradient-wise aggregation takes place at the aggregation server. Specifically, BatchCrypt jointly uses batch encoding, quantization, and analytical quantization modeling for improving computation and communication efficiency. One of the main advantage of BatchCrypt is the ability to preserve the aggregated model quality similar to cross-silo federated learning based on homomorphic encryption.\par
	After quantization and encryption, there is a need to use effective channel coding scheme with low overhead and better performance (step 5). The feasible way can be to use Short Block-Length Codes designed for Ultra-Reliable Low Latency Communications (URLLC) \cite{shirvanimoghaddam2018short, sybis2016channel}. The key design aspects of URLLC codes are latency, reliability, and flexibility. To achieve low latency communication, we should use short packets. Another way to achieve low latency is to use more bandwidth to minimize the latency. Therefore, use of short-block length codes are preferred for URLLC using short packets. Turbo codes, convolutional codes, Bose, Chaudhuri, and Hocquenghem (BCH), and Low-density parity-check (LDPC) codes can be used for packet error rate improvement. However, the complexity associated with Turbo codes makes them difficult to apply to IoT devices with computation resource constraints \cite{8409125}. In \cite{shirvanimoghaddam2018short}, it is shown that BCH codes exhibits highest reliability under optimal decoding among Turbo codes, LDPC codes, polar codes, and convolutional codes. Therefore, one can use BCH codes for federated learning-based IoT networks where ultra-reliable communication is necessary. After channel coding (step $5$), the local learning model updates are transmitted to the aggregation server. \par
	There are many ways to aggregate the local learning models, such as at edge server or cloud server or using miners of a blockchain network \cite{bao2019flchain} and \cite{hua2020blockchain}. Aggregation at the edge and cloud server is simple by using averaging in case of FedAvg \cite{khan2019federated} and then sending back the global learning model updates to end-devices. In case of aggregation via blockchain miners, first of all the miners will receive local model updates from the end-devices. Next, they will share the learning models among themselves and run a blockchain consensus algorithm to update their blocks. After updating the blocks, a global model aggregation takes place at every miner in a distributed manner without the need for centralized aggregation server. Although using blockchain-based miners for global model aggregation in a distributed manner suffers from significant computational complexity due to running a consensus algorithm, it offers the advantage of robustness without the need for a centralized aggregation server. A centralized aggregation server malfunctions due to a security attack or physical damage. Next, we discuss various federated learning schemes depending mainly on various aggregation schemes.

	\subsection{Federated Learning Schemes}
	In this subsection, we present various federated learning schemes. These schemes use different strategies to account for the challenges that are involved in distributed learning systems. The two main challenges in federated learning are:
	\begin{itemize}
		\item System heterogeneity refers to the variations in end-devices features (i.e., computational resource ($CPU-cycles/sec$), backup power, etc.) and wireless channel uncertainties.
		\item Data heterogeneity refers to unbalanced local datasets and non-IID data among end-devices.
	\end{itemize}  
	To show how federated learning model is trained, consider a set  $\mathcal{N}$ of $N$ devices. Every device $n$ has a local dataset $d_{n}$ of $k_{n}$ data points. The goal of federated learning is to minimize the global loss function $f_\textrm{FL}$, i.e., $\underset{\boldsymbol{\omega}}{\text{min}} \frac{1}{K} \sum_{n=1}^{N}\sum_{k=1}^{k_{n}}{f_{k}(\boldsymbol{\omega})}$. Where $\omega$ and $K$ represent the global model weights and total data points of all devices involved in learning, respectively. The nature of the loss function $f_\textrm{FL}$ depends on problem. For instance, prediction error can be the loss function for prediction of a time series, whereas classification error for classification tasks. One way to train a global machine learning model for a set of devices with local dataset is through stochastic gradient descent (SGD) \cite{chen2016revisiting}. However, typically SGD involves one device in training of a local learning model during one communication round that may take long time for convergence. Therefore, one must modify the SGD algorithm for federated setting. In federated SGD, for all end-devices (i.e., fraction of clients $C=1$) and learning rate $\eta$, every client computes $g_{n}=\bigtriangledown F_{n}(\boldsymbol{\omega})$ (where $F_{n}=\frac{1}{K_{n}}\sum_{k=1}^{k_{n}}{f_{k}(\boldsymbol{w})}$ is computed once using their local dataset $d_{n}$). Then all the local models are aggregated and the update $\boldsymbol{\omega}^{t+1} \leftarrow \boldsymbol{\omega}^{t}- \eta \sum_{n=1}^{N}{\frac{k_{n}}{{K}}g_{n}}$ is applied. Although federated SGD can enable distributed machine learning without migrating the end-devices local datasets to the centralized server for training, its performance can be further improve if we perform more local iterations before sending the local model updates to an aggregation sever. Overview of FedAvg is given in Algorithm~\ref{algo:FedAVg} that uses multiple local iterations at devices before sending model updates to the aggregation server\cite{mcmahan2017communication}. \par
	\begin{algorithm}[t!]
		
		%\scriptsize
		%	\SetAlgoLined
		\caption{FedAvg \cite{mcmahan2017communication}}
		\label{algo:FedAVg}
		
		% % % % % % % % % % % % % % % % % % % % %
		\begin{algorithmic}[1]
			\State \textbf{\emph{Aggregation Server}}
			\State Weights initialization $\boldsymbol{\omega}^{0}$
			\For {t=0, 1,..., Global Rounds-1}
			\State \text{Select $\mathcal{N}_{s} \leftarrow m=max(C.N)$ clients randomly.}
			\For {For every device $n \in \mathcal{N}_{s}$,  parallel run.}
			\State $\omega_{n}^{t+1} \leftarrow DeviceUpdate (n, \omega_{n}^{t})$
			\State $\boldsymbol{\omega}^{t+1} \leftarrow  \sum_{n=1}^{N_{s}}{\frac{k_{n}}{{K}}\omega_{n}^{t+1}}$
			\EndFor
			\EndFor 
			\State \emph{\textbf{DeviceUpdate}(n,$\omega$)}
			\State \text{$\mathcal{B} \leftarrow$ Split $d_{k}$ into batches.}
			\For {e=0, 1, .. , Local iterations-1}
			\For{$b \in \mathcal{B}$}
			\State $\omega \leftarrow \omega - \eta \bigtriangledown l(\omega, b)$
			\EndFor
			\EndFor 
		\end{algorithmic}
	\end{algorithm} \par
	
	Although the FedAvg scheme \cite{mcmahan2017communication} was the first work to enable distributed learning in environments with main focus on heterogeneous systems parameters, and data distributions, it seems difficult to provably get convergence \cite{kairouz2019advances}. This is because FedAvg is based on a simple averaging of local model updates from the end-devices at aggregation server that might not produce better results for heterogeneous environments. The work in \cite{mcmahan2017communication} has shown that hyper parameters (i.e., local iterations, global iterations, learning rate, etc.) play an important role in convergence of FedAvg. However, for high number of local iterations, the dissimilar local objectives might lead end-devices toward their local optima rather than the global one. Therefore, we can say that FedAvg seems difficult to provably converge for heterogeneous scenarios. Furthermore, setting higher local iterations may cause some of the devices not to complete computing their local learning models within maximum allowed time. Therefore, these devices will not be able to participate in learning process for the given communication round. To cope with these limitations, FedProx was introduced in \cite{li2018federated} that is based on adding a proximal term to local model loss function term in FedAvg.
	\begin{equation}
		\label{eq:0}
		\begin{aligned}
			\underset{\boldsymbol{\omega}}{\text{min}}~h_{n}(\omega;\omega^{t})= F_{n}(\omega)+\frac{\mu}{2} \parallel \omega -\omega^{t} \parallel^{2}
		\end{aligned}
	\end{equation}\newline
	In \eqref{eq:0}, the proximal term offers the advantage of better addressing the statistical heterogeneity by restricting the local updates more to a global model, and thus devices heterogeneous data impact will be minimized on the local model. Both FedAvg and FedProx tried to effectively consider the effect of heterogeneous system parameters (i.e., local computational resource) and data heterogeneity (i.e., unbalanced datasets, non-independent and identical distributions), they did not study the effect of wireless channel uncertainties on the performance federated learning. For enabling federated learning over IoT networks, we must carefully handle loss in performance of federated learning due to wireless channel uncertainties. Chen \textit{et al.} in \cite{chen2019joint} studied the effect of wireless channel uncertainties on the performance of federated learning. They presented a framework for joint learning and communication. Moreover, they formulated a problem to minimize a federated learning cost by jointly optimizing power allocation, resource block allocation, and user selection. In another work, Tran \textit{et al.} analyzed the performance of federated learning over wireless networks \cite{tran2019federated}. An optimization problem to minimize both energy consumption and computation time was formulated. Moreover, they provided extensive numerical results to analyze the performance of federated learning over wireless networks.        
	
	\subsection{Local Learning Models}
	\label{Local Learning Models}
	We present various local learning models used for computing local learning models in federated learning. The type of local learning model used strictly depends on the IoT application considered. CNN, FNN, LSTM, and SVM, among others can be used for local model computation.  Table~\ref{tab:locallearningmodels} lists various machine learning algorithms for IoT applications \cite{ham2014linear, yadav2019deep, ndikumana2020deep,du2017stacked, thar2018deepmec, singh2020machine, hodo2016threat, 8474384, liu2020privacy, hewamalage2019recurrent, htwe2020botnets, bhatia2020fog}. Typically, a neural network has an input layer, hidden layers, and the output layer. Increasing the number of hidden layers generally improves the performance but at the cost of computational complexity. On the other hand, the selection of a particular local learning model strictly depends on the application. For instance, CNN is normally considered for IoT tasks that are based on image classification. One can also consider simple fully connected FNN with low complexity for image classification. However, the low complexity local learning model might not perform very well. Coping with this issue, CNN with sufficient layers should be preferably used. It must be noted here that increasing the number of hidden layers does not always guarantee performance improvement \cite{kairouz2019advances}. Therefore, we must properly choose network size. Furthermore, activation functions (i.e., determines output of a neural network) must be properly used in various neural networks. Activation functions can be broadly categorized into: binary step functions, linear activation functions, and non-linear activation functions. Within non-linear activation functions, there can be sigmoid, hyperbolic tangent (tanh), rectified linear unit (ReLU), soft-max, among others \cite{activation_functions}. A summary of activation functions that can be used in various local learning model is given in Table~\ref{tab:activationlocallearningmodels}. For federated learning, selection of a local learning model with low complexity and better performance is of significant interest. In contrast to the centralized machine learning, federated learning involves a massive number of end-devices with lower computational capacity than the centralized server used by centralized machine learning. Furthermore, there are energy limitations of the IoT end-devices. Keeping in the mind the aforementioned points, there is a need to use efficient local learning model for federated learning. To effectively choose the neural network size, one can use neural architecture search (NAS) to find the optimal architecture out of possible available architectures \cite{xie2018snas}. More details on how NAS can be employed for local learning models at the end-devices will be given in Section~\ref{device_design}.

	\begin{table}[]
		%\rowcolors{2}{gray!25}{white}
		\caption {Local learning models used for IoT applications.} \label{tab:locallearningmodels} 
		\begin{center}
			\begin{tabular}{p{2 cm}p{1.2cm}p{4cm}}
				\toprule
				
				\textbf{Local learning model} & \textbf{Reference}&  \textbf{Primary IoT application} \\ \midrule
				SVM & \cite{singh2020machine}  & Network slicing.\\
				\cline{2-3}
				& \cite{ham2014linear}    &  Malware detection.
				\\ \midrule
				
				CNN & \cite{yadav2019deep} & Disease diagnosis based on image classification. \\ \cline{2-3}
				&  \cite{ndikumana2020deep}  & Caching for infotainment-enabled smart cars.  \\ \midrule

				LSTM &  \cite{du2017stacked} & Traffic prediction in Vehicle-to-Vehicle Communication.\\ \cline{2-3}
				
				&  \cite{thar2018deepmec} & Edge caching.\\ \midrule
				
				FNN & \cite{hodo2016threat}  & Distributed denial of service attacks detectio.n \\ \midrule
				
				K-means &\cite{8474384}
				& Clustering of sensor networks to minimize packet error rate. \\ \cline{2-3}
				
				&  \cite{liu2020privacy}
				& Proactive caching for IoT networks. \\ \midrule

				RNN &\cite{hewamalage2019recurrent}  & Time series forecasting.\\ \midrule
				
				Naive Bayes &\cite{htwe2020botnets}  & Botnets attack detection for IoT.   \\ \cline{2-3} 
				
				&\cite{bhatia2020fog}  & Anomaly detection for IoT-Fog-Cloud computing model. \\ 
				
				\bottomrule 
			\end{tabular}
		\end{center}
	\end{table} 
	
	\begin{table*}[]
		%\rowcolors{2}{gray!25}{white}
		\caption {Summary of commonly used activation functions for local learning models.} \label{tab:activationlocallearningmodels} 
		\begin{center}
			\begin{tabular}{p{2.3 cm}p{3cm}p{3.5cm}p{3.5cm}p{3.5cm}}
				\toprule
				
				\textbf{Activation function} & \textbf{Equation}& \textbf{Derivative}&  \textbf{Diagram} &  \textbf{Limitation} \\ \midrule
				
				\multirow{4}{*}{Unit step function} &  $  f(x) =\begin{cases} 
					0 & \text{if $x<0$} \\
					0.5 & \text{if $x=0$} \\
					1 & \text{if $x>0$} \\
				\end{cases}$ & $f^{'}(x)=\delta(t)$  & \raisebox{-.5\totalheight}{\includegraphics[width=0.08\textwidth]{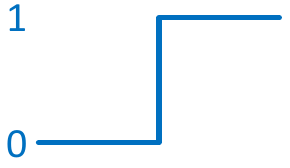}} & Does not allow multi-value outputs (i.e., classification of inputs into several categories). \\ \midrule
				
				Linear function & $f(x)=x$ & $f^{'}(x)=1$ & \raisebox{-.5\totalheight}{\includegraphics[width=0.08\textwidth]{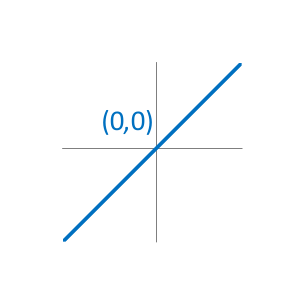}} & The last layer will be the linear sum of previous layers independent of the number of layers are in a network. Therefore, a neural network with a linear activation function is simply a linear regression model. \\ \midrule
				
				Sigmoid function &  $f(x)=\frac{1}{1+e^{-x}}$  & $f^{'}(x)=f(x)(1-f(x)) $&\raisebox{-.5\totalheight}{\includegraphics[width=0.12\textwidth]{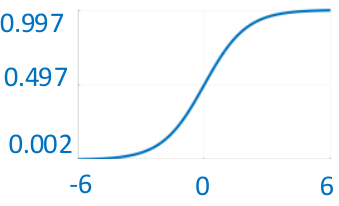}} & The change in prediction is not significant for lower and higher values of input. Therefore, it might cause a vanishing gradient problem. Moreover output is not centered at zero and it is computationally expensive. \\ \midrule
				
				Hyperbolic tangent &$f(x)=\frac{e^{x}-e^{-x}}{e^{x}+e^{-x}}$  & $f^{'}(x)=1-f^{2}(x)$ &\raisebox{-.5\totalheight}{\includegraphics[width=0.12\textwidth]{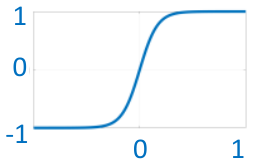}} &  Similar to sigmoid, the change in prediction is not significant for lower and higher values of input. Additionally, it is computationally expensive. \\ \midrule
				
				Rectified linear unit & $  f(x) =\begin{cases} 
					x & \text{if $x\geq0$} \\
					0 & \text{if $x<0$}\\
				\end{cases}$  & $f^{'}(x)=\begin{cases} 
					1 & \text{if $x\geq0$} \\
					0 & \text{if $x<0$}\\
				\end{cases}$ &\raisebox{-.5\totalheight}{\includegraphics[width=0.07\textwidth]{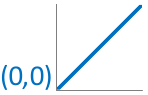}} & For zero or less than zero inputs, the function becomes zero and does not learn.  \\ 
				\bottomrule 
			\end{tabular}
		\end{center}
	\end{table*}

	Next to selection of the local learning model, there must be effective technique to optimize the local model weights update to enable faster convergence of the federated learning model. Mostly, SGD was considered for local model weights updating \cite{kairouz2019advances}. In SGD, the weights are updated at end-devices using partial derivative of loss function w.r.t weights, which are then multiplied by the learning rate. The learning rate represents the step size of gradient descent. The mini-batch SGD is given by.
	\begin{equation}
		\label{eq:1}
		\begin{aligned}
			\omega=\omega - \eta\frac{\partial F}{\partial\omega_{t}}
		\end{aligned}
	\end{equation}\newline
	\begin{equation}
		\label{eq:1}
		\begin{aligned}
			\frac{\partial F}{\partial\omega}\approx \frac{1}{m}\sum_{i \in \mathcal{B}}\frac{\partial f_{i}}{\partial\omega},
		\end{aligned}
	\end{equation}\\
	where $m$ denotes the number of elements in a batch. The batch size significantly affects the performance global federated learning model. For FedAvg, using client fraction $C=0.1$ and batch size $10$ generally have shown better results compared to full batch size for MNIST dataset \cite{mcmahan2017communication}. Other than client fraction $C$ and batch size, learning rate significantly affects the performance of the federated learning global model. Therefore, we must wisely chose the client fraction $C$, batch size, and learning rate.

	\subsection{Frameworks}
	\label{Frameworks}
	In this subsection, we discuss various frameworks developed for implementing federated learning over IoT networks \cite{he2020fedml, Dusit_FL_survey}:   
	\par
	\begin{itemize}
		\item \textbf{PySyft:} PySyft is a python framework that can be used to implement secure and private deep learning \cite{ryffel2018generic}. End-devices private data is decoupled during the training process by using Encrypted Computation (like Multi-Party Computation (MPC) and Homomorphic Encryption (HE)), differential privacy, and federated learning. PySyft is primarily based on Pytorch and it retains native Torch interface \cite{Dusit_FL_survey}. For simulation of federated learning using PySyft, end-devices are created as virtual workers. The data is divided into virtual workers. Next, a PointerTensor is used for specification of data storage location and owner. For global federated learning model aggregation, local model updates can be fetched from virtual workers.
		\item \textbf{FedML:} FedML is an open research framework to facilitate development of federated learning \cite{he2020fedml}. FedML has two main components, such as FedML-core and FedML-API. FedML-core performs separation of model training and distributed communication into two modules, such as distributed communication module and training module. The responsibility of distributed communication module is to carry out low-level communication among various clients, whereas training module is implemented using PyTorch. New algorithms in FedML can be easily implemented using client-oriented programming interface. The primary advantage of FedML is a TopologyManager that can provide a support for many network topologies to implement various federated learning algorithms.    
		
		\item \textbf{TensorFlow Federated:} TensorFlow Federated (TFF) is an open source framework developed by Google for performing experiments on distributed datasets using machine learning and other computations \cite{ingerman2019introducing}. TFF interfaces can be divided into two types, such as federated core API, and federated learning API. Federated learning API provides high-level interfaces set that enable us to use the existing TensorFlow models. Federated core API is set of lower-level interfaces that can enable new federated learning algorithms using TensorFlow and distributed communication operations. 
		\item \textbf{LEAF:} LEAF is a framework for federated settings that can support various applications, such as meta-learning, multi-task learning, and federated learning \cite{caldas2018leaf}. LEAF consists of three parts, such as open-source datasets suit, system and statistical metrics array, and reference implementations set. LEAF offers the modular design that has the advantage of easier implementation of diverse experimental scenarios. The datasets suit has Federated Extended MNIST, Sentiment140, Shakespeare, CelebA, Reddit, and synthetic dataset that is based on modification of synthetic dataset of \cite{li2019fair} for making it more challenging for meta-learning methods. Initial implementations of LEAF are Mocha \cite{bagdasaryan2020backdoor}, FedAvg \cite{mcmahan2017communication}, and minibatch SGD. 
		
		\item \textbf{Paddle FL:} Paddle FL is an open source framework developed mainly for industry applications \cite{ma2019paddlepaddle}. Paddle FL can easily replicate various federated learning algorithms for large scale distributed clusters. Paddle FL will offer federated learning implementation in natural language processing, computer vision, and so on. Furthermore, support for implementation of transfer learning and multi-task learning in federated learning will be provided \cite{Paddle}. Using elastic scheduling of training job on Kubernetes and large scale distributed training of PaddlePaddle's, paddle FL can be easily deployed on full-stack open sourced software.    
	\end{itemize}\par
	All of the above-mentioned frameworks present implementation of various federated learning schemes, but they did not effectively considered the effect of wireless channel uncertainties and limited communication resources. IoT applications will suffer from significant performance degradation due to wireless channel uncertainties. Therefore, we should propose novel federated learning frameworks for IoT networks that will enable us to effectively handle wireless channel uncertainties. Furthermore, we can use channel coding schemes to improve the performance of the federated learning over IoT networks. Therefore, the framework for federated learning over IoT networks must provide support for channel coding schemes.    
	
	% % % % % % % % % % % % % % % % % % % % % % % % % %
	\begin{table*}[]
		\caption {Evaluation metrics: explanation, dimension, and advantages.} \label{tab:performance_parameters} 
		\begin{center}
			\begin{tabular}{p{3cm}p{6cm}p{6cm}}
				\toprule
				\textbf{Metric}    & \textbf{Explanation}  &  \textbf{Benefits for federated learning} \\ \midrule
				\textbf{$P_1$: }\textbf{Security and privacy}  & This metric deals with malicious user attacks during learning model parameters transmission between end-devices and the aggregation server. Moreover, the malicious aggregation server and end-device can infer other end-devices sensitive information from their local learning model updates. Therefore, we must propose federated learning schemes that offer both security and privacy.& \begin{itemize} \item Users' privacy protection. \item Trustful verification of learning model updates. \item Secure exchange of learning model updates.  \end{itemize}\\ \midrule
				\textbf{$P_2$: }\textbf{Scalability}   & Scalability captures to the ability of a federated learning system to incorporate more users during the training process for better accuracy.   & \begin{itemize} \item High federated learning accuracy. \item Massive connectivity during federated learning process. \item Better accuracy due to participation of more users. \end{itemize} \\ \midrule
				\textbf{$P_3$: }\textbf{Quantization}  & Quantization refers to the need for minimizing the size of local learning model updates to reduce federated learning convergence time. & \begin{itemize} \item Fast federated learning convergence. \item Better accuracy due to participation of more users for fix communication resources. \end{itemize}  \\ \midrule
				\textbf{$P_4$: }\textbf{Robustness}  & This refers to the ability of a federated learning algorithm to perform the learning process successfully in face of a possible malfunction or failure of the edge/cloud server.  &  \begin{itemize} \item Cost optimization. \item Accurate federated learning model. \end{itemize}\\ \midrule
				%\textbf{$P_5$: }\textbf{Context-awareness}  &    It deals with  & \begin{itemize}  \item  \end{itemize}  &  \begin{itemize} \item \end{itemize}  \\ 
				%\midrule
				\textbf{$P_5$: }\textbf{Sparsification}  & Sparsification refers to the selection of most suitable devices among a set of massive numbers of devices according to a specific criteria.  &  \begin{itemize} \item Lower federated learning convergence time. \item High federated learning accuracy. \end{itemize}\\ 
				\bottomrule
			\end{tabular}
		\end{center}
	\end{table*}
	
	\section{State-Of-The-Art}
	\label{sec:advances}
	This section discusses the recent works in federated learning towards enabling IoT-based smart applications. We critically investigate and rigorously evaluate these recent works, whose summary is given in Tables \ref{tab:recent_advances_evaluation1} and \ref{tab:recent_advances_evaluation2}. To rigorously evaluate the recent works, we derive several key metrics from the literature \cite{wang2019adaptive, edgeAI, nishio2019client, feraudo2020colearn, zhao2020federated, qu2020decentralized, liu2019lifelong, yu2019federated, kang2020reliable, conway2019demonstration, savazzi2020federated, chen2020fedhealth, chen2020federated,ye2020federated,  du2020federated} as follows.
	\begin{itemize}
		\item \textbf{Security and privacy:} Although federated learning was developed to preserve the users' privacy, it still faces privacy challenges. For instance, the edge/cloud server can infer the end-devices' sensitive information using their learning model updates, and thus causes the end-device privacy leakage. Furthermore, the malicious user can infer the devices' sensitive information from their local learning models during their exchange with the aggregation server. Beyond inferring the end-devices' sensitive information, the malicious user can alter the learning model updates, and thus prolong the convergence time of a federated learning model. On the other hand, an unauthorized user can access the edge/cloud server and perform false data injection. Therefore, it is necessary to enable federated learning with security and privacy.    
		
		\item \textbf{Scalability:} For a limited communication resource, scalability refers to the ability to incorporate more devices in the federated learning process. Generally, incorporating more devices in federated learning can lead to more accurate results. Scalability in federated learning can be enabled via different schemes such as resource optimization, selection of high-computational power devices, and compression schemes for learning model parameters transmission. An efficient resource optimization scheme for fixed communication resources can enable more devices to participate in the federated learning process, and thus offers better performance. Typically, a set of devices can participate in the federated learning process \cite{Dusit_FL_survey}. Therefore, for synchronous federated learning algorithms that allow fixed local model computation time for all the participating devices, selecting the devices' high-computation power (CPU-cycles/sec) will enable more devices to participate in the federated learning process. %Another way to increase the participation of devices in federated learning process is to consider effective compression schemes for learning model parameters transmission.         
		
		\item \textbf{Quantization:} The concept of quantization refers to schemes used to reduce the size of the local learning model updates. Minimizing the number of local model updates size results in high throughput, which subsequently minimizes the federated learning convergence time.  
		
		\item \textbf{Robustness:} This deals with the ability of the system to successfully carry out the process of federated learning during edge/cloud server failure. Traditional federated learning based on a centralized edge/cloud server for global aggregation suffers from interruption if the aggregation server stops working due to malfunctioning.  
		
		%\cite{hancox2020robustness}, for more information.
		
		%\item \textbf{Context-awareness:}

		\item \textbf{Sparsification:} This metric deals with the selection of a set of most suitable devices under communication resource constraints for federated learning. Under communication resource constraints, only a subset of a massive number of devices will be allowed to participate in federated learning. The selection criteria for suitable devices can be less noisy datasets, large size datasets, and data uniqueness. Selecting the most suitable device set results in low convergence time for federated learning. Different from quantization-based schemes that reduce the size of local learning model parameters for all devices before sending to the aggregation server, sparsification-based schemes select only a set of most relevant devices according to a particular criterion.     
	\end{itemize}\par

\begin{figure*}[!t]
		\centering
		\captionsetup{justification=centering}
		\includegraphics[width=18cm, height=7cm]{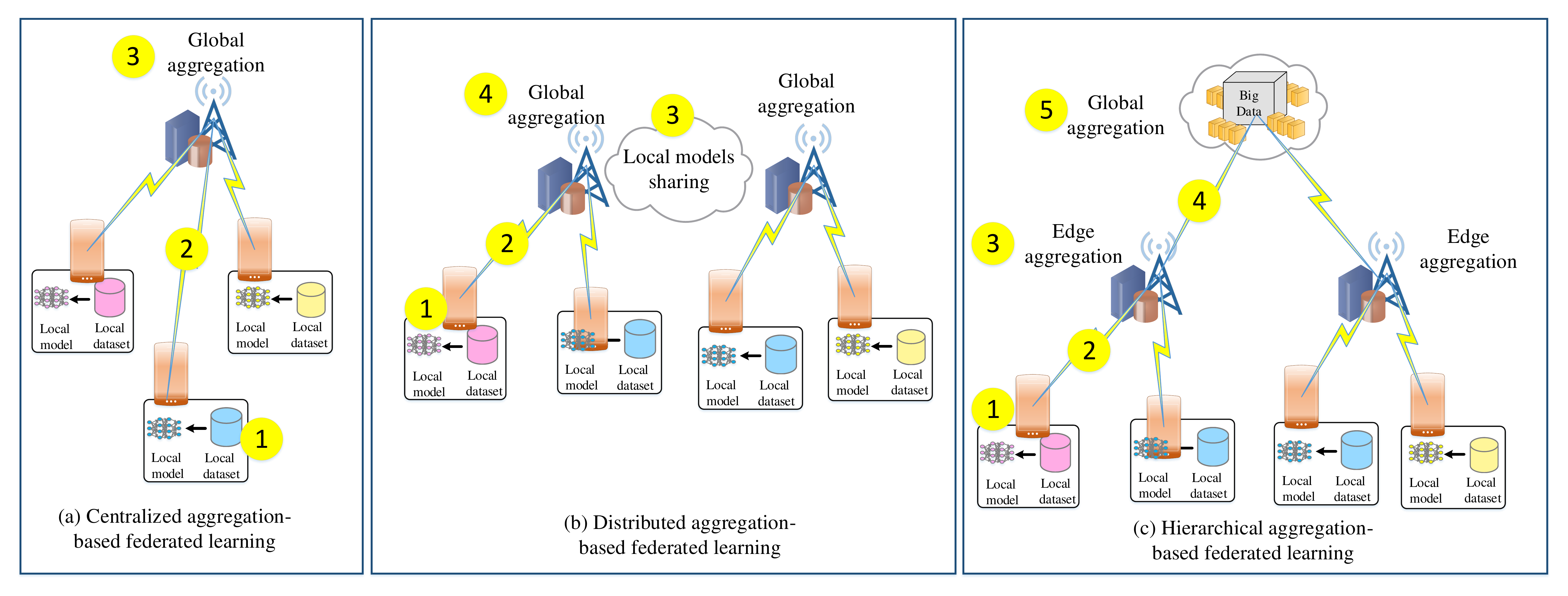}
		\caption{Federated learning based on: (a) centralized aggregation, (b) distributed aggregation, and (c) hierarchical aggregation.}
		\label{fig:advances_comparison}
	\end{figure*}

	%%%%%%%%%%%%%%%%%%%%%%%%%%%%%%%%%%%%%%%%%%%%%%%%%%%%%%%%%%%%%%%%%%%%%%%%
	A summary of the evaluation metrics with advantages is given in Table~\ref{tab:performance_parameters}. We use (\cmark) if the evaluation metric is fulfilled by recent advances and (\xmark) otherwise. A summary of these recent works is given in Table~\ref{tab:detailsadvances}. Furthermore, Table~\ref{tab:recent_advances_evaluation1} and Table~\ref{tab:recent_advances_evaluation2} evaluate these works using precise metrics (discussed later). We categorize the recent advances mainly into advances based on centralized aggregation, advances based on distributed aggregation, and advances based on hierarchical aggregation. Fig.~\ref{fig:advances_comparison} shows the overview of various aggregation fashion for federated learning. In centralized aggregation, a single edge or cloud server acts for the aggregation of local learning models of all devices. Distributed aggregation involves multiple aggregation servers that receive local learning model updates from their associated devices. The received local learning models from various devices are shared among various aggregation servers which are followed by global aggregation. On the other hand, hierarchical aggregation involves aggregations (e.g., at edge servers) before a central global aggregation takes place. Next, we present various advances of federated learning for IoT networks. \par

	\subsection{Advances Based on Centralized Aggregation}
	\label{Advances Based on Centralized Aggregation}
\begin{figure}[!t]
		\centering
		\captionsetup{justification=centering}
		\includegraphics[width=8cm, height=9cm]{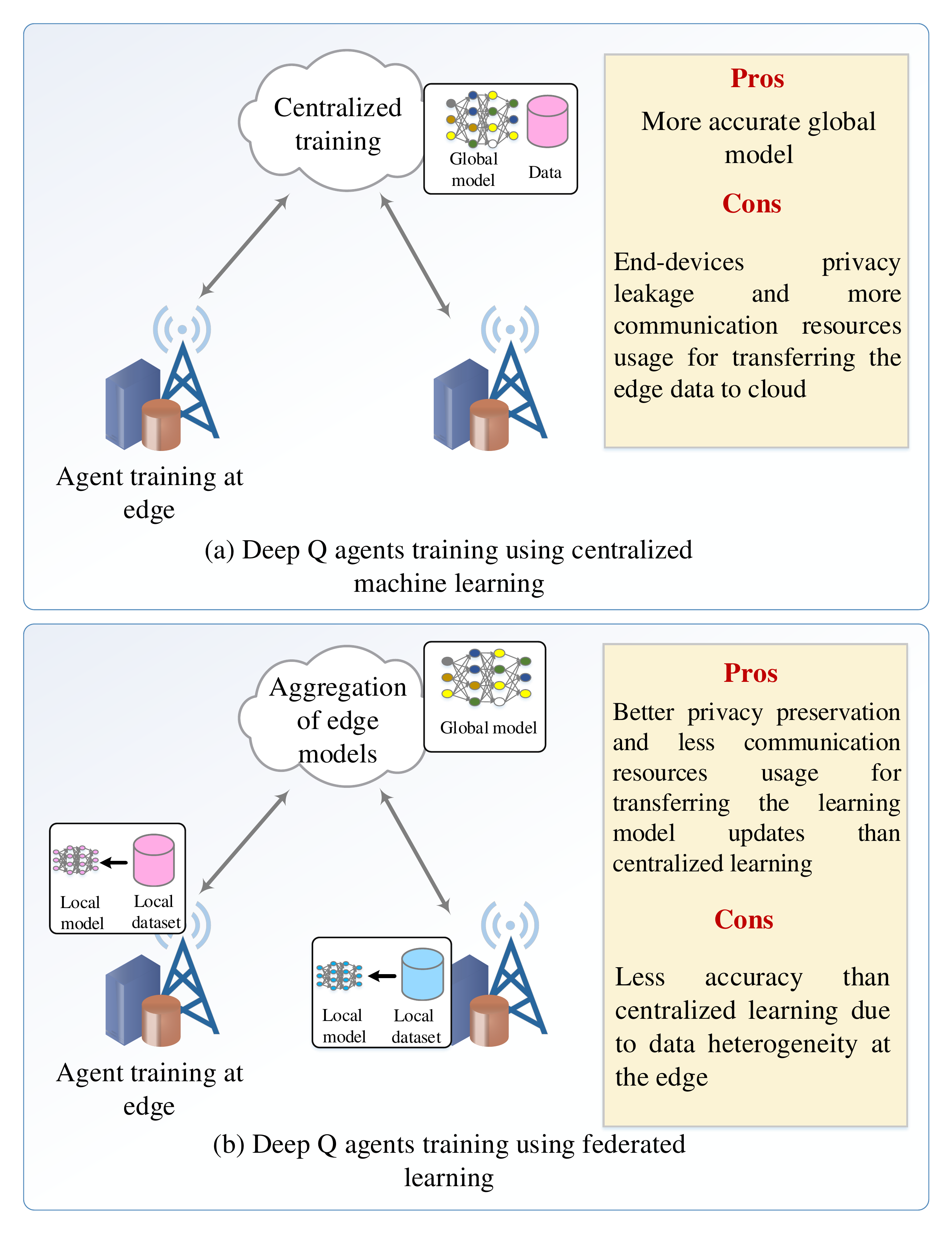}
		\caption{Deep Q-Learning agents training using (a) centralized training, (b) federated learning \cite{edgeAI}.}
		\label{fig:In_EDGE_AI}
	\end{figure}

	Many works such as \cite{wang2019adaptive, edgeAI,nishio2019client, feraudo2020colearn, yu2019federated, kang2020reliable, conway2019demonstration, chen2019joint, chen2020fedhealth}, and \cite{ye2020federated} considered federated learning based on a centralized aggregation of end-devices local learning models for various IoT applications. In all these works, end-devices compute their local learning models and send the learning model updates to a centralized aggregation server for computing the global federated learning model. The centralized aggregation server can be either located at the network edge or a remote cloud. In \cite{edgeAI}, Wang \textit{et al.} proposed an intelligent framework as shown in Fig.~\ref{fig:In_EDGE_AI}, namely, In-Edge-AI for edge computing-enabled IoT networks. The authors provided a detailed discussion on how to use Deep Q-Learning agents for two tasks, computation offloading and edge caching. Different use cases of deep Q-learning agent deployment were discussed with their pros and cons. Motivated by the notion of unbalanced, non-IID data, and limited communication resources, federated learning was proposed for the training of deep Q-learning agents. Simulation results showed that federated learning has significantly reduced the training cost for the proposed In-Edge-AI framework. Finally, the authors presented open research challenges regarding the practical deployment of their framework. Although the authors in \cite{edgeAI} proposed a detailed framework regarding the implementation of Deep Q-Learning agents based on federated learning, the authors did not properly addressed the issues of security and privacy in their framework. A malicious end-device or an aggregation server can easily infer the end-devices sensitive information from local learning models. Therefore, we must tackle the issues of security and privacy for the In-Edge AI framework. Additionally, Deep Q-Learning agents based on federated learning result in slight loss of performance compared to Deep Q-Learning agents using traditional, centralized machine learning. One possible reason for this performance degradation can be the simple averaging of local learning models. There exist significant variations in the wireless channel and local learning models that must be taken into account while averaging the learning model updates from the edge servers where agents are deployed. One way can be weighted averaging by assigning higher weights to poorly performing nodes so as to minimize the impact of better performing devices on the global model. In another work \cite{wang2019adaptive}, the authors also considered federated learning at the network edge. They analyzed the convergence bound for federated learning using gradient descent. A novel theoretical convergence bound was derived that considered non-IID nature of the data. Furthermore, based on the derived convergence bound, a control algorithm was proposed that minimizes learning loss for a fixed resource budget. The control algorithm performed learning loss minimization by learning model characteristics, system dynamics, and data distribution. The performance of the proposed control algorithm was evaluated for real datasets using hardware prototype and simulations results. Although the work in \cite{wang2019adaptive} presented extensive theoretical convergence analysis for convex loss functions, it does not provide analysis for non-convex loss functions which is more practical for various IoT learning tasks. Additionally, efficient management of resources such as local computation resources and wireless resources must be performed for federated learning over edge networks. \par
	In \cite{chen2019joint}, a framework for joint learning and communication for wireless federated learning was presented. First, the authors extensively studied the performance of federated learning for wireless networks. Then, joint wireless resource allocation, power allocation, and user selection optimization problem was formulated for minimizing the federated learning loss function. For the expected convergence rate of federated learning, a closed-form expression was derived to account for the wireless factors on learning performance. Using the derived closed-form expression, optimal transmit power allocation was performed for a fixed resource block allocation and user selection. Next, resource block allocation and user selection were performed to minimize the federated learning loss function. The framework of \cite{chen2019joint} can offer scalability by enabling the participation of more users using efficient resource allocation. Moreover, it can offer sparsification by enabling the participation of suitable devices. Although the proposed framework offered several advantages, its performance against the security attacks can be further enhanced by using encryption schemes. Beyond wireless security, the centralized aggregation server might fail due to physical damage, and thus can be suffered from robustness issues.\par
\begin{figure}[!t]
		\centering
		\captionsetup{justification=centering}
		\includegraphics[width=8cm, height=7.5cm]{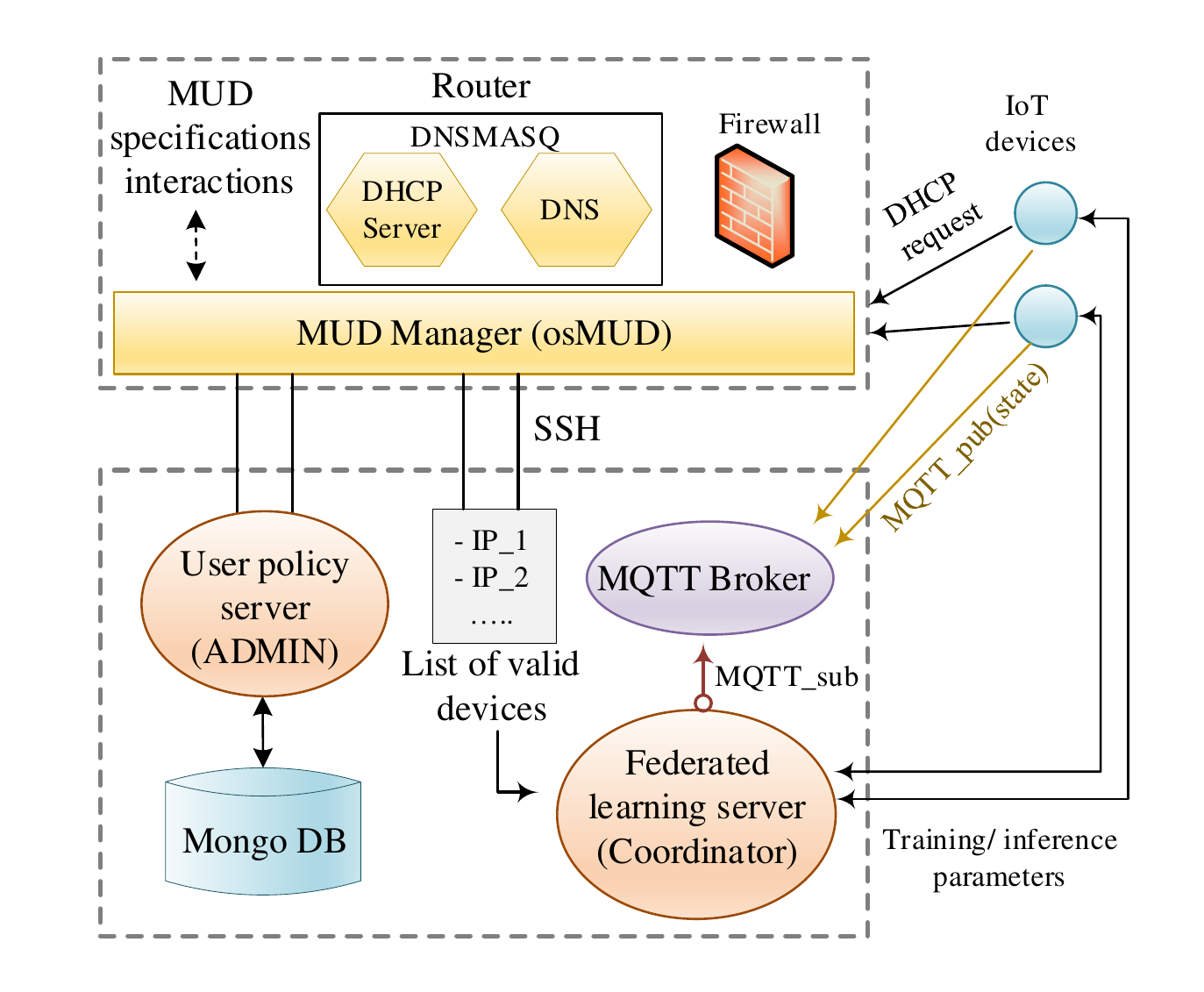}
		\caption{CoLearn framework \cite{feraudo2020colearn}.}
		\label{fig:CoLearn}
	\end{figure}

	Beyond the theoretical works on the deployment of federated learning at the network edge in \cite{wang2019adaptive, edgeAI, chen2019joint}, the work in \cite{feraudo2020colearn} proposed an architecture, namely, CoLearn for federated learning at edge networks. CoLearn was proposed to address the challenges of malicious devices' presence and asynchronous participation of resource-constrained IoT devices \cite{feraudo2020colearn}. CoLearn uses federated learning framework \emph{PySyft} and open-source Manufacturer Usage Description (MUD). PySyft is a python framework that can be used to implement federated learning \cite{ryffel2018generic}. PySyft uses Pytorch while retaining native Torch interface. Fig.~\ref{fig:CoLearn} shows the interaction between different components of CoLearn architecture. The CoLearn mainly consists of MUD manager and user policy server, federated learning server, and IoT devices. The IoT devices transmit their MUD-URLs in their DHCP requests to the router. The MUD manager sends valid IoT devices to the federated learning coordinator to minimize the risk of attack. Moreover, the federated learning coordinator performs interaction with devices using WebSockets and MQTT. To enable more flexibility in the design, the user policy server enables a network administrators to enforce rules other than defined by the manufacturer. The proposed architecture showed significant advantages, its performance can be further improved by applying light-weight authentication algorithms at the IoT devices. Another study in \cite{conway2019demonstration} demonstrated federated learning over emulated wide-area communications network having features of intermittent, heterogeneous, and dynamic availability of resources. The network is emulated using CORE/EMANE and consists of both mobile and fixed nodes. Any node can be selected as a server that initiates the process of federated learning. The specifications i.e., as the data type, expected prediction labels, and model architecture are broadcasted to all nodes during the task initialization phase. All the nodes wait for a certain time after the task initialization is done and then start the federated learning process. Any node in a network can leave or enter the learning process. The outdated local learning model parameters of the nodes are ignored and are not considered in the federated learning process. The proposed emulated federated learning environment has advantages of considering intermittent and dynamic behavior of the network resources that make it more feasible for practical applications. In \cite{conway2019demonstration}, the authors provided an implementation of asynchronous federated learning for a scenario with both mobile and static nodes. However, a random selection of any node to act as an aggregation server might not always result in better performance. There must be an effective criterion for the selection of nodes to acts as a server. One possible criterion can be a selection of a static node as a server. This will enable higher connectivity to other static nodes compared to mobile nodes. However, this approach might not yield promising results when the number of mobile nodes is significantly greater than the static nodes. Therefore, we must propose effective criteria for the selection of nodes as an aggregation server that will result in higher connectivity with other devices involved in federated learning. \par

	\begin{figure}[!t]
		\centering
		\captionsetup{justification=centering}
		\includegraphics[width=8cm, height=9cm]{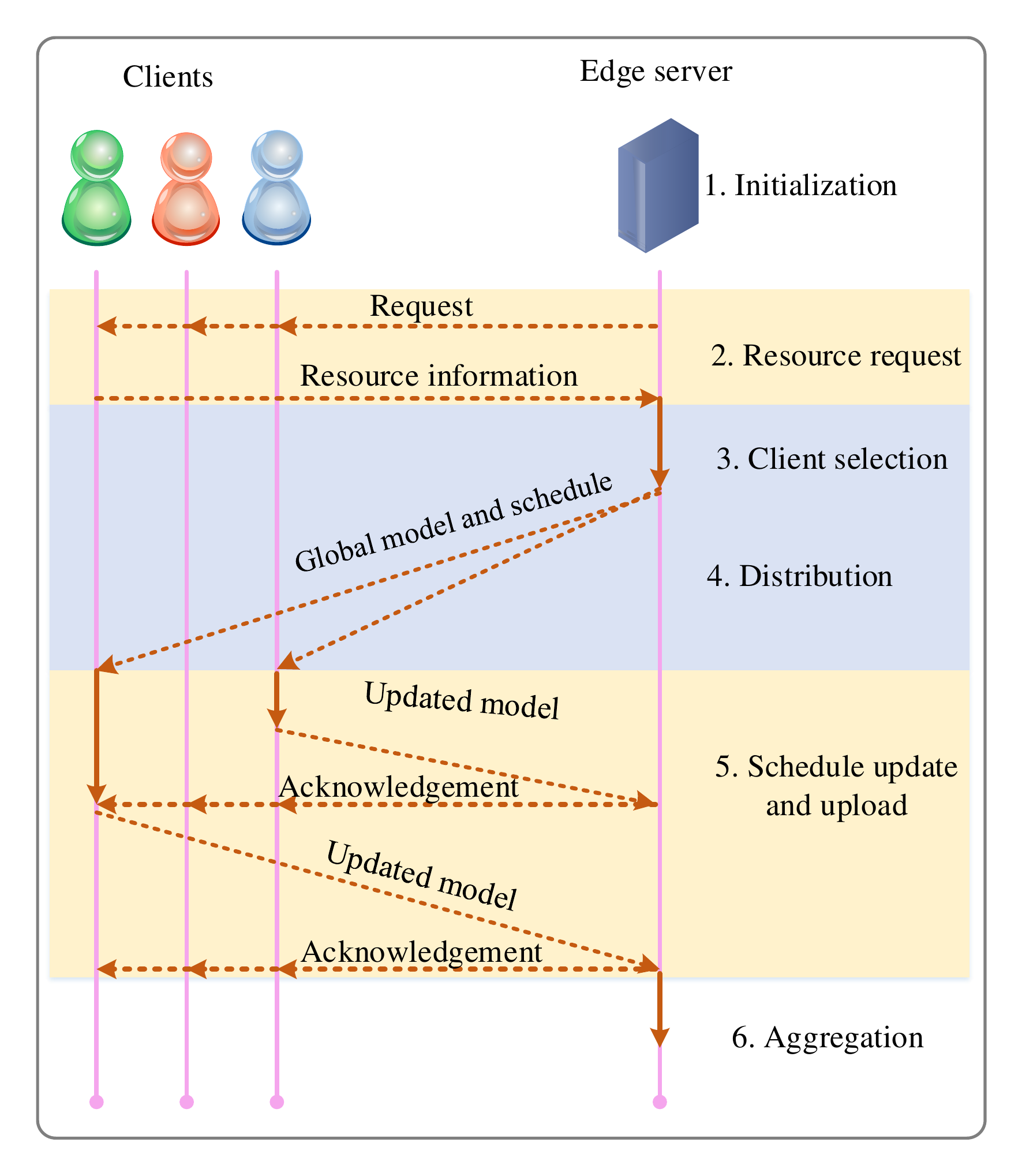}
		\caption{FedCS protocol steps \cite{nishio2019client}.}
		\label{fig:FedCS}
	\end{figure}
	The aforementioned works in \cite {edgeAI, wang2019adaptive, chen2019joint, feraudo2020colearn}, and \cite{conway2019demonstration} provide theoretical analysis pertaining to the deployment of federated learning as well general architecture with implementation for IoT networks. In contrast, the work in \cite{nishio2019client} particularly focused on a federated learning client selection protocol, namely, \emph{FedCS} (shown in Fig.~\ref{fig:FedCS}) that offers an efficient selection of clients for federated learning process based on their available resources. Typically, client selection is performed randomly by a federated learning algorithm that might suffer from an increase in overall federated learning convergence time. For instance, selecting the nodes with the low computational resource (CPU-cycles/sec) results in an increase in overall federated learning time for synchronous federated learning. In FedCS, first of all, the edge computing server requests all the end-devices for their information such as data size, computational capacities, and wireless channel states. Based on the information, a set of nodes is selected for the federated learning process. Then, the edge server distributes the learning model updates to the selected nodes. Finally, all the selected nodes for learning update their local models and send them to the edge server. Simulation results showed that the FedCS mostly attains comparable accuracy within less time compared to traditional federated learning protocols for CIFAR-10 and Fashion-MNIST datasets. The FedCS protocol uses end-devices private information (e.g., the number and classes of its data) by aggregation server, and thus it might suffer from end-devices privacy leakage issue. Additionally, FedCS tried to involve a higher number of end-devices in learning for better performance. However, it does not consider power allocation that can be optimally performed to minimize the transmission latency so that a large number of devices will be able to participate in the synchronous federated learning process. Synchronous federated learning will consider only end-devices local learning models that arrive within a maximum limit of waiting time. On the other hand, a single aggregation edge server is considered in FedCS at BS. There may be multiple base stations (BSs) and an aggregation server will be at the remote cloud. For this kind of scenarios, one must use computing resource efficient devices and a BS association scheme for minimizing the transmission latency. \par 
	
	\begin{table*}[!t]
		%\rowcolors{2}{gray!25}{white}
		\caption {A summary of state-of-the-art advances.} \label{tab:detailsadvances} 
		\begin{center}
			\begin{tabular}{p{1 cm}p{1.2cm}p{1.5cm}p{1.2 cm}p{1.2 cm}p{1.3 cm}p{3cm}p{3cm}}
				\toprule 
				\textbf{Reference}  & \textbf{Local learning model} &\textbf{Dataset}& \textbf{Data Distribution}& \textbf{Aggregation fashion}& \textbf{Analytical convergence analysis}  &\textbf{Primary objective} &\textbf{Limitations}   \\ \midrule
				%1
				Wang \textit{et al.},~[29] & FNN & Xender’s trace [101] & Non-IID & Centralized aggregation & \xmark  & Federated learning was used for training of Deep Q-agents in a resource optimized way. & Edge-AI framework has security and privacy concerns.  \\ \midrule
				%2
				Wang \textit{et al.},~ [88] & SVM, CNN & MNIST, Fashion MNIST &  IID, non-IID & Centralized aggregation & \cmark &  Optimized number of local iterations at network edge. & No convergence analysis for non-convex loss functions  \\ \midrule
				
				%3
				Chen \textit{et al.},~[64] & FNN & MNIST, Synthetic dataset & Non-IID & Centralized aggregation & \cmark & A joint learning and communication framework for federated learning at edge was proposed.& No convergence analysis for non-convex loss functions\\ \midrule

				%4
				Feraudo \textit{et al.},~[89] & FNN & IoT Botnet identification dataset & IID & Centralized aggregation & \xmark  &  A CoLearn framework for federated learning at network edge was proposed. & CoLearn framework has security concerns. \\ \midrule
				
				%5
				Conway-Jones \textit{et al.},~[95] & SVM & MNIST & IID, non-IID & Centralized aggregation & \xmark &  Emulated federated learning in a resource constrained environment. & Lack of efficient criteria for selecting node as an aggregation server\\ \midrule
				
				%6
				Nishio \textit{et al.},~[45]  & CNN & Fashion MNIST, CIFAR-10 & IID, non-IID & Centralized aggregation & \xmark & A client selection protocol for performance improvement in FedAvg was proposed.& FedCS protocol did not consider transmit power allocation than can further improve performance. \\ \midrule
				
				%7
				Kang \textit{et al.},~[94] & Not available  & MNIST & IID, non-IID & Centralized aggregation & \xmark  &  A framework to minimize false local learning model injection was proposed.& High latency due to running blockchain consensus algorithm \\ \midrule

				%8
				Chen \textit{et al.},~[97] & CNN & UCI Smartphone [102] & Non-IID & Centralized aggregation & \xmark &  Federated healthcare framework using transfer learning was proposed.& The framework did not effectively address the issue of data heterogeneity.\\ \midrule
				
				%9
				Yu \textit{et al.},~[93] & CNN & Pascal VOC 2007, Pascal VOC 2012 [103] & IID, non-IID & Centralized aggregation & \xmark  & Federated learning-based object detection scheme was proposed. & Federated optimization scheme can be modified to improve the performance for extreme non-IID case.\\ \midrule
				
				%10
				Ye~\textit{et al.},~[99] & CNN & MNIST, BelgiumTSC [104] & IID & Centralized aggregation &  \xmark & Federated learning framework for vehicular edge computing was proposed.& The framework has robustness issues in case of failure of a centralized aggregation server. \\ \midrule
				%11
				Chen \textit{et al.},~[98] & CNN & CIFAR-10 & IID, non-IID & Centralized averaging & \xmark  & Federated learning-based augmented reality framework was proposed.& FedAvg can be replaced by other federated optimization schemes (e.g., FedProx) to further improve the performance.\\ \midrule
				%12
				Qu \textit{et al.},~[91] & CNN & Fashion-MNIST, CIFAR-10 & IID & Distributed aggregations & \xmark  &  A robust architecture, namely, FL-Block was proposed. &  High latency and energy consumption associated with the framework \\ \midrule
				%13
				Savazzi \textit{et al.},~[96]& CNN, 2-NN & Real time industrial IoT setup data & Non-IID & Distributed aggregations & \xmark & A framework using cooperation among devices to enable federated learning without using centralized aggregation server.& The framework did not consider resource and power allocation that can further improve its performance.\\ \midrule
				%13
				
				%14
				Zhao \textit{et al.},~[90] & MLP & MNIST & IID, non-IID & Hierarchical aggregation & \xmark & A hierarchical architecture for federated learning using for radio access was proposed.& No theoretical convergence analysis for their proposed scheme \\ \midrule
				
				%15
				Wang~\textit{et al.},~[105] & Not available  & CIFAR-10 & IID, non-IID & Hierarchical aggregation &  \cmark &   Hierarchical federated learning framework was proposed.& A criterion for grouping of devices based on wireless parameters is missing. \\ \midrule

				%16
				Liu~\textit{et al.},~[31] &  CNN & MNIST, CIFAR-10 & IID, non-IID & Hierarchical aggregation &  \cmark &   Hierarchical edge-cloud-based federated learning framework was proposed.& Robustness concerns due a centralized global aggregation server \\

				\bottomrule \\
				%\multicolumn{5}{c}{\noindent * N. A refers to unavailability of local learning model type in the given paper.}
			\end{tabular}
		\end{center}
	\end{table*}

	\begin{table*}[!t]
		\centering
		\caption {Primary contribution area, key contributions, and evaluation of the recent advances.} \label{tab:recent_advances_evaluation1} 
		\begin{tabular}{|>{\centering\arraybackslash}m{3cm}|p{1.8cm}|p{7cm}|l|l|l|l|l|}
			\hline
			
			\multirow{2}{*}{Primary contribution area} & \multirow{2}{*}{Reference}&  \multirow{2}{*}{Key contributions} & \multicolumn{5}{|l|}{Evaluation parameters} \\ 
			\cline{4-8} 
			&       &       & P1 & P2 & P3 & P4 & P5 \\
			\hline
			\multirow{3}{*}{Edge AI}
			& Wang \textit{et al.},~ \cite{wang2019adaptive}  &\begin{itemize}  \item A control algorithm is proposed to minimize learning loss under resource budget constraints. \item Performance of the proposed algorithm was evaluated using real time datasets.  \item A theoretical convergence bound was derived for federated learning at network edge. \item The proposed control algorithm learns learning model characteristics, system dynamics, and data distribution. \end{itemize} & \centering{\xmark} &\xmark & \xmark & \xmark  & \xmark \\ 
			\cline{2-8}
			& Wang \textit{et al.},~\cite{edgeAI}  & \begin{itemize} \item An EdgeAI framework was proposed. \item Intelligent computation offloading and edge caching were considered.   \item Used double deep Q-learning networks. \item Federated learning was considered for reducing the transmission overhead during the training process. \end{itemize}&\xmark & \cmark & \xmark & \xmark &  \xmark \\
			\cline{2-8}
			& Nishio \textit{et al.},~\cite{nishio2019client}  & \begin{itemize} \item Proposed a client selection protocol using edge server. \item The edge server manages the resources for communication between clients and edge server. \item The proposed FedCS mostly has attained accuracy within less time for CIFAR-10 and Fashion-MNIST datasets.  \end{itemize}&\xmark & \cmark & \xmark & \xmark &  \cmark \\

			\cline{2-8}
			& Feraudo \textit{et al.},~\cite{feraudo2020colearn}  & \begin{itemize} \item A federated learning architecture for edge networks called as CoLearn was proposed. \item  To enable more flexibility in their architecture, the user policy server was considered that enables network administrator to enforce rules other than defined by manufacturer. \end{itemize}&\cmark & \xmark & \xmark & \xmark &  \xmark \\
			\cline{2-8}
			&  Zhao \textit{et al.},~\textit{et al.}~\cite{zhao2020federated} & \begin{itemize} \item A federated learning-enabled intelligent fog radio access networks framework was proposed. \item Considered hierarchical federated learning that offered two stages of learning model aggregation such as local at fog node and global at cloud. \item Neural network compression and learning model parameters compression are considered. \item Key enabling technologies for enabling federated learning-enabled intelligent fog radio access networks were presented.   \end{itemize}&\cmark & \cmark & \xmark & \xmark &  \xmark \\

			\cline{2-8}
			
			& Qu \textit{et al.},~\cite{qu2020decentralized}& \begin{itemize} \item Proposed a blockchain-enabled federated learning (FL-Block) for fog computing scenarios to offers robustness against the poisoning attacks. \item FL-Block used decentralized fashion for aggregation of local learning model rather than centralized server to offer robustness. \item Additionally, decentralized privacy scheme was proposed for FL-Block.   \end{itemize}& \cmark & \xmark & \xmark & \cmark & \xmark  \\

			\hline

			\multirow{3}{*}{Smart object detection}
			& Yu \textit{et al.},~\cite{yu2019federated}  & \begin{itemize} \item Proposed a smart object detection scheme. \item Considered three types of data distribution such as IID. Non-IID, and extreme IID for analysis. \item Abnormal weights suppression was used to trim the weights far from average normal weights during training. \end{itemize}&\xmark & \xmark & \xmark & \xmark &  \xmark \\
			\hline
			\multirow{3}{*}{Augmented/Virtual reality}
			&Chen \textit{et al.},~\cite{chen2020federated}  & \begin{itemize} \item  A system model for augmented reality applications with object detection and classification problem formulation was presented. \item To solve the formulated problem, the authors proposed a framework based on mobile edge computing enabled by federated learning.   \end{itemize}&\xmark & \xmark & \xmark & \xmark &  \xmark \\
			%\item The authors validated their proposal for image classification task using CIFAR-10 datasets for both Non-IID and IID settings.
			\hline
		\end{tabular}
	\end{table*}

	%%%%%%%%%%%%%%%%%%%%%%%%%%%%%%%%%%%%%%%%%%%
	%%%%%%%%%%%%%%%%%%%%%%%%%%%%%%%%%%%%%%%%%%%
	%%%%%%%%%%%%%%%%%%%%%%%%%%%%%%%%%%%%%%%%%%%
	
	\begin{table*}[!t]
		\centering
		\caption {Primary contribution area, key contributions, and evaluation of the recent advances.} \label{tab:recent_advances_evaluation2} 
		\begin{tabular}{|>{\centering\arraybackslash}m{3cm}|p{1.8cm}|p{7cm}|l|l|l|l|l|}
			\hline
			
			\multirow{2}{*}{Primary contribution area} & \multirow{2}{*}{Reference}&  \multirow{2}{*}{Key contributions} & \multicolumn{5}{|l|}{Evaluation parameters} \\ 
			\cline{4-8} 
			&       &       & P1 & P2 & P3 & P4 & P5 \\
			\hline
			
			\multirow{3}{*}{Mobile networks}
			& Kang \textit{et al.},~\cite{kang2020reliable} & \begin{itemize} \item Proposed a reliable federated learning scheme based on consortium blockchain for mobile networks. \item Using the reputation-based metric, a worker (reliable end-devices) selection scheme based on consortium blockchain was proposed. \item Several future research directions were provided.  \end{itemize}&\cmark & \xmark & \xmark & \xmark &  \xmark \\
			\cline{2-8}
			&  Conway-Jones \textit{et al.},~\cite{conway2019demonstration} & \begin{itemize} \item  Federated learning over emulated wide area communications network having features of intermittent, heterogeneous, and dynamic  availability of resources was studied. %\item The emulated network has both fixed and mobile nodes. 
				\item Intermittent and dynamic behavior was considered which might made the design suitable for use in various practical applications. \end{itemize}&\xmark & \xmark & \xmark & \xmark &  \xmark \\
			\cline{2-8}
			&  Savazzi \textit{et al.},~\cite{savazzi2020federated} & \begin{itemize} \item  Presented federated learning scheme that uses cooperation between devices to avoid use of centralized edge/cloud server for global model aggregation. \item Two strategies such as consensus-based federated averaging and consensus-based federated averaging with gradients exchange, were proposed to enable effective distributed optimization. %\item The proposed scheme offered robustness and scalability. 
			\end{itemize}&\xmark & \cmark & \xmark & \cmark &  \xmark \\
			\cline{2-8}
			&  Chen \textit{et al.},~\cite{chen2019joint} & \begin{itemize} \item A framework for joint learning and communication for wireless federated learning was presented. \item A problem for joint power allocation, resource block allocation, and user selection problem was formulated.
				\item A close-form expression for the expected convergence rate of federated learning was derived to account for the effect of wireless channel on federated learning. 
				% \item The proposed framework offers scalability and sparsification.  
			\end{itemize}&\xmark & \cmark & \xmark & \xmark &  \cmark \\
			\cline{2-8}
			
			& Wang~\textit{et al.},~\cite{wang2020local} & \begin{itemize} \item A hierarchical framework for federated learning was presented. \item Detailed convergence analysis were presented.
			\end{itemize}&\xmark & \cmark & \xmark & \xmark &  \xmark \\
			\cline{2-8}
			& Liu~\textit{et al.},~\cite{liu2019client}  & \begin{itemize} \item An edge-cloud-hierarchical federated learning framework was presented. \item Convergence analysis with simulation results were presented.
			\end{itemize}&\xmark & \cmark & \xmark & \xmark &  \xmark \\
			\hline
			\multirow{3}{*}{Smart health-care}
			& Chen \textit{et al.},~\cite{chen2020fedhealth}  & \begin{itemize} \item  A framework  namely, FedHealth, based on federated transfer learning was proposed for  smart  health-care. \item Homomorphic encryption was considered for secure transfer of learning model updates between the end-users and aggregation server. \item Transfer learning was used by FedHealth framework to enable effective personalized models for end-users.   \end{itemize}&\cmark & \xmark & \xmark & \xmark &  \xmark \\
			
			\hline
			\multirow{3}{*}{Vehicular networks}
			%&Lu \textit{et al.},~\cite{lu2020federated}& \begin{itemize} \item Proposed a privacy preserving federated learning-based vehicular network architecture. \item Addition of noise to local learning model was considered to further improve privacy preservation in federated learning. \item Random sub-gossip updating added enhanced privacy preservation in addition to noise addition to local learning models. \end{itemize}& \cmark & \xmark & \xmark & \cmark & \xmark  \\
			
			% \cline{2-8}
			& Ye~\textit{et al.},~\cite{ye2020federated} & \begin{itemize} \item Proposed a selective model aggregation-based federated learning framework for image classification task in vehicular edge computing. \item To overcome the limitation of FedAvg that is based on random selection of clients for learning process, the authors proposed a selective clients selection based on contract theory. \end{itemize}&\xmark & \xmark & \xmark & \xmark & \cmark \\

			\hline
		\end{tabular}
	\end{table*}
	
	To apply federated learning for IoT networks, we must take into account the reliability of end-devices involved in federated learning. A malicious end-device can send the wrong local learning model updates to the aggregation server, and thus delay the convergence of the global federated learning model. Additionally, the global federated learning model might not converge due to wrong local learning model updates. Therefore, there is a need to address this challenge. To do so, Kang \textit{et al.} proposed a reliable federated learning scheme based on consortium blockchain for mobile networks \cite{kang2020reliable}. The proposed scheme tackles the issue of unreliable updates in federated learning process. A non trusty end-device can send unreliable updates to the aggregation server, and thus degrades the global federated learning accuracy. The authors introduced reputation-based metric for reliable federated learning over mobile networks. Using the reputation-based metric, a worker (i.e., reliable end-devices) selection scheme based on consortium blockchain is proposed. The scheme consists of five steps, task publishment, worker selection, reputation collection, federated learning, and reputation updating. In task publishment, task publishers broadcast task with requirements to mobile devices. Mobile devices fulfilling requirements and willing to join a certain task in turn send joining request to publisher. Next in worker selection scheme, the publisher verifies devices and allow only valid devices participation. The workers with reputation values (computed using subjective logic model) greater than the certain threshold are selected. Then, the federated learning step is performed to yield global federated learning model updates. Finally, the reputation values of end-devices are updated to yield the latest updates after blockchain consensus algorithm is finished. Although the proposed scheme offers protection against malicious users in federated learning, the delay associated with the blockchain consensus algorithm might not be desirable. Additionally, the energy consumption associated with the blockchain consensus algorithm must be minimized. For instance, PoW consumes significantly high energy and suffers from high latency but it offers more decentralization. Therefore, depending on the IoT application and requirements, one must develop blockchain consensus algorithms that will consume low energy (e.g., delegated proof of stake) and offer low-latency (e.g., Byzantine fault tolerance) \cite{majeed2021blockchain}.   \par 
	%In \cite{8851408}, the authors used a federated learning-based algorithm to minimize the breaks in presence events in wireless virtual reality networks. The breaks in presence events in wireless virtual reality are due to a sudden drop in throughput. Blocking the line-of-sight communication path can be one of the reasons for a sudden drop in throughput. Furthermore, behavioral metrics (e.g. user awareness) affect the breaks in presence. User's awareness depends on user actions and perceptions. Therefore, to minimize the breaks in presence, the authors formulated a problem to jointly consider user's awareness, virtual reality video quality, delay of virtual reality video. To minimize the breaks in presents, the authors proposed a federated echo state learning. In a federated echo state learning scheme, all the BSs train their models using partially available data (i.e., orientations and locations) of their associated users. Next, the locally trained model at each BS is shared with the other BSs to yield a global echo state federated learning model. The authors further investigated the effect of an increase in memory for storing end-devices data at the BSs on federated learning prediction accuracy. \par         
	The works in \cite{yu2019federated,chen2020fedhealth, ye2020federated}, and \cite{chen2020federated} applied federated learning various IoT applications. In \cite{chen2020fedhealth}, a framework, namely, FedHealth, based on federated transfer learning was proposed for smart health-care. The purpose of using transfer learning in FedHealth is to reduce the distribution divergence among various domains. Additionally, FedHealth uses homomorphic encryption to enable secure transfer of learning model updates between the aggregation server and end-users. FedAvg has been used as a federated optimization scheme in FedHealth. Although the global model obtained at cloud is available to all end-users, the personalized model for each user might not perform well. Therefore, to improve the performance of the federated learning model for end-users, the authors used transfer learning to learn personalized models. For validation of the FedHealth framework, the public human activity recognition dataset, namely, UCI smartphone was used. The FedHealth framework considered FedAvg as a federated optimization scheme, and thus, its performance can be further improved using FedProx that better considers heterogeneity among end-users. In \cite{yu2019federated}, Yu \textit{et al.} proposed a federated learning-based object detection scheme to overcome the privacy leakage issue of deep learning-based object detection schemes. The authors considered three kinds of data distributions such as IID, non-IID, and extreme non-IID. IID has local datasets of similar distribution, whereas non-IID uses a subset of images of a certain type for the same client. In the case of extreme non-IID, the only object of interest is marked (labeled) in contrast to the non-IID case which involves marking all the objects. The authors used abnormal weight suppression in their proposed federated object detection model to enable trimming of the weights (due to non-IID and extreme non-IID local datasets) that are far from normal weights mean. The federated learning-based object detection model was validated for both IID and non-IID data which revealed degraded performance for non-IID data. Furthermore, for extreme non-IID datasets, performance degradation is the highest.  \par
	Ye \textit{et al.} in \cite{ye2020federated} proposed a selective model aggregation-based federated learning framework for image classification task in vehicular edge computing. The proposed framework consists of a centralized server and vehicular clients. The vehicular clients equipped with various sensors to capture images, train their local learning models for image classification tasks after receiving a request from the centralized server. The locally learned models are then sent to the centralized server for global aggregation. To overcome the limitation of FedAvg that is based on a random selection of clients for the learning process, the authors proposed a selective client selection. Selecting clients with low image quality and poor performance degrades the global federated learning model accuracy. Therefore, we must select clients with better performance. However, determining the client local information (i.e., image quality and computational power) at the centralized is server is not available. Coping with this challenge of information asymmetry, a set of devices with fine quality images is selected using contract theory-based scheme. The authors evaluated their scheme for Belgium TSC and MNIST datasets which showed better performance than FedAvg. \par
	Chen \textit{et al.} in \cite{chen2020federated} studied augmented reality based on federated learning-enabled mobile edge computing. Specifically, the authors focused on the solution of computation efficiency, low-latency object detection, and classification problems of AR applications. A system model for augmented reality applications with object detection and classification problem formulation was presented. One way to solve the problem is through the use of centralized machine learning. However, it suffers from high communication resources consumption costs associated with transferring the data from end-devices to the centralized server for training. Moreover, the end-devices of augmented reality-based applications might not want to share their data with the centralized server due to privacy concerns. To solve the formulated problem, the authors proposed a framework based on mobile edge computing enabled by federated learning. Finally, the authors validated their proposal for image classification tasks using CIFAR-10 datasets for non-IID and IID settings. \par

	\subsection{Advances Based on Distributed Aggregation}
	\label{Advances Based on Distributed Aggregation}
	The works in \cite{qu2020decentralized} and \cite{savazzi2020federated} considered federated learning based on a distributed aggregation of end-devices local learning models for various IoT applications. Here, end-devices compute their local learning models and sent the learning model updates to various distributed aggregation servers. The distributed aggregation servers share the local learning models of their associated end-devices and finally, perform global model aggregation. In contrast to the centralized aggregation server-based federated learning, distributed aggregation servers based federated learning do not require the use of a centralized aggregation server. Qu \textit{et al.} proposed a blockchain-enabled federated learning
	(FL-Block) for fog computing scenarios to offers robustness against the poisoning attacks \cite{qu2020decentralized}. FL-Block uses decentralized fashion for aggregation of local learning models rather than the centralized server to offer robustness. A single centralized server might get malfunctioned or accessed by an unauthorized malicious user. The process of federated learning in FL-Block starts with the local learning model by all the devices. The local learning models are transmitted to the fog servers where cross-verification takes place. Next, the block is generated by the winning miner and block propagation takes place. Finally, global model updates are computed via the aggregation of local learning model parameters at every fog server and sent to all the devices involved in learning. The authors analyzed the latency of the proposed framework. Additionally, a decentralized privacy-preserving scheme is proposed for FL-Block. The authors provided extensive simulation results to validate their proposal. Specifically, they have shown that the proposed scheme performs superior in terms of learning latency for the scenario without malfunctioning of the miners than malfunctioned miners case. Although the FL-Block offers a key feature of robustness, the consensus algorithms used by blockchain typically require high latency to reach consensus. Therefore, novel schemes will be required for federated learning based on blockchain. On the other hand, the end-devices involved in the learning framework of \cite{savazzi2020federated} might not be able to access the fog server due to strict communication resource constraints. To cope with this issue, one can use device-to-device communication that can reuse the channel resources already in use by other cellular users \cite{khan2019edge}. Other than channel resources reuse, cooperation between devices can enable federated learning without the need for a centralized aggregation server. To do so, the work in \cite{savazzi2020federated} presented a federated learning scheme that used cooperation between devices to avoid the use of a centralized edge/cloud server for global model aggregation. To effectively enable all the devices in a network to contribute in distributed optimization of the global federated learning model, two strategies are proposed: consensus-based federated averaging and consensus-based federated averaging with gradients exchange. In consensus-based federated averaging, all the devices receive learning model parameters from other nodes which are subsequently used for updating their own models. On the other hand, consensus-based federated averaging with gradient exchange uses additional steps including consensus-based federated averaging to improve convergence speed, but at the cost of extra complexity. Finally, the authors validate their proposal via experimental results for industrial IoT setup. The advantage of the approach considered in \cite{savazzi2020federated} is the non-usage of a centralized server and thus, offers more robust federated learning. Furthermore, the proposed scheme can offer scalability by incorporating more nodes in a dense D2D network. Although the proposed scheme offered several advantages, one must perform efficient resource allocation for dense D2D networks. Optimization theory, heuristic schemes, and game theory can be used for resource allocation. Other than resource allocation, one can propose effective power control schemes to further improve the packet error rates, and thus increase the global federated learning model accuracy. This power control scheme will enable users to optimally allocate the transmit power while fulfilling the power constraints of the system. \par 
	%In \cite{lu2020federated}, Lu \textit{et al.} proposed a federated learning-based vehicular network architecture that mitigates vehicular network data leakage. A vehicular network scenario was considered with roadside units (RSUs) to cache the data from other RSUs, vehicles, and BSs. The two main parts of their architecture are collaborative data leakage detection and intelligent transformation. Intelligent transformation mainly occurs at the vehicles to enable the transformation of raw data into a form that mitigates the risk of leakage. To do so, they used federated learning for the transformation of vehicular data into learned data models which can be further used in many applications (i.e., resource allocation). On the other hand, data leakage detection is performed at the RSU because they are acting as data aggregators. Mainly, the process of federated learning for vehicular networks consists of local learning models computation. Then, the addition of noise to local models is performed to avoid reconstruction of the end-devices data by a malicious user. Next, random sub-gossip updating takes place which involves the random selection of end-devices set by devices. The selecting device only sends its learning model updates to devices in its set. This way of selecting end-devices randomly prevents end-devices to get identity information about other nodes easily. Finally, distributed aggregation takes place at different vehicles until a global federated learning model of desirable accuracy is obtained. \par

	\subsection{Advances Based on Hierarchical Aggregation}
	\label{Advances Based on Hierarchical Aggregation}
	Considering the aforementioned advantages and disadvantages of both centralized aggregation and distributed aggregation-based federated learning, we can consider the hierarchical fashion of aggregation for federated learning. In hierarchical federated learning, few aggregations of local learning models (e.g. at the edge servers) take place prior to central aggregation (e.g., at cloud server). Few works \cite{zhao2020federated, wang2020local}, and \cite{liu2019client} consider hierarchical federated learning for IoT networks. A framework for federated learning-enabled intelligent fog radio access networks was proposed in \cite{zhao2020federated}. The authors identified the key issues (i.e., high communication cost and security and privacy issues) of a centralized machine learning-enabled fog radio access networks. To overcome these limitations, the federated learning-based framework was proposed. Both traditional federated learning and hierarchical federated learning were considered. In the hierarchical federated learning, two kinds of aggregation take place, one at fog nodes and the other one at a remote cloud. Furthermore, to enable participation of more end-devices in the training process, local model neural network architecture compression and parameter compression are considered. Finally, key enabling technologies for enabling federated learning-enabled intelligent fog radio access networks and open research challenges were presented. Furthermore, the works in \cite{wang2020local} and \cite{liu2019client} proposed a hierarchical federated learning framework for mobile networks. Their proposals considered local aggregations prior to global aggregation at the cloud. It must be noted here that hierarchical federated learning can be used to improve performance. End-devices can communicate with the edge by reusing the communication resources already in use by other cellular users. However, careful design considerations, such as frequency of aggregations at the edge server before a global aggregation takes place, must be taken. In hierarchical federated learning based on FedAvg, there will be two kinds of weights divergences, such as the edge (i.e., due to client-edge divergence) and cloud (i.e., due to edge-cloud divergence), where aggregations take place. The upper bound on the weights divergences is dependent on the number of edge aggregations and local learning model iterations \cite{wang2020local}. For a fixed product of local iterations and the number of edge aggregations, generally increasing the number of edge aggregations will improve the learning performance for non-IID data. One thing must be noted here that the performance of hierarchical federated learning also depends on data distribution and the federated optimization (i.e., FedAvg and FedProx) scheme used. Therefore, properly handling the data heterogeneity using an effective federated optimization scheme and the numbers of local iterations and edge aggregations will cause performance improvement for hierarchical federated learning. \par

	\subsection{Lessons Learned and Recommendations}
	We have discussed recent advances of federated learning towards enabling IoT networks. We have learned several lessons and future enhancements:
	\begin{itemize}
		\item From \cite{chen2019joint}, we derived that federated learning must use some effective compression technique due to limited communication resources and a massive number of IoT devices. Several compression techniques such as gradient compression and model broadcast compression can be used \cite{kairouz2019advances}. The gradient compression minimizes the size of local learning models that are transmitted to the edge/cloud server. Although compression results in size reduction of the bits used to represent the learning model updates, it might result in data loss. Therefore, compression techniques with low loss in data will need to be developed. 
		
		\item We derived from \cite{savazzi2020federated} that there should be robust federated learning scheme for IoT networks. Vanilla federated learning is based on a single centralized aggregation server, which can be easily attacked by a malicious user or it stops working due to physical damage. These issues result in the interruption of the federated learning process. Therefore, we must tackle these issues to enable robust federated learning over IoT networks. To address the robustness issue, a novel design based on distributed aggregation servers will be required.   
		
		% \item We learned that there is need to propose dynamic incentive mechanisms for both edge/cloud server-based federated learning and blockchain-based federated learning to enable IoT-based smart service. Numerous IoT-based smart services are characterized by significantly different unique requirements. For instance, consider autonomous driving and smart keyboard keyword suggestion.     

		\item We derived from \cite{nishio2019client} and \cite{ye2020federated} that heterogeneity and noise-aware federated learning protocols must be proposed. The set of devices involved in federated learning have different computational resource (CPU-cycles/sec) and dataset sizes. Furthermore, the distribution of their datasets is not always identical. Such a heterogeneous nature of end-devices poses significant challenges to the design of a federated optimization algorithm. For instance, FedAvg was developed without more carefully considering devices' heterogeneous system parameters. To cope with this limitation, FedProx was developed to offer a more practical design. FedProx is based on the addition of a scaled proximal term to FedAvg local device loss function. However, the scaling constant might be difficult to adjust for different applications. On the other hand, we must assign communication resources adaptively (i.e., more communication resources to a device with high local model computation time and vice versa) to optimize the overall global federated learning model time. Furthermore, the noise present in the local devices datasets significantly affects the local learning model accuracy. Therefore, novel federated optimization schemes that consider devices heterogeneity and local datasets noise in the devices selection phase will be required.

		\item From \cite{qu2020decentralized}, we derived that distributed aggregation-based federated learning for IoT networks must share the learning model updates in a secure and trustful manner. Recently, blockchain-based federated learning has been proposed for secure and trustful sharing of learning model updates between different miners. Although blockchain can provide many features, it has an inherent issue of high latency associated with consensus algorithms. On the other hand, federated learning consists of a number of global iterations (i.e., communication rounds). Therefore, using blockchain during the training of a federated learning model might not be desirable due to the high latency associated with the blockchain consensus algorithm. To cope with the high latency issue, there is a need to develop low complexity consensus algorithms.

	\end{itemize}

	\section{Taxonomy of Federated Learning for IoT}
	\label{Taxonomy}
\begin{figure*}[!t]
		\centering
		\captionsetup{justification=centering}
		\includegraphics[width=18cm, height=23cm]{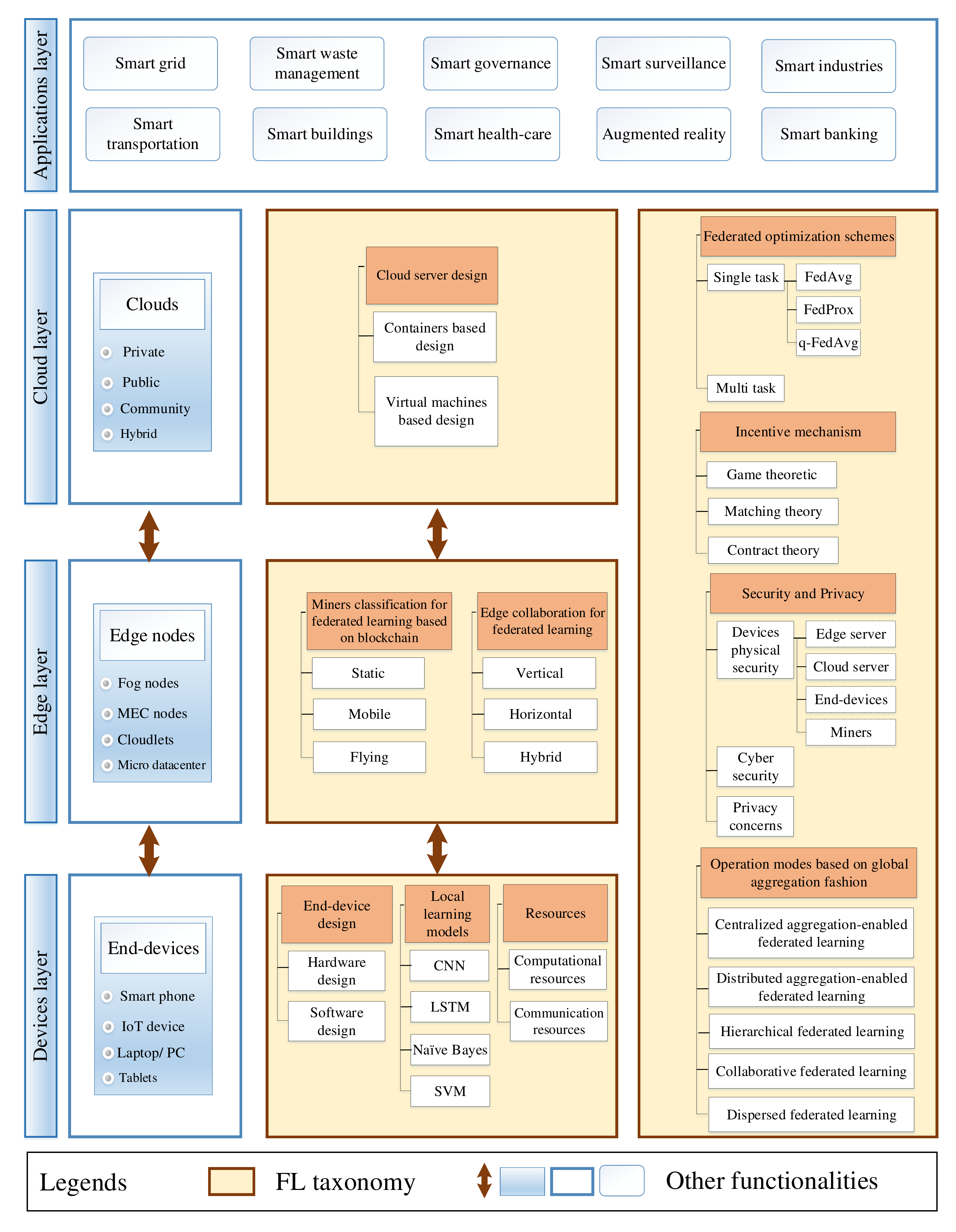}
		\caption{Taxonomy of federated learning for IoT networks.}
		\label{fig:taxonomy}
	\end{figure*}

	Federated learning over wireless networks involves many players, such as end-devices, edge/cloud servers, and miners \cite{Dusit_FL_survey}. These players use wireless channel and core network resources in addition to computing resources for successfully enabling complex interaction between themselves \cite{khan2019federated}. Moreover, end-devices must be given some incentives to motivate their participation in federated learning process. Keeping in mind the aforementioned facts, we derive a taxonomy (illustrated in Fig.~\ref{fig:taxonomy}) using operation modes based on global aggregation, resources, local learning models, incentive mechanism, federated optimization schemes, end-device design, miners classification, cloud server design, edge collaboration, security, and privacy, as parameters. A detailed discussion with insights about every parameter of the taxonomy will be given later in this section. In Fig.~\ref{fig:taxonomy}, we place various parameters of taxonomy at three layers: end-devices layer, edge layer, and cloud layer, to provide more clear insights about their relationship to federated learning-enabled IoT networks. End-devices design and local learning models are specific to local devices and it will perform local model training for federated learning. Next to local model training, we can do aggregation either at a remote cloud or edge servers. Cloud servers at the cloud layer can be implemented using container-based design or virtual machine-based design. To perform aggregation at the edge layer, we can use simply an edge server (edge-assisted aggregation) or miners implemented at the edge. The use of miners for blockchain-based federated learning will be explained in more detail later in this section. Moreover, we can use collaboration among edge servers for federated learning global model computation (more details will be given in Section~\ref{Edge Collaboration}). On the other hand, incentive mechanism design, security, and federated learning optimization/learning schemes involve end-devices, edge servers, and cloud servers. Therefore, they are positioned by considering all the three layers, such as the devices layer, edge layer, and cloud layer. Next, we describe various parameters of taxonomy for federated learning over IoT networks in detail.\par
	
	\subsection{Security and Privacy}
	Federated learning over IoT networks have security and privacy concerns. Although federated learning was introduced to enable users' privacy, it has still privacy leakage concerns\footnote{In this sub-section, we use the keyword "privacy leakage" to end-devices sensitive information leakage due to the information inferring capability of a malicious user or aggregation server using their learning model updates. On the other hand, accessing end-devices due to weak authentication schemes and altering the end-devices are considered as a security concern.}. A malicious end-device and aggregation server can infer end-devices private information \cite{nasr2019comprehensive, privacy_1}. To address this issue, a differential privacy scheme can be used that is based on the addition of noise to local learning model parameters before sending it to the global aggregation server \cite{Dusit_FL_survey, zhao2020local,lu2019differentially, truex2019hybrid, wei2020federated, DFL_privacy_1}. In local differential privacy, each end-device performs differential private transformation by adding noise which reduces the accuracy of the federated learning model. Noise for local differential privacy can be added via various ways, such as the Gaussian mechanism and the Laplace mechanism. Laplace mechanism adds noise of Laplace distribution \cite{phan2017adaptive}. However, adding a Laplace noise directly to the local learning models might cause significant performance degradation especially for larger values of the noise \cite{mcsherry2007mechanism, ma2020safeguarding}. To cope with this limitation, one can use differential privacy based on an exponential mechanism. An exponential mechanism is based on mapping the local learning model updates to values within a range with a probability that follows an exponential distribution. Although exponential mechanism generally performs better than Laplace mechanism, it might suffer from performance degradation \cite{dwork2014algorithmic}. Another way to improve the performance of local differential privacy is to use a Gaussian mechanism that adds noise according to a zero-mean isotropic Gaussian distribution \cite{dwork2014algorithmic}. The addition of Gaussian noise has the main advantage of being additive in nature. Adding a Gaussian noise to Gaussian distributed local learning model weights will result in another Gaussian distribution whose analysis is simple. In \cite{wei2020federated}, the authors showed that generally adding more noise to local learning model parameters can increase the privacy preservation capability, but at the cost of loss in accuracy and long convergence time. Therefore, we must make a tradeoff between the convergence and privacy preservation level. Another way to avoid prolonging the convergence time would be to use secure multiparty computation-based schemes, achieved by the encoding of local learning model updates. However, this approach will cause more communication overhead and might not be fully effective against the privacy leakage attacks. Therefore, to resolve these issues, one can use a hybrid approach that combines both local differential privacy via the addition of minimum possible noise and secure multiparty computation-based scheme via low-overhead encryption \cite{truex2019hybrid}. \par
	In addition to privacy leakage, federated learning for IoT networks has a few security concerns. Generally, security in federated learning for wireless networks can be divided into two main categories: devices physical security and cybersecurity. Devices' physical security refers to restricting the physical access to the end-devices \cite{khan2019edge}. Enabling end-devices, servers, and miners with physical security is challenging. On other hand, cybersecurity refers to network security, information security, and applications security \cite{Cyber_security_1, Cyber_security_2}. To ensure the integrity of data stored at end-devices, one must use effective and light-weight authentication schemes due to the fact that a malicious user can physically access the end-devices with fewer efforts \cite{lee2014lightweight,lara2020lightweight, arasteh2016new}. Similarly, the edge servers have distributed nature and enabling its physical security is very challenging. To enable its security, one must develop new light-weight authentication schemes to provide access to only registered and trustworthy users \cite{9045654, 8190883}. The main reason for the development of light-weight authentication schemes for federated learning-enabled IoT networks is due to power constraints of the most IoT devices. In contrast to the edge, the cloud has centralized nature and can be easily attacked by a malicious user \cite{kaushik2018cloud}. Specifically, the cloud-based federated learning has serious security concerns. A malicious user can easily access the centralized cloud server and introduce error in the global federated learning model. Therefore, similar to edge, we must use effective authentication schemes for cloud security \cite{li2015towards, jiang2016security, barreto2015security}. It is to be noted here that, the authentication scheme used at the cloud can be of higher complexity than that of edge server due to presence of more computational capacity at the cloud. Other than end-devices, edge servers, and cloud security, federated learning is susceptible to malicious user attacks during the exchange of learning model updates between end-devices and the aggregation server. To address this issue, one can use an effective encryption scheme \cite{truex2019hybrid}. It must be noted here that using an effective authentication scheme can preserve privacy as well. A malicious user or aggregation server can not easily infer the end-devices sensitive information from their encrypted local learning model updates. Generally, complexity of an encryption algorithm increases with performance. Therefore, we must make a trade-off between performance and communication overhead and complexity. \par    
	
	%From this section, we learned several lessons. There is a need to develop federated optimization schemes that offer secure exchange of learning model parameters between the end-devices and edge/cloud server. Although blockchain-based federated learning scheme has been proposed in \cite{} to offer enhanced robustness and security. However, extra overhead and computational complexity associated with solving consensus algorithm is not desirable. Furthermore, running a blockchain consensus algorithm consume significant amount of energy \cite{}. Therefore, we must devised novel consensus algorithms for blockchain-based federated learning to offer learning with low latency and energy consumption. 
	
	\begin{figure*}[!t]
		\centering
		\captionsetup{justification=centering}
		\includegraphics[width=18cm, height=11.5cm]{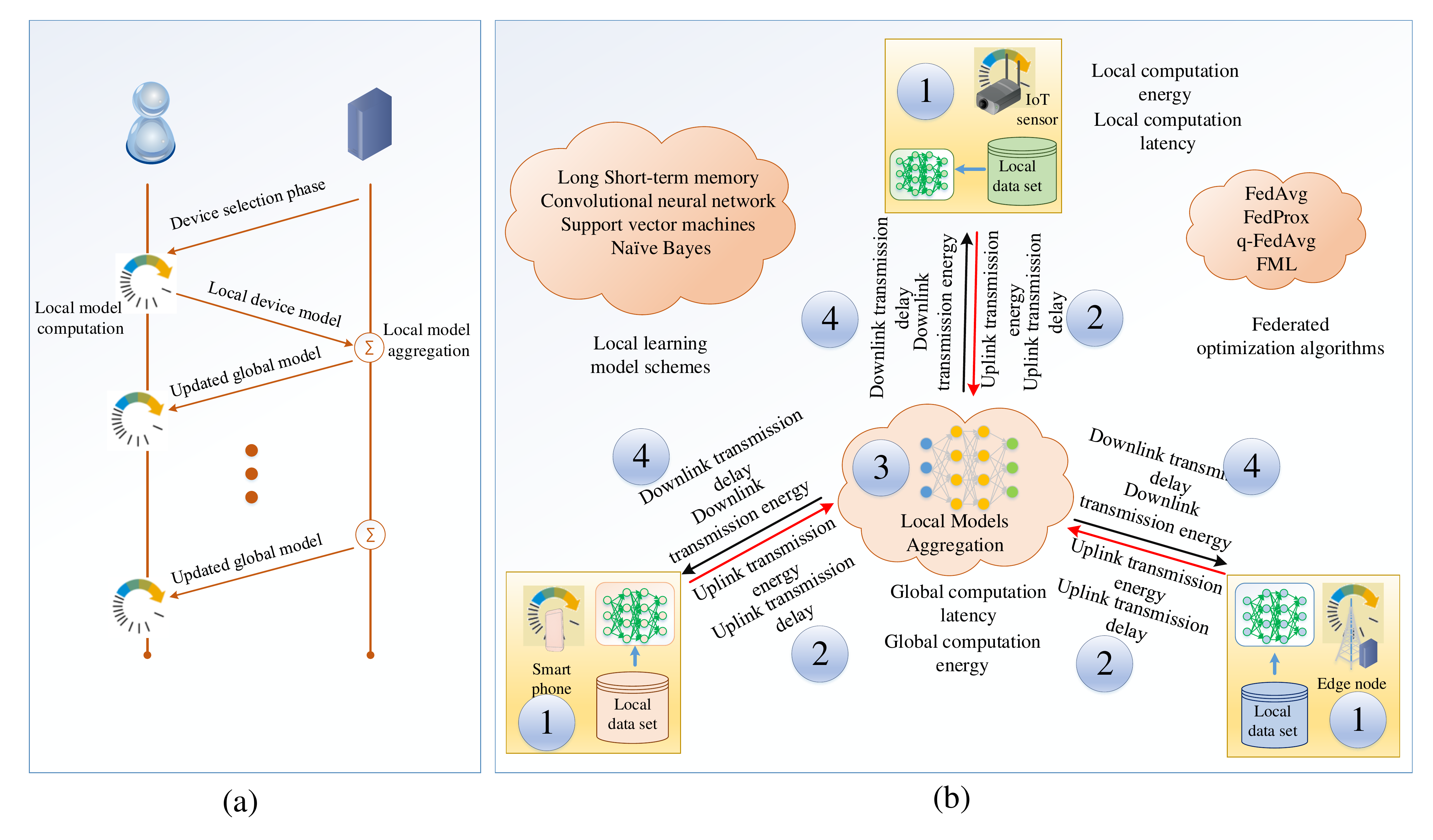}
		\caption{Federated learning (a) sequence diagram, (b) overview of resources, local learning algorithms, and federated optimization schemes \cite{khan2019federated}.}
		\label{fig:seq}
	\end{figure*}

	\subsection{Resources}
	Enabling efficient federated learning over an IoT-network requires optimization of both computational and communication resources, as shown in Fig.~\ref{fig:seq} \cite{zhang2020federated, khan2019federated, niknam2020federated}. The computational resource can be local computational resources of the end-devices or global computational resources of the aggregation server \cite{nagar2019privacy, qu2019proof, jere2020federated, zhan2020experience}. To compute the local learning model, the end-device performance (i.e., local learning accuracy and computation time) is mainly determined by its dataset size, available energy, and computational resource (CPU-cycles/sec). The computation time of a local learning model for a fixed accuracy strictly depends on the dataset size and the device computational resource (CPU-cycles/sec). The energy consumption of the device having operating frequency $f$, dataset size $D$, and $I_{l}$ local iterations, is given by \cite{9013160, ndikumana2019joint, khan2020self}:
	\begin{equation}
		\label{eq:1}
		\begin{aligned}
			E_\textrm{local} = I_{l}\left(\rho \zeta D f^{2}\right),
		\end{aligned}
	\end{equation}\newline
	where $\rho$ and $\zeta$ are the CPU dependent constant parameter and CPU-cycles required to process a single dataset point, respectively. The local learning model computation time is given by:
	\begin{equation}
		\label{eq:2}
		\begin{aligned}
			T_\textrm{local} = I_{l}\left( \frac{ \zeta D}{f} \right). 
		\end{aligned}
	\end{equation}\newline
	From (\ref{eq:1}) and (\ref{eq:2}), it is clear that there exists a tradeoff between simultaneously minimizing local learning model computation time and energy consumption. For a fixed local model accuracy and device operating frequency, the dataset size determines the computation time of the local learning model. For fixed dataset points and local model accuracy, the local model computation time can be decreased by increasing the operating frequency of the device. However, energy consumption increases proportionally to the square of the local device operating frequency. Therefore, we must make a trade-off between the end-device operating frequency and energy consumption while computing a local learning model. A set of devices involved in federated learning shows significant heterogeneity in terms of computation resource, dataset size, operating frequency, and available energy. For a fixed local model computation time, the end-devices' heterogeneous parameters result in local learning models of variable accuracy. To account for variable local learning model accuracy of end-devices due to heterogeneity, we can use the notion of relative local accuracy $\theta$ \cite{pandey2019crowdsourcing}. Smaller values of relative local accuracy show better local learning model accuracy and vice versa.\par
	After computing the local learning model at every end-device, the learning model parameters are sent to the aggregation server. Depending on the federated learning scheme, the aggregation of the local learning model can be carried out either at a centralized aggregation server or distributed nodes. For a vanilla federated learning, the global model aggregation takes place at the edge server/cloud \cite{khan2019federated}. The global model updates are sent back to the end-devices. On the other hand, the local learning model updates of all the end-devices are transferred between miner-enabled BSs for a blockchain-based federated learning model \cite{kim2019blockchained}. The miners exchange the local learning model parameters of all the devices through distributed ledger in a trustworthy way. After a consensus is reached among the miners, all the miners send the block which contains local learning models of all the devices to their corresponding end-devices where the global model aggregation takes place. The transmission time for transferring of model updates between the BSs and end-devices depends on the radio access scheme and channel variations \cite{wadu2020federated, amiri2020federated}. For a data rate $R_{n}, \forall{n \in \mathcal{N}}$ , the time taken during sending of learning model parameters of size $\nu$ to the aggregation server can be given by:
	\begin{equation}
		\label{eq:3}
		\begin{aligned}
			T_\textrm{trans}=\sum_{n \in \mathcal{N}}{\left(\frac{\nu}{R_{n}}\right)}.
		\end{aligned}
	\end{equation}\par
	On the other hand, the energy consumed during transmission using transmit power $P_{n}, \forall{n \in \mathcal{N}}$ of the learning model parameters from end-devices to the aggregation server is given by:
	\begin{equation}
		\label{eq:4}
		\begin{aligned}
			E_\textrm{trans} = \sum_{n \in \mathcal{N}}{\left(\frac{\nu P_{n}}{R_{n}}\right)}. 
		\end{aligned}
	\end{equation}\par
	From the above discussions regarding computational resource and communication resource consumption, one can define computational and communication costs. The computational cost can be the number of local iterations for local model computation, whereas communication cost can be the number of global communication rounds between the end-devices and aggregation server. For a fixed global federated learning model accuracy, computational and communication costs have contradictory relations with each other. An increase in computational cost decreases the communication cost and vice versa. For a relative local accuracy $\theta$ and global federated learning model accuracy $\epsilon$, the global iterations can be given by \cite{konevcny2016federated2}:
	\begin{equation}
		\label{eq:5}
		\begin{aligned}
			I_{g} = \frac{\alpha \log \left(\frac{1}{\epsilon}\right)}{1-\theta},
		\end{aligned}
	\end{equation}\par
\begin{figure}[!t]
		\centering
		\captionsetup{justification=centering}
		\includegraphics[width=8cm, height=6cm]{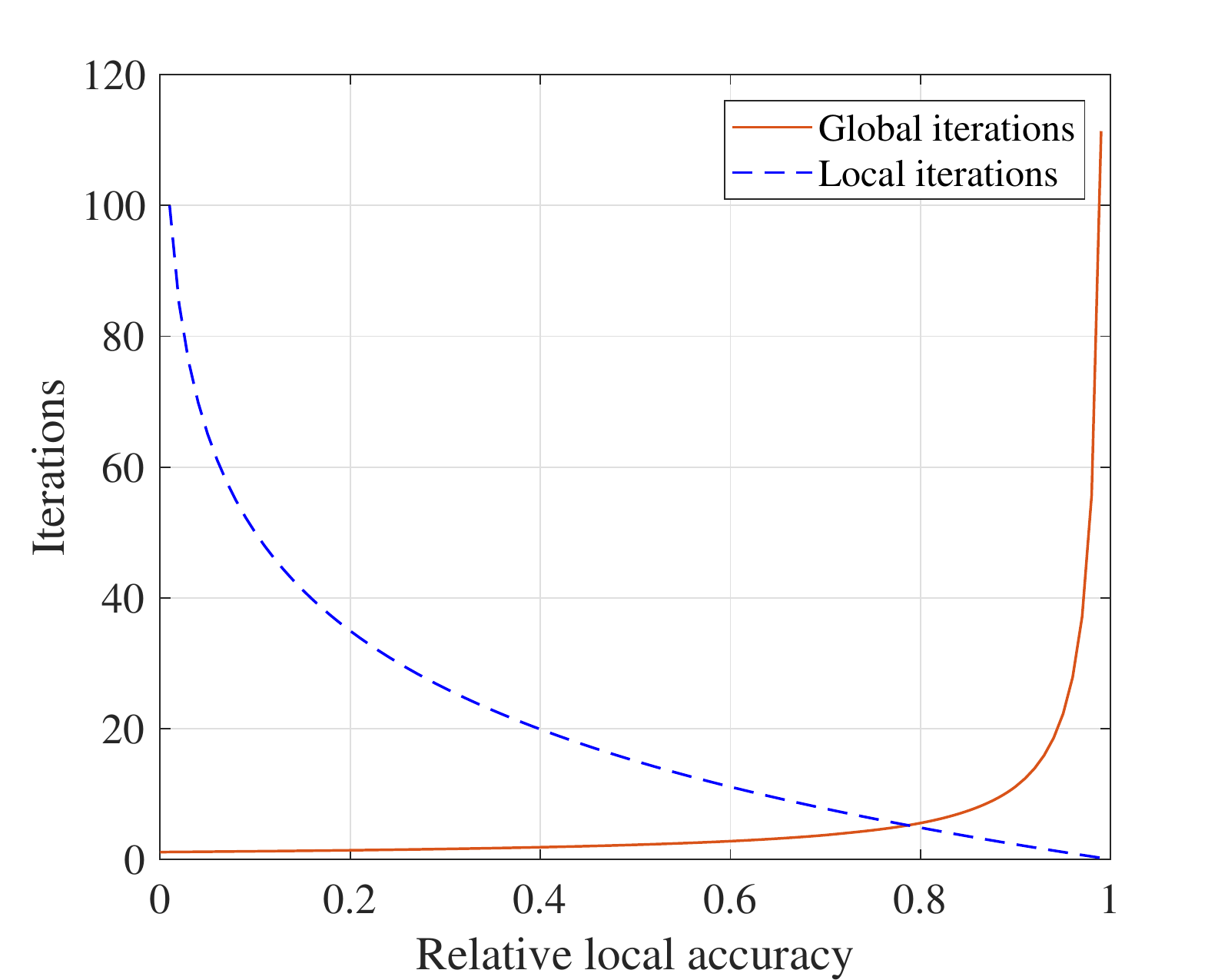}
		\caption{Global and local iterations vs. relative local accuracy.}
		\label{fig:iterations}
	\end{figure}

	From (\ref{eq:5}), it is clear that for a fixed global accuracy, more local iterations are required to minimize the relative local accuracy $\theta$ (i.e., to maximize the local learning model accuracy). The local iterations can be given by \cite{pandey2019crowdsourcing}:
	\begin{equation}
		\label{eq:6}
		\begin{aligned}
			I_{l} = \gamma \log \left(\frac{1}{\theta}\right),
		\end{aligned}
	\end{equation}\newline
	where $\gamma$ is constant, which depends on end-device local sub problem and dataset \cite{konevcny2017semi}. Fig.~\ref{fig:iterations} shows the variations in global and local iterations vs. relative local accuracy for a fixed global accuracy. To achieve lower values of relative local accuracy, more local iterations and few global iterations are needed for a fixed global accuracy and vice versa.  
	
	From the aforementioned discussions, it is evident that there must be some effective mechanism for the selection of local devices operating frequencies for federated learning. The devices' operating frequencies determine the energy consumption and local model computation delay for a fixed number of dataset points and fixed local learning model accuracy \cite{khan2019federated}. On the other hand, there exist significant variations in the wireless channel and different devices have different transmission delays for sending local learning models to edge/cloud server \cite{ang2020robust}. Therefore, the joint optimization of energy and latency for federated learning over wireless networks must be considered. In \cite{tran2019federated}, joint optimization of energy and latency for single task federated learning was considered. On the other hand, multitask federated learning has attractive applications in smart industries. Therefore, multitask federated learning schemes that jointly optimize energy and latency will be required.

	%\item Effect of variations in learning rate on the proposed optimization problem
	
	%\item  Furthermore, the effect of packet loss rate on the optimization problem must be considered. 

	\begin{table*}[]
		%\rowcolors{2}{gray!25}{white}
		\caption {Local learning models with discussion used in various federated learning-enabled IoT applications.} \label{tab:locallearning2} 
		\begin{center}
			\begin{tabular}{p{1 cm}p{1cm}p{1.2cm}p{4.6cm}p{8.7cm}}
				\toprule 
				\textbf{Reference}   & \textbf{Local learning model} & \textbf{Primary contribution}  & \textbf{Local model architecture} &\textbf{Discussion}\\ \midrule
				Wang \textit{et al.},~ \cite{wang2019adaptive} & SVM, CNN & Edge AI & The network has $9$ layers with $5 \times 5 \times 32$ convolutional, $2 \times 2$ MaxPool, local response normalization, $5 \times 5 \times 32$ Convolutional, local response normalization, $2\times 2$ MaxPool, $z \times 256$ fully connected, $256 \times 10$ fully connected, and softmax activation. The other model is squared-SVM. & The authors considered various data distributions for analysis with the aim to find the optimal number of local iterations. Their analysis shows that number of optimal local iterations depends on the type of dataset, data distribution, and local learning model. Meanwhile, their proposed adaptive algorithm generally achieved better performance than traditional federated learning \cite{mcmahan2017communication} by optimally adjusting local learning model iterations. Although theoretical analysis for convex loss functions are provided, the work did not consider theoretical analysis for non-convex loss functions of many practical scenarios. \\ \midrule
				Wang \textit{et al.},~\cite{edgeAI} & FNN &  Intelligent edge caching and computational offloading & Single-layer fully-connected FNN, that has 200 neurons. & The federated learning-based double deep Q-network performed near to centralized double deep Q-network. Meanwhile, transmission cost is also low for federated learning-based double deep Q-network compared to centralized learning-based double deep Q-network. However, its performance is little degraded than the deep Q-network based on centralized training, which may be further improved by considering more complex local learning model (e.g., convolutional neural network with many layers).  \\ \midrule
				Nishio \textit{et al.},~\cite{nishio2019client} & CNN & Client selection protocol &  The model has six convolution layers with $32$, $32$ , $64$, $64$, $128$, and $128$ channels. All the convolutional layer use $3\times 3$ filters. All the layers are are followed by ReLU and batch normalization. After every two layers $2 \times 2$ max pooling is used. These convolutional layers are followed by three fully connected layers with $382$ and $192$ units. The last layer has $10$ units with softmax activation. & The proposed client selection protocol showed performance improvement than the random client selection of FedAvg \cite{mcmahan2017communication} with an increase in deadline that accounts for overall global model computation time using both IID and non-IID settings for fashion MNIST and CIFAR-$10$ datasets. Although the client selection protocol offered significant performance improvement, it did not consider the scenario of multiple BSs with aggregation server at a remote cloud.  \\ \midrule
				Feraudo \textit{et al.},~\cite{feraudo2020colearn} & FNN & Framework for intelligent edge & FNN model with two hidden layers having $50$ and $30$ neurons. & From the analysis of this work, it is revealed that their framework generally results in performance improvement for an increase in number of local iterations. The proposed framework does not effectively tackle the security issues of the IoT devices. One must use effective authentication schemes for IoT devices to avoid injection of false local learning model updates. Injecting false local learning model updates will prolong the global federated learning convergence time. \\ \midrule
				Qu \textit{et al.},~\cite{qu2020decentralized} & CNN & Edge AI & The model has six convolution layers with $32$, $32$ , $64$, $64$, $128$, and $128$ channels. All the convolutional layer use $3\times 3$ filters. All the layers are are followed by ReLU and batch normalization. After every two layers $2 \times 2$ max pooling is used. These convolutional layers are followed by three fully connected layers with $382$ and $192$ units. The last layer has $10$ units with softmax activation. & A block-FL framework using blockchain was proposed to avoid data poisoning. For both CIFAR-10 and fashion MNIST, the authors evaluated the performance of proposed scheme by increasing the round time for global model computation. Generally, increasing the number of round time causes an increase in global federated learning accuracy. Although FL-block can offer us with the robust operation, it has additional latency due to blockchain consensus algorithm.  \\ \midrule
				%  Liu \textit{et al.},~\cite{liu2019lifelong} &Reinforcement learning & Smart mobile robotics \\ \midrule
				%  Yu \textit{et al.},~\cite{yu2019federated} & Convolutional neural network & Smart object detection\\ \midrule
				Savazzi \textit{et al.},~\cite{savazzi2020federated} & CNN, FNN & Mobile networks & 
				The 2NN model uses fully connected layer of $32$ hidden nodes (dimension $512 \times 32)$ followed by a ReLU layer and a second fully connected layer of dimension $32 \times C$.The second CNN model consists of a convolutional layer ($8 \times 16 \times 1$) followed by max pooling ($5\times 5$) and a fully connected layer of dimension $168 \times C $. &  A fully decentralized federated learning approach was proposed to minimize global communication rounds and improve convergence. Two schemes such as cooperative federated learning based FedAvg (CFA) and cooperative federated learning based FedAvg with gradient exchange (CFA-GE) were proposed. CFA-GE has shown less validation loss than CFA. Additionally, increasing the number of devices in cooperation also improves performance due to the fact that more devices participate in learning process. Although the proposed scheme can offer several advantages, it will result in additional complexity for performing cooperation between the devices.     \\ \midrule
				Chen \textit{et al.},~\cite{chen2020fedhealth} & CNN & Healthcare & The model has $2$ convolutional layers, $2$ pooling layers, and $3$ fully connected layers. A convolution size of $1 \times 9$ was used. & A federated transfer learning framework was proposed for end-devices with their own local datasets and a public dataset at cloud. The convolutional layers and max-pooling layers are kept frozen (not updated) in model transfer at end-devices due to low-level features extraction characteristic of these layers regarding the end-user activity. Using transfer learning in such as fashion at end-devices results in significant performance improvement. Their proposal out performed support vector machines and k-nearest neighbors algorithm. Furthermore, the proposed framework can be extended with incremental learning for more personalized healthcare. \\ \midrule
				Chen \textit{et al.},~\cite{chen2020federated} & CNN & Augmented reality & The network has a convolutional layer (i.e., $3 \times 6 \times 5$), maxpool layer (i.e., $2 \times 2$ ), convolutional layer (i.e., $6 \times 16 \times 5$ ), and followed by a fully connected network. The activation function used was ReLU. & This work considered a convolutional neural network model and CIFAR-10 dataset. From the analysis, it is revealed that an increase in local iterations will cause a proportional performance improvement. However, the increase will not be more for higher number of local iterations (i.e., for $15$ and $20$). Simple averaging in the proposed scheme can be replaced by weighted aggregation scheme that will take into account the wireless fairness issues.\\ 
				% Ye~\textit{et al.},~\cite{ye2020federated} & Convolutional neural network & Vehicular network\\ 

				\bottomrule 
			\end{tabular}
		\end{center}
	\end{table*} 
	
	\subsection{Local Learning Models}
	In typical federated learning, two types of nodes such as end-devices and centralized edge/cloud servers are involved. First of all, a local learning model is trained at every device using its local dataset. The choice of local learning model strictly depends on the type of IoT task and end-devices computational power with backup energy. For a particular kind of local learning model (e.g., CNN), selecting the number of hidden layers with activation functions determines its performance. Generally, increasing the number of layers improves the local learning performance. However, considering large models that consist of many hidden layers might not always perform well. The reason for this can be the small size of the considered local dataset. It may happen with deep learning models used for local model computation to cause performance degradation with an increase in a number of hidden layers for small local datasets. Therefore, depending on the type of local dataset and task, we should select wisely the number of hidden layers for the local learning model. Additionally, the activation functions must be chosen properly as already mentioned in Section~\ref{Local Learning Models}. Various machine learning schemes: Naive Bayes, SVM, CNN, and LSTM can be used for local model training \cite{Dusit_FL_survey}. Summary of local learning models with their IoT application is given in Table~\ref{tab:locallearning2} \cite{wang2019adaptive, edgeAI, nishio2019client, feraudo2020colearn, qu2020decentralized,   savazzi2020federated, chen2020fedhealth, chen2020federated}. Additionally, Table~\ref{tab:locallearning2} provides details about the local learning model architecture and insights about the performance of federated learning for various applications using these local learning models. The CNN models are used in \cite{ savazzi2020federated, chen2020fedhealth, chen2020federated} for smart object detection, mobile networks, healthcare, augmented reality, and vehicular networks. Other works \cite{wang2019adaptive, edgeAI, nishio2019client, feraudo2020colearn, qu2020decentralized} mainly focused on edge intelligence to enable various IoT applications. Typically, end-devices have low computational power and limited backup power. Therefore, running complex deep learning algorithms on end-devices with low computational might not be desirable. To address this issue, one must use some effective local computation complexity reduction approach to minimize the local learning model computation time \cite{kairouz2019advances}. We must modify the existing local learning model algorithms to reduce their computational complexity. However, generally reducing the complexity of the deep learning network results in performance degradation. Therefore, we must make a trade-off between local learning model accuracy and complexity.\par

	\begin{table*}
		
		\caption {Incentive mechanism for federated learning.} \label{tab:incentive mechanism} 
		\centering
		\begin{tabular}{p{2.5cm}p{2.5cm}p{3cm}p{3cm}p{3.5cm}}
			\toprule 
			
			\textbf{Reference} &  \textbf{Incentive model} & \textbf{Motivation} & \textbf{Device utility/ bid} & \textbf{Aggregation server utility} \\
			\midrule
			
			Pandey \textit{et al.},~\cite{pandey2019crowdsourcing} & Stackelberg game &  To enable a high-quality global federated learning model with wireless communication efficiency. & Difference between reward
			(i.e., \$/accuracy level) and cost (i.e., communication and computation cost) & Concave function of $\log(1/global~accuracy)$ \\
			
			\midrule
			Le \textit{et al.},~\cite{le2020auction} & Auction theory  & To minimize social cost which denotes true cost of end-devices bids. & Difference between reward and cost. & Minimizes cost of bids. \\
			
			\midrule
			
			Kang\textit{et al.},~\cite{8832210} & Contract theory & To motivate participation of reliable end-devices in federated learning. &  Difference between reward and energy consumption. & Maximize the profit that is given by the difference between total time for global iteration and reward given to end-devices. \\
			
			\bottomrule 
		\end{tabular}
		
	\end{table*}

	\subsection{Incentive Mechanism}
	In order to ensure large-scale adoption of federated learning, there is a need to devise novel incentive mechanisms. Since there are different ways to implement federated learning (e.g., centralized edge/cloud server-based vanilla federated learning and blockchain-based federated learning), there is a need to design different incentive mechanisms that can be appropriately used for the federated learning implementation of interest.\par 
	For centralized edge/cloud server-based federated learning, the two players of the federated learning are devices and edge/cloud server, which interact with each other to train the global federated learning model. Numerous ways can be used to model such type of interaction between the devices and edge/cloud server. Several approaches (summary is given in Table~\ref{tab:incentive mechanism}) have been proposed regarding the design of incentive mechanisms for federated learning \cite{pandey2019crowdsourcing, le2020auction, 8832210}. These approaches can be classified into game theory, contract theory, and auctions theory. In \cite{pandey2019crowdsourcing}, a crowdsourcing framework was proposed for federated learning to enable a high-quality global model with wireless communication efficiency. A two-stage Stackelberg game-based incentive mechanism was adopted to jointly maximize the utility of the aggregation server and the end-devices. The utility of the aggregation server was modeled as a concave function of $\log(1/global~model~accuracy)$, whereas end-device utility was modeled as a difference between reward (i.e., $\$$/accuracy level) and cost (i.e., communication and computation cost). Le \textit{et al.} presented an auction-based incentive mechanism for federated learning over cellular networks \cite{le2020auction}. Every end-device submits a set of bids to the BS. The bid consists of claimed cost and resources (i.e., local accuracy level, CPU cycle frequency, transmission power, and sub-channel bundle). The authors formulated a social cost minimization problem representing the true cost of user bids. To deal with NP-hard social cost minimization problem, randomized auctions based mechanism was proposed. Next, the BS determines the set of all winner profiles. Finally, the winning end-devices participate in the federated learning process. On the other hand, \cite{8832210} proposed an incentive mechanism for reliable federated learning. The authors defined a reputation-based metric to determine the trustworthiness and reliability of end-devices participating in the federated learning process. Then, a reputation based end-device selection scheme using a multi-weight subjective model was proposed. Moreover, blockchain was considered for secure end-devices reputation management. Finally, they proposed an incentive mechanism based on contract theory to motivate end-devices participation in federated learning. \par
	On the other hand, blockchain-based federated learning consists of sets of devices and miners \cite{kim2019blockchained, nagar2019privacy}. First, the devices compute their local learning models and then send the learning model parameters to their associated miners. The miner verifies and exchanges the learning model parameters with other miners and run a consensus algorithm. Once the miners complete the consensus algorithm, a block consisting of verified local learning models is added to the blockchain. The end-device associated with the miners download the block containing the verified local learning models of different devices. Finally, the global federated learning model aggregation is performed at the end-devices. This process of federated learning offers robustness and security but suffers from additional overhead compared to the traditional federated learning approach. Designing an incentive algorithm for such kind of federated learning scheme involves end-devices and miners. The reward of the end-devices must be based on their dataset size and local learning model accuracy. The miners will provide the reward to the end-devices, whereas miners will receive their reward from the blockchain service providers responsible for federated learning model management.  \par   
	
	%An auctions-based, contract theory, and Stackelberg game-based approaches can be used for modeling of incentive-based federated learning \cite{pandey2019crowdsourcing, le2020auction, kang2019incentive}. A typical incentive mechanism based on a game theory involves the announcement of a reward from the edge/cloud server. The cloud/ edge server wants to maximize its own utility in terms of accuracy of global federated learning model accuracy. The end-devices involved in federated learning try to maximize their own benefit. The end-device utility can be the accuracy of the local learning model within the pre-defined fixed training duration. The local learning model accuracy is directly proportional to the incentive received by the end-device from the edge/cloud server. \par
	
	%In another work \cite{zhan2020learning}, the authors proposed  deep reinforcement learning-based incentive mechanism for federated learning over IoT networks. The system model consist of a set of edge nodes running local learning models and a remote cloud running aggregation server. 

	\subsection{Miners Classification for Federated Learning based on Blockchain}
\begin{figure}[!t]
		\centering
		\captionsetup{justification=centering}
		\includegraphics[width=8cm, height=10cm]{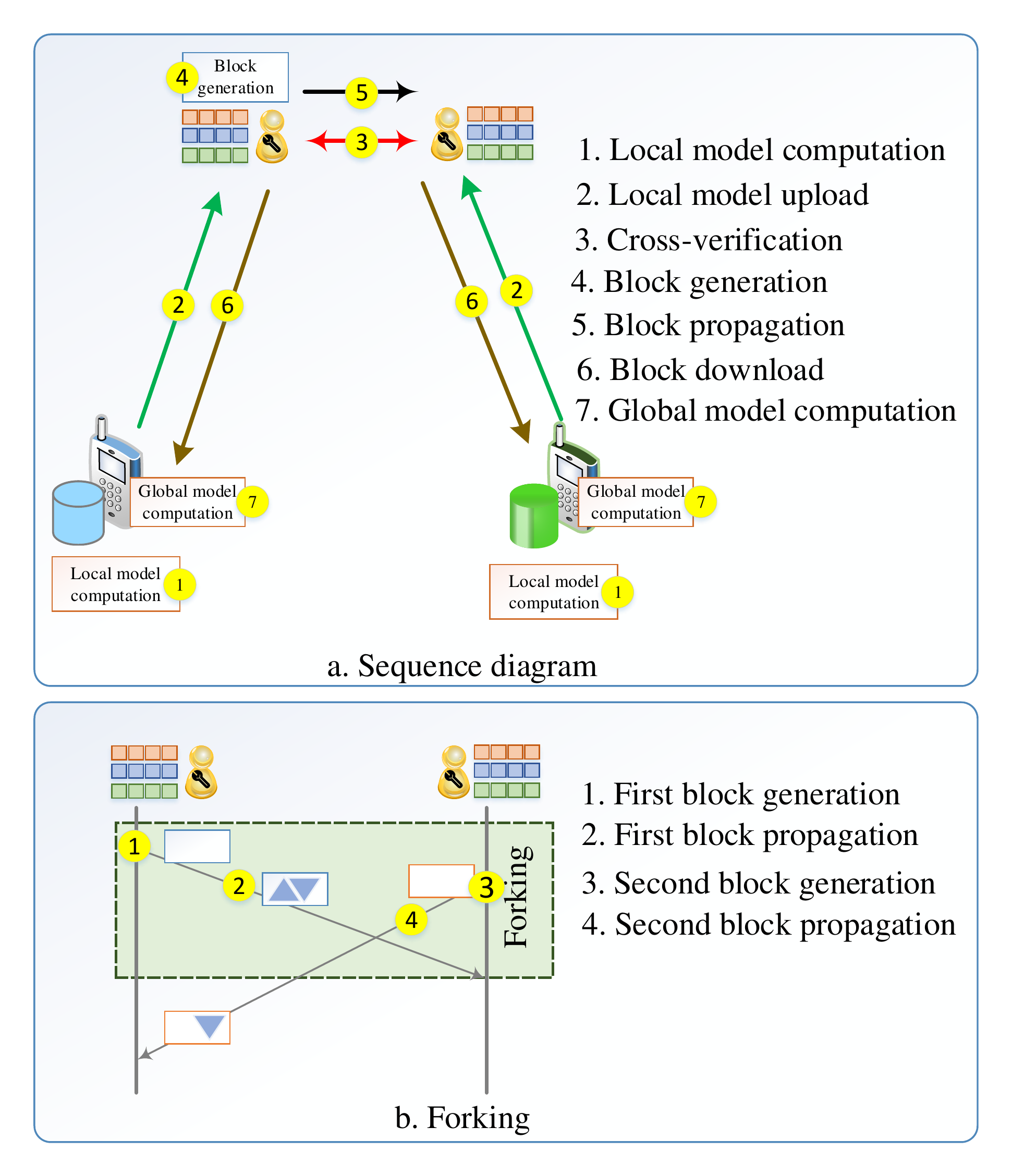}
		\caption{Blockchain-based federated learning: (a) sequence diagram, and (b) forking effect \cite{kim2019blockchained}.}
		\label{fig:BFL}
	\end{figure}

	Federated learning based on blockchain uses miners for secure and trustful exchange of learning model parameters \cite{kim2019blockchained, sharma2020blockchain, qu2020decentralized}. The miners can be either static (i.e., BS) \cite{kim2019blockchained}, mobile (i.e., autonomous cars) \cite{lee2020performance}, and flying (unmanned aerial vehicles (UAV)) \cite {alladi2020applications}. There are two main challenges in using miners for blockchain-based federated learning, such as wireless for miners and blockchain consensus design for miners. Wireless for miners deals with the wireless communication design for efficient communication, whereas blockchain consensus for miners deals with the efficient consensus algorithm design. The primary purpose of miners is to enable a trustful exchange of local learning models as shown in Fig.~\ref{fig:BFL} \cite{kim2019blockchained}. First of all, local learning models are computed for all devices and sent to their corresponding miners. The miners verify the trustfulness of the local learning model to avoid injection of wrong local learning models from a malicious user. Following cross verification, a block of a distributed ledger is generated. The block has two parts body and header. Body stores the local learning model updates and header contains control information, such as pointer to the previous block, output of the consensus algorithm, and block generation rate. Every miner has its block that stores local learning models of its associated devices. Next, every miner generates a hash value repeatedly until their hash value becomes less than the target value. After the miner succeeds in computing the hash value using proof of work (PoW), its block is considered as a candidate block which is transmitted to all other miners. All other miners that receive the newly generated block from the winning miner should stop computing their PoWs. It must be noted that the other miners may broadcast their candidate block after computing a hash value before receiving a candidate block from the other winning miner, and thus there will be two candidate blocks within a blockchain network. This will cause some of the miners to add this second candidate block, and thus will result in global federated learning model error. This event is called forking. After block propagation, the block is downloaded by all the devices and global aggregation is performed. Blockchain-based federated learning avoids the use of a centralized server, and thus offers robustness but at the cost of computational complexity associated with blockchain consensus algorithm. Additionally, more communication resources will be consumed compared to traditional federated learning. Other than these issues, it is necessary to tackle the issue of forking. One way can be to minimize the block generation rate and block transmission delay. Another possible solution can be sending of candidate block by a winning miner with a certain probability. This will reduce the change of forking event, but at the cost of block generation delay. Additionally, setting this probability will be scenario dependent, such as the number of miners and consensus algorithm used. Therefore, we must take proper attention to the design of blockchain-based federated learning.\par
	
	\begin{table}
		
		\caption {Comparison of miners used for blockchain-based federated learning.} \label{tab:miner comparison} 
		\centering
		\begin{tabular}{p{2.2cm}p{1.5cm}p{1.5cm}p{1.5cm}}
			\toprule 
			
			\textbf{Miner type} &  \textbf{Available computational power} & \textbf{Forking probability} & \textbf{Block propagation delay}  \\
			\midrule
			Static miner (edge server-based BS) & Highest & Lowest & Lowest \\
			\midrule 
			Flying miner (UAVs) & Low & High & Highest \\
			\midrule 
			Mobile miner (autonomous car) & High & Low &  Low \\

			\bottomrule 
		\end{tabular}
		
	\end{table}
	For BS-based miners, the association of end-devices has lower complexity than mobile and flying miners. Moreover, BS-based miners can communicate with each other using high-speed optical fiber backhaul links. Using fast back-haul links for transferring blocks among miners as discussed earlier will minimize the forking effect. For flying miners based on UAVs, the forking effect will be more prominent due to the large propagation delay between the miners. Furthermore, the mobility of the flying miners will add further complications to the design due to the fact the miner might get out of the communication range of the other miners. There is a need to optimize radio access network resources for BS-based miners to minimize the blockchain-based federated learning convergence time. On the other hand, mobile miners require more careful design consideration than static BS-based miners. Mobile miners can communicate with each other via fast backhaul or directly using wireless links. Comparison of various miners that can be used for blockchain-based federated learning are given in Table~\ref{tab:miner comparison}. For mobile miners' communication via a wireless channel, there is a need to carefully design resource allocation schemes for carrying out the transfer of learning model updates between miners. Similar to mobile miners, the UAV-based flying miners require a careful design for federated learning. For UAV-based miners, we need to address two kinds of resource allocation problems such as (a) communication of end-devices with UAVs and (b) communication between UAVs. Furthermore, the three-dimensional motion of UAVs imposes more design complexity \cite{zeng2020federated, fan2018optimal,mozaffari2018beyond, mozaffari2019tutorial}.

	\begin{figure*}[!t]
		\centering
		\captionsetup{justification=centering}
		\includegraphics[width=18cm, height=13cm]{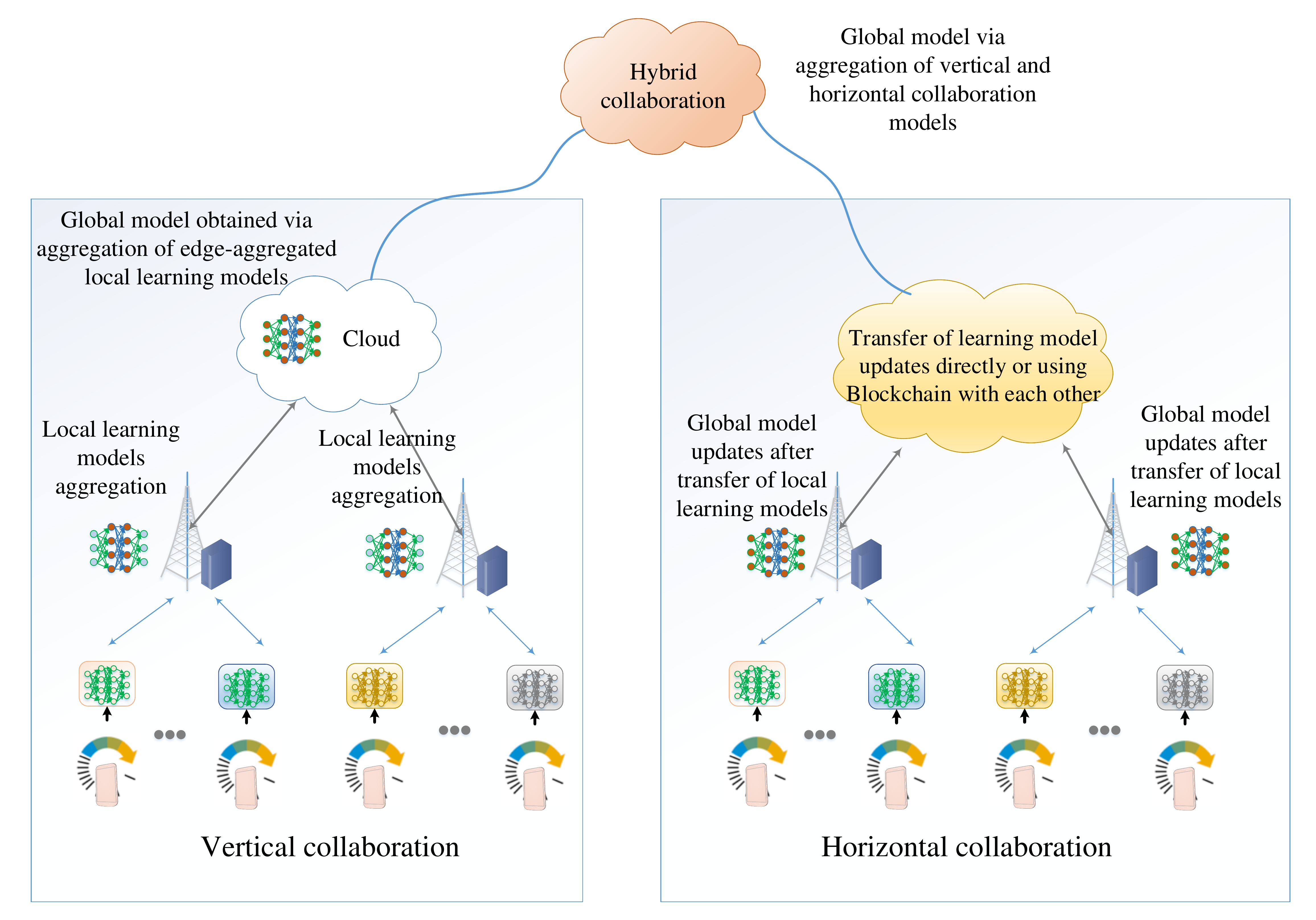}
		\caption{Edge collaboration categories.}
		\label{fig:edge_collaboration}
	\end{figure*}

	\subsection{Edge Collaboration}
	\label{Edge Collaboration}
	Training of a cloud-based federated learning model for a massive number of IoT devices results in significant overhead of communication resources \cite{khan2019federated}. To cope with this issue, one can consider collaboration between edge servers and a cloud. In \cite{liu2019client}, a client-edge-cloud hierarchical federated learning framework has been proposed. The framework resolves high communication resources consumption issue associated with the training of cloud-based federated learning models for a massive number of users via collaboration between edge servers and a cloud server. The client-edge-cloud hierarchical federated learning framework performs two levels of aggregations at edge servers and a cloud. Access to the edge servers is easier and requires fewer communication resources than a remote cloud. Therefore, performing edge based-aggregations and then cloud-based aggregation will offer efficient usage of communication resources than the federated learning scheme based on communication between end-devices and cloud. Edge collaboration can be (a) horizontal, (b) vertical, and (c) hybrid (as shown in Fig.~\ref{fig:edge_collaboration}), depending on the nature of application and available communication resources. \par
	Although federated learning based on collaboration between edge servers and edge-cloud servers can improve the performance in terms of accuracy, it will face weight divergence issues. For collaboration among edge servers and a cloud, there will be two kinds of weights divergences, such as (a) between end-devices and edge server and (b) between edge servers and cloud. These weight divergences depend on the federated learning scheme used and data heterogeneity. For FedAvg using non-IID data and a fixed product of local iterations and edge aggregations prior to global aggregation, increasing the number of edge aggregations will improve performance \cite{wang2020local}. According \cite{wang2020local}, the reason for this weights divergence is more for a higher number of local iterations compared to edge aggregations for a fixed product of local iterations and edge aggregations prior to cloud aggregation. It must be noted here that we should perform extensive analysis for federated learning based on hybrid collaboration as shown in Fig.~\ref{fig:edge_collaboration}. Additionally, using a federated optimization scheme different than FedAvg might result in different observations than those of \cite{wadu2020federated}. Therefore, there is a need to properly adjust the federated learning parameters and select an optimization scheme for federated learning based on collaboration. 
	
	%Later in Section~\ref{ }, we introduce a novel concept of dispersed federated learning-based mainly on horizontal collaboration.      

	\subsection{Cloud Server Design}
	To design federated learning based on the cloud, there is a need to properly plan the design of cloud-based aggregation servers. The cloud-based server will coordinate with devices ranging from hundreds to millions \cite{bonawitz2019towards}. Additionally, the size of collected updates ranges from kilobytes to tens of megabytes. Therefore, to successfully carry out the federated learning process in an economical way, there is a need to optimally use cloud resources. To truly realize cloud as a dynamic micro data center, there is a need to evolve both hardware and software. Other than hardware evolution, we must effectively isolate functionality from the hardware. Micro-services concept can be used in such a type of isolation. The key requirement of micro-services is a similar environment at all run-time locations. To run micro-services, there can be two ways, i.e., containers-based design \cite{park2017design,wu2017container,kang2016container } and virtual machine-based design \cite{zhong2016virtual, lagar2009snowflock}. Containers can be preferably used compared to virtual machines due to their low overhead \cite {khan2019edge}. Therefore, we can use containerization and hardware evolution for economical cloud server design to enable federated learning for millions of devices.

	\begin{comment}
	\begin{table*}[]
	%\rowcolors{2}{gray!25}{white}
	\caption {Comparison of federated optimization/learning schemes.} \label{tab:locallearning} 
	\begin{center}
	\begin{tabular}{p{3.5 cm}p{6cm}p{2.5cm}p{1.5cm}p{1.5cm}}
	\toprule 
	\textbf{Federated learning scheme}   & \textbf{Local model computation} & \textbf{Aggregation} &\textbf{Fairness} & \textbf{Tasks} \\ \midrule
	FedAvg & $F_{n}=\frac{1}{K_{n}}\sum_{k=1}^{k_{n}}{f_{k}(\boldsymbol{w})}$ & $\frac{1}{K}\sum_{k=1}^{k_{n}}{F_{k}(\boldsymbol{w})}$  & No & Single task \\ \midrule
	FedProx & $F_{n}(\omega)=\frac{1}{K_{n}}\sum_{k=1}^{k_{n}}{f_{k}(\boldsymbol{w})}+\frac{\mu}{2} \parallel \omega -\omega^{t} \parallel^{2}$ & $\frac{1}{K}\sum_{k=1}^{k_{n}}{F_{k}(\boldsymbol{w})}$  & No &  Single task\\ \midrule
	q-FedAvg &  &  & Yes & Single task \\ \midrule
	FML &  &   & No & Multi task \\ 
	\bottomrule 
	
	\end{tabular}
	\end{center}
	\end{table*} 
	
	\end{comment}

	\subsection{Federated Optimization Schemes}
	The process of federated learning consists of local model computation by end-devices, which are then sent to the aggregation server. Finally, after global aggregation, the global model parameters are sent back to end-devices. Such a process continues in an iterative fashion until convergence. The goal of federated learning to minimize global loss function that is application dependent. To minimize the global loss function, we need federated optimization schemes. Mainly there are two types of federated optimization schemes depending on the number of tasks: single task federated optimization and multi-task federated optimization. In single task federated learning, the global federated learning model is trained only for a single task. Several works considered the training of the global federated learning model for a single task \cite{mcmahan2017communication, li2018federated}. On the other hand, multi-task federated learning involves the training of multiple models for different tasks \cite{smith2017federated}. \par
	FedAvg and FedProx were proposed for training of a single-task federated learning models. FedAvg was based on simple averaging of local learning models at the aggregation server. However, there exist statistical and system heterogeneity among the end-devices. Therefore, simple averaging in FedAvg might not work well. To cope with this challenge, FedProx was proposed \cite{li2018federated}. The difference in computation steps between FedProx and FedAvg lies in the local device objective function. FedProx uses an additional proximal term for minimizing the impact of heterogeneity of end-device on the global federated learning model. Although FedProx was developed to account for statistical and system heterogeneity, it faces some practical implementations challenges \cite{khan2020dispersed}. The weighted proximal term in FedProx is difficult to choose for various applications. Furthermore, the answer to the question, that can a FedProx provably improve the convergence rate is not clear \cite{kairouz2019advances}. A scheme $q$-FedAvg was proposed to cope with fairness issues due to system heterogeneity (i.e., local model accuracy and wireless resources) \cite{li2019fair}. The $q$-FedAvg modifies the objective function of FedAvg by assigning the end-devices with poor performance higher weights than better-performing end-devices. In $q$-FedAvg, the devices with higher values of local loss functions are given more importance using large values of $q$. Specifically, the amount of fairness is determined by the values of $q$. In another work \cite{wang2019adaptive}, a scheme offering a trade-off between the local model update and global model aggregation with the aim of reducing the loss function under resource budget constraints was proposed. In \cite{li2018federated, li2019fair, wang2019adaptive}, all the schemes considered a single task global federated learning model. To offer multi-task learning, federated multi-task learning (FML) was proposed in \cite{smith2017federated}. Additionally, the FML scheme considered joint optimization of computational and communication resources while fulfilling the fault tolerance constraint. \par
	
	The iterative exchange of learning model parameters in the federated optimization schemes via wireless channels requires careful design. It will be necessary to develop a novel federated optimization algorithm that considers significantly the effect of packet errors introduced due to channel uncertainties on the global federated learning model. We must define some threshold regarding the packet error rate to determine whether the devices local learning model parameters should be considered or not in the global federated learning model computation. The devices the higher packet error rates must not be considered in training of the global federated learning model computation. One of possible way to improve the packet error rate is the use of channel coding schemes (e.g., turbo codes, convolutional codes, and linear block codes). Different channel coding schemes have different performance and overhead with complexity. Generally, turbo codes outperforms linear block codes but at the cost of communication overhead. Another feasible way can be the use of URLLC codes, such as BCH codes. Therefore, we must make a trade-off while selecting a channel coding scheme for federated learning.\par
	
	\begin{figure}[!t]
		\centering
		\captionsetup{justification=centering}
		\includegraphics[width=8cm, height=8cm]{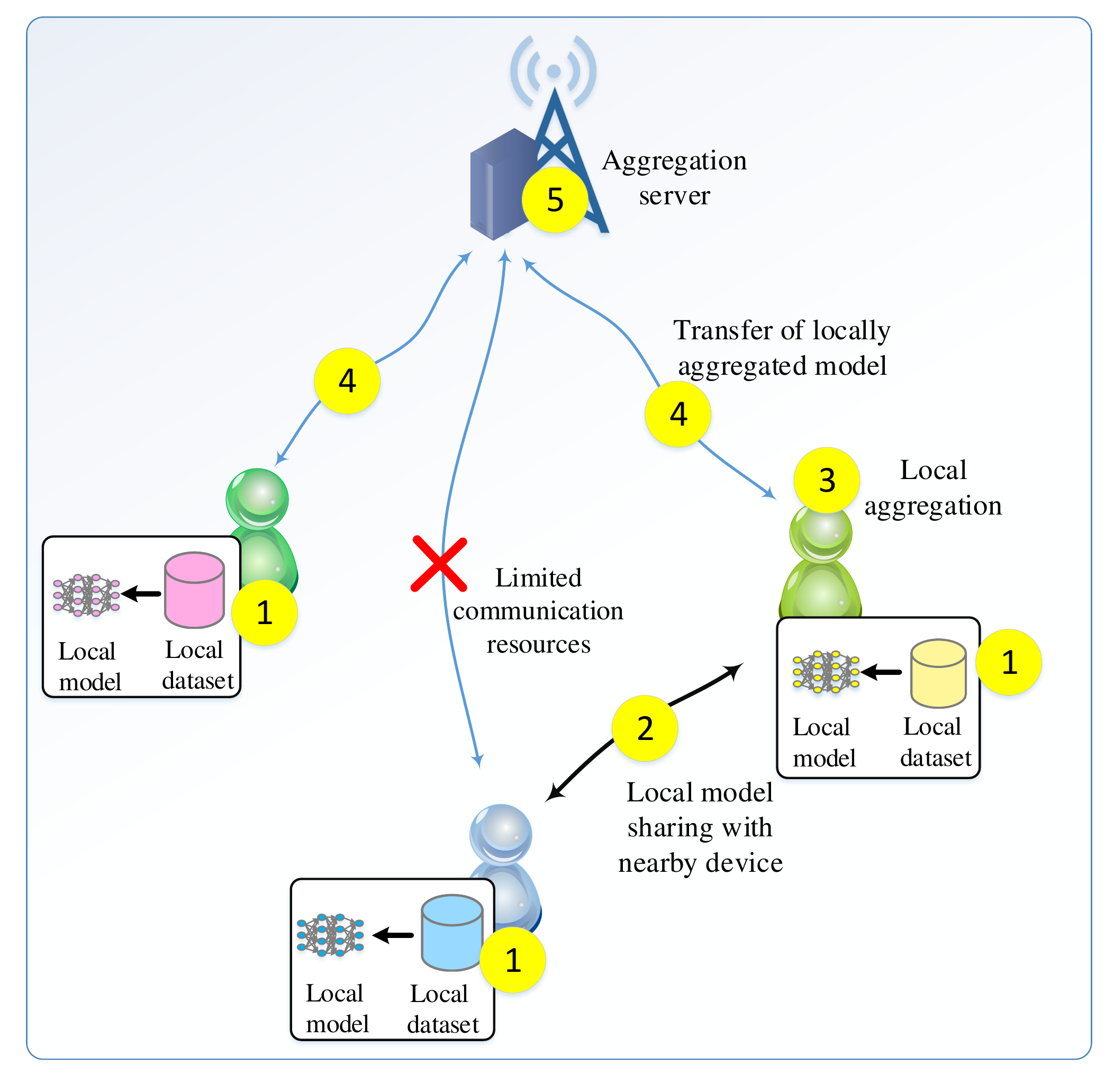}
		\caption{Collaborative federated learning \cite{chen2020wireless}.}
		\label{fig:collaborativeFL}
	\end{figure}

	\begin{figure}[!t]
		\centering
		\captionsetup{justification=centering}
		\includegraphics[width=8cm, height=6.5cm]{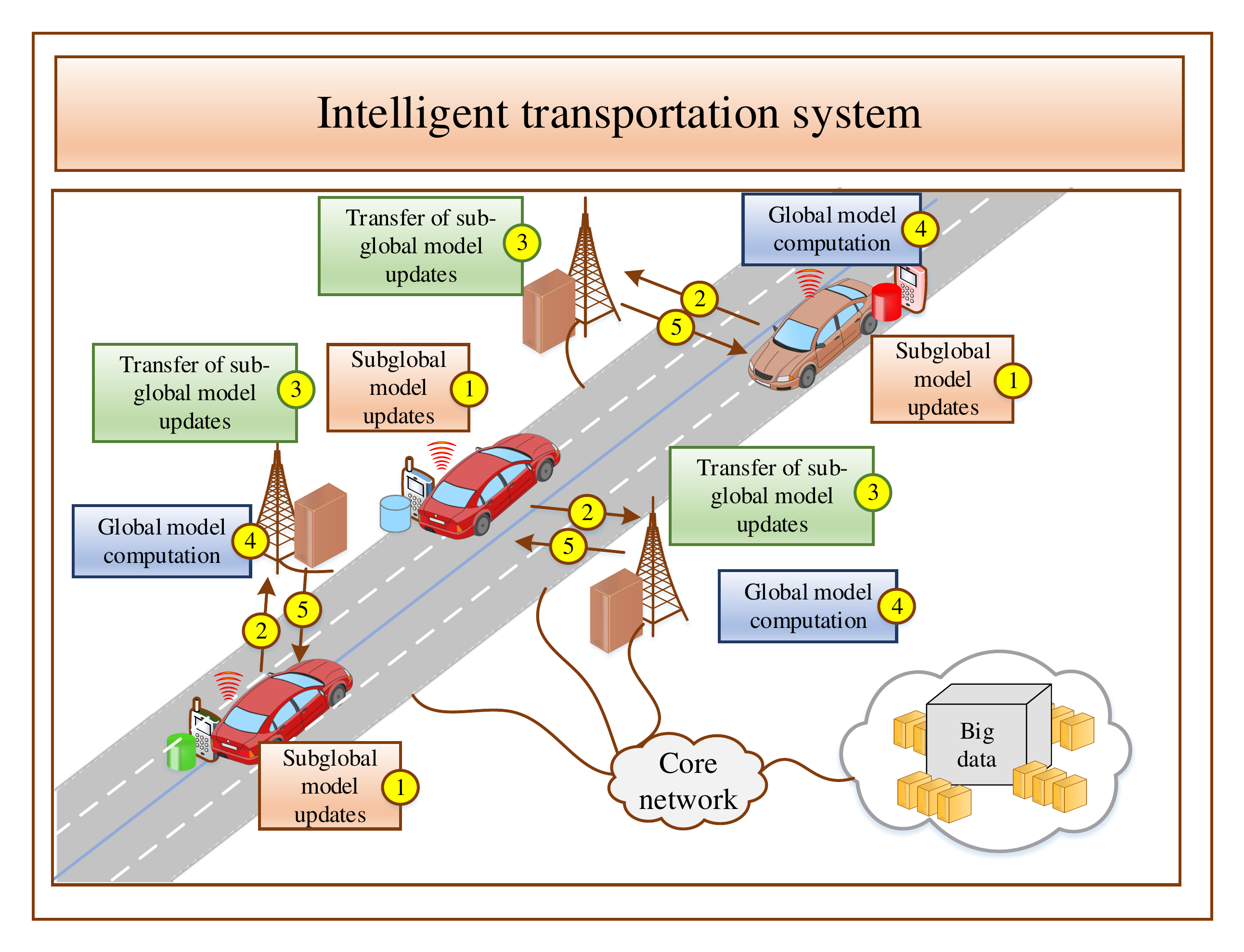}
		\caption{Intelligent transportation use case of dispersed federated learning.}
		\label{fig:usecases}
	\end{figure}

	\subsection{Operation Modes Based on Global Aggregation Fashion}
	We can categorize federated learning into five categories depending on the fashion of global aggregation as centralized aggregation-enabled federated learning, distributed aggregation-enabled federated learning, collaborative federated learning, hierarchical federated learning, and dispersed federated learning \cite{mcmahan2017communication,chen2020wireless, khan2020dispersed}. In centralized federated learning, global aggregation takes place only at the centralized edge/cloud server, whereas decentralized aggregation-enabled federated learning avoids the use of a centralized aggregation by performing aggregations at distributed servers. However, hierarchical federated learning involves aggregation of local learning models (e.g., at edge servers) prior to global aggregation at the cloud. In distributed aggregation-enabled federated learning, end-devices are associated with multiple aggregation servers at BSs. The BSs then share the received local learning models from their associated devices with other devices. Finally, aggregation takes place at BSs in a distributed manner. This mode of federated learning can offer robustness but at the cost of additional communication resources associated with sharing of local learning model updates between BSs. Moreover, it will add delay due to sharing of local learning models between BSs. On the other hand, hierarchical federated learning combines both distributed and centralized aggregation-enabled federated learning. Several aggregations of local learning models take place at edge, which is followed by sending the edge aggregated models to the cloud for global aggregation. Hierarchical federated learning offers advantage of using low cost communication for connecting end-devices with the edge servers. At the network edge, one can use easily reuse the already occupied frequency bands by other cellular users while protecting them by keep interference level below the maximum allowed limit.\par
	For scenarios that have massive number of IoT devices and constrained communication resource, it will be difficult for devices to participate in the federated learning process for the aforementioned federated learning schemes. To address this challenge, one can use collaborative federated learning \cite{chen2020wireless}. The collaborative federated learning was introduced mainly for ensuring participation of devices that are unable to access the centralized BS due to communication resources constraints as shown in Fig.~\ref{fig:collaborativeFL}. In collaborative federated learning, the devices with insufficient communication resources send their model to other nearby devices with sufficient communication resources. The receiving devices perform aggregation of other devices' local learning models, with their own models and send the aggregated model parameters to the centralized BS. Some of the devices send their own local learning models directly to the centralized BS. The centralized BS finally performs global aggregation and sends back the global model parameters to the end-devices. Although collaborative federated learning offers benefits of enabling more devices to participate in the federated learning process, there is a need for successful communication between devices for sharing the learning model updates. The primary challenge in enabling collaborative federated learning lies in the network formation that will define the communication among devices for local aggregations as shown in Fig.~\ref{fig:collaborativeFL}. We can use two types of operation for collaborative federated learning, such as synchronous and asynchronous. In a synchronous fashion, the aggregation server must receive a learning model (i.e., local learning models and locally aggregated models) within a maximum allowed time prior to aggregation. To do so, resource-constrained devices must send their local learning models to nearby devices according to the network formation topology within the allowed time. Next, after local aggregation at receiving devices, the locally aggregated models are sent to the centralized aggregation server. It must be noted that the time from local learning model computation to global aggregation in synchronous collaborative federated learning must be less than or equal to the maximum threshold. On the other hand, asynchronous collaborative federated learning has more freedom of computing locally aggregated models. Here, the resource-constrained devices can send their local learning models to nearby devices where local aggregation takes place. If the locally aggregated models are received by the centralized aggregation server later than the some fixed time, then it will consider the local aggregation model in the next round rather than discarding that happens in synchronous collaborative federated learning. Although asynchronous collaborative federated learning allows freedom of allowing more devices in the learning process, it will offer design challenges. Performance degradation in asynchronous collaborative federated learning might occur due to less number of participating end-devices in one global round. Additionally, asynchronous collaborative federated learning might not converge fast for extreme non-IID conditions with most of the end-devices having a distinct class of data. To resolve this challenge, one can use synchronous collaborative federated learning by using fewer local iterations to reduce the local model training time so as to reduce the overall time consumed (i.e., local training and local aggregations along with transmission delay) before the centralized server aggregation. \par
	
	\begin{table*}[]
		%\rowcolors{2}{gray!25}{white}
		\caption {Comparison of federated learning schemes based different aggregation fashions.} \label{tab:locallearning} 
		\begin{center}
			\begin{tabular}{p{3.5 cm}p{2.7cm}p{2.5cm}p{2.5cm}p{2.5cm}}
				\toprule 
				\textbf{Federated learning scheme}   & \textbf{Robustness} & \textbf{Communication resources consumption} &\textbf{Implementation complexity} & \textbf{Global model computation latency} \\ \midrule
				& Robustness deals with the successful operation of federated learning scheme in case of failure of aggregation server due to physical damage or security attack. &This parameter refers to the use of communication for federated learning global model computation. & This parameter deals with the implementation complexity of the algorithm used for computing a global federated learning model. & Global model computation latency is the overall latency in computing global federated learning model for one global round.  \\ \midrule
				Centralized aggregation-based federated learning & Lowest & Low (for edge-based federated learning) & Lowest & Low \\ \midrule
				Distributed aggregation-based federated learning & High & Highest (for blockchain-based federated learning using wireless miners) & Highest  (for blockchain-based federated learning using wireless miners) & High\\ \midrule
				Collaborative federated learning & Low & Low & High (for mobile nodes) & Low \\ \midrule
				Hierarchical federated learning & Low & Low  & Low (edge and cloud based) & Low \\ \midrule
				Dispersed federated learning & High & Low & Low (edge servers based) & Low \\ 
				\bottomrule \\
				\multicolumn{5}{c}{\noindent * One sub-global and one-edge aggregation are considered for dispersed federated learning and hierarchical federated learning}\\
				\multicolumn{5}{c}{used for comparison, respectively.}\\
				
			\end{tabular}
		\end{center}
	\end{table*}

	The aforementioned centralized aggregation-enabled federated learning, hierarchical, and collaborative federated learning faces a robustness challenge. Additionally, these schemes will suffer from privacy leakage due to the end-devices sensitive information inferring capability of a malicious aggregation server/ end-devices. One can enable differential privacy via addition of artificial noise to the learning model parameters prior to sending them to the aggregation server \cite{Dusit_FL_survey}. However, simultaneously achieving convergence time and high privacy level for federated learning is generally not possible. We must make a tradeoff between the federated learning convergence time and privacy level. Additionally, enabling federated learning with privacy preservation schemes adds extra complexity, which might not be desirable. To cope with the aforementioned issues of federated learning over IoT networks, in \cite{khan2020dispersed}, the novel concept of \emph{dispersed federated learning} was proposed. In dispersed federated learning, sub-global federated learning models are computed within different groups, in a fashion similar to that in federated learning. Then, the sub-global models are transferred between different groups. Next, sub-global models are aggregated to yield the global federated learning model. This process of dispersed federated learning takes place in an iterative manner until desirable global federated learning accuracy is achieved. For dispersed federated learning, we can use reuse communication resources within groups used for sub-global model computation to enable communication efficient operation. Additionally, there is a need to efficiently allocate resources for further performance enhancement. In contrast to federated learning global aggregation server, a global aggregation server used to aggregate sub-global models can not easily infer the end-devices private information, and thus offer enhanced privacy preservation. However, within group used for sub-global computation, the sub-global aggregation server can infer the end-devices private information. To add more privacy preservation to dispersed federated learning, there is a need to develop a novel privacy preservation schemes within groups of dispersed federated learning.\par
	Dispersed federated learning can be easily adopted in many practical IoT applications. An intelligent transportation system suffers from frequent addition of data updates which motivated use of federated learning in contrast to centralized machine learning. For instance, according to statistics, autonomous driving cars generate $4,000$ gigaoctet everyday \cite{FL_industry_1}. Therefore, transferring the whole data for training is not a feasible solution due to high communication overhead. Instead, we can use federated learning that is based on sending of only learning model updates \cite{liang2019federated}. However, autonomous driving cars suffer from high mobility which causes frequent handoffs. An IoT-based smart device inside an autonomous driving car might not be able to maintain seamless connectivity with the RSU during the training process for federated learning \cite{yaqoob2019autonomous}. To tackle this issue, we can use a dispersed federated learning for autonomous driving cars. Fig.~\ref{fig:usecases}  shows an overview with a sequence diagram for a smart transportation use case. First of all, a sub-global model is computed within every autonomous car in an iterative manner similar to traditional federated learning model computation. The set of devices within autonomous driving car exchange their local learning modes with the edge computing-based access point. After a particular number of sub-global iterations, the sub-global model updates are sent to the associated RSU by every autonomous driving car. The next step is sharing of the sub-global model updates between different RSUs. The transfer of sub-global model updates is possible using (a) blockchain-based transfer, (b) direct transfer, and (c) light-weight authentication-based transfer. A blockchain-based transfer provides secure transfer and trustful verification of sub-global model updates. However, it suffers from inherent disadvantage of high latency associated with consensus algorithm. The direct transfer of sub-global model updates between different RSUs might suffer from false data injection, and thus prone to error due to malicious attacks. To cope with the aforementioned issues of false data injection and high-latency, we can use light-weight authentication-based scheme for secure transfer of sub-global model updates.     
	
	\subsection{End-Device Design for Federated Learning}
	\label{device_design}
	On-device learning is the key essence of federated learning. Enabling end-devices with local machine learning requires careful design requirements. Every device is characterized by a fixed maximum computational power (CPU-cycles per) and limited backup power. For fixed dataset points, to minimize local learning model computation time, there is a need to run the device at high CPU-cycles/sec. However, device power consumption increases with an increase in operating frequency. Therefore, there is a need to trade-off between the end-devices operating frequencies and power consumption. The key metrics generally considered for performance evaluation of embedded machine learning are cost, throughput/latency, energy consumption, and learning accuracy. For a fixed dataset, the performance of end-devices involved in federated learning depends on both hardware design and software design \cite{khan2019federated}. On the other hand, for a fixed hardware and software design, the performance matrices are strongly dependent on the used dataset and application. \par      
	Various datasets and applications offer a variety of complexities to the embedded machine learning design. For instance, MNIST \cite{mcmahan2017communication} dataset used for image classification task has lower complexity than GoogLeNet \cite{szegedy2015going} dataset that is used for classification of $1000$ image classes. Therefore, there is a need for efficient hardware design with programmability for end-devices to handle various applications offering different complexities. The programmable nature enables the use of generic hardware for different models. The requirement of programmability and high dimensionality causes an increase in data movement and computation power consumption due to more generated data, respectively \cite{sze2017hardware}. The programmability deals with storage and reading the weights, which results in significant energy consumption \cite{horowitz20141}. Application-specific manycore processors are attractive candidates to enable different designs while fulfilling area, temperature, and energy constraints \cite{ceze2016arch2030, jouppi2017datacenter}. However, various challenges exist in the implementation of these systems. First of all, fine-tuning is required for implementing effective run-time resource managers. Second, the use case significantly modifies the design objectives and deployment platform. Third, the evaluation of these designs is expensive and slow. Machine learning can be used to provide highly optimized design. Kim \textit{et al.} \cite{kim2018machine}, proposed machine learning to offer computationally efficient hardware design optimization for manycore systems.\par
\begin{figure}[!t]
		\centering
		\captionsetup{justification=centering}
		\includegraphics[width=8cm, height=9cm]{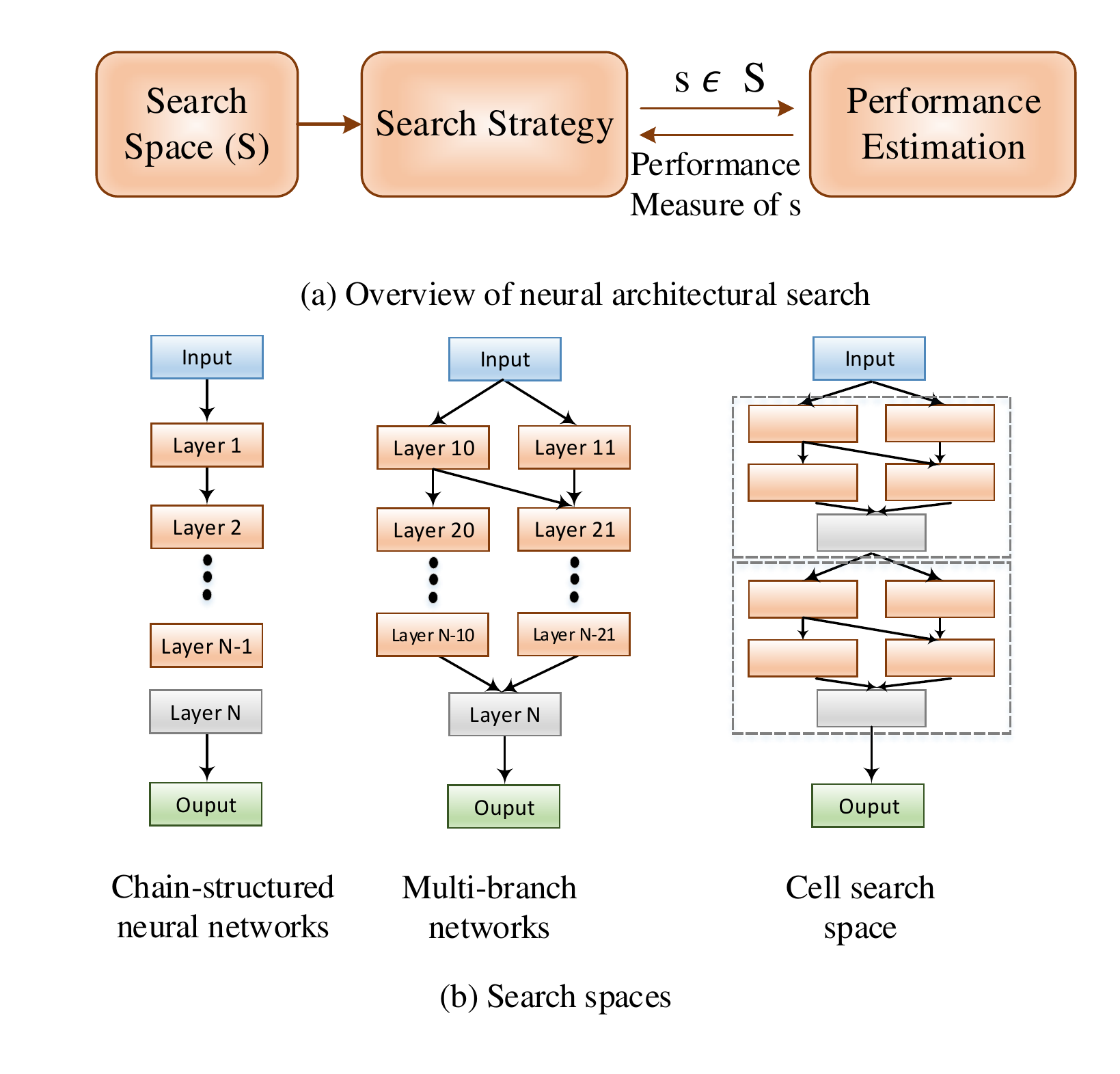}
		\caption{Neural architecture search: (a) Overview, and (b) Search spaces \cite{NAS_1}.}
		\label{fig:NAS}
	\end{figure}

 
	For fixed hardware design, we can search different neural architectures for optimal out of available ones. NAS allows us to find the optimal architecture out of many available architectures for a particular dataset and application \cite{liu2018progressive, pham2018efficient,xie2018snas,ying2019bench}. NAS can be considered to be a subfield of automated machine learning that deals with the complete process from dataset to its preferable machine learning models without knowing expert-level knowledge of machine learning \cite{NAS_1}. An overview of a NAS is given in Fig.~\ref{fig:NAS}a \cite{elsken2018neural}. Several possible neural architectures can be chain-structured, multi-branch networks, and cell search space, are shown in Fig.~\ref{fig:NAS}b. The chain-structured neural networks consist of a sequence of layers with the first layer as input and last layer as output. Each layer receives an input from the preceding layer. In such a type of neural search space, we can adjust the number of layers, operation at each layer (i.e., convolution, depth-wise separable convolutions, or pooling), and hyperparameters (i.e., number of filters and kernel size and strides for a convolutional layer). The hyperparameters are dependent on operation layer, and thus this search space is conditional. On the other hand, multi-branch networks have nodes that can take their inputs as function of previous layers' output as shown in Fig.~\ref{fig:NAS}b. The multi-branch networks has more degree of freedom than chain-structured neural networks. To provide more freedom of design, we can use the cell search space approach that involves searching in cells.\par

	\subsection{Lessons Learned and Recommendations}
	We have provided a taxonomy for federated learning towards enabling IoT-based smart applications. Some of the lessons and future enhancements learned are as follows.   
	
	\begin{itemize}
		\item It is necessary to adaptively adjust local iterations and global iterations strictly depending on the nature of the application. The end-device in an IoT network with high mobility might not have seamless connectivity with the aggregation server all the time. In these scenarios, it is recommended to consider a high number of local iterations. For a fixed global federated learning model accuracy, an increase in the number of local iterations will require less global iterations. Therefore, a higher number of local iterations with few global iterations are desirable in high mobility networks. A practical example such networks is vehicular networks that have high mobility vehicles. 
		
		\item A massive number of IoT nodes show a significant amount of heterogeneity in terms of local dataset sizes and distributions, available backup power, computation resource, and communication resources. All of these factors significantly affect the performance of federated learning algorithms (i.e., FedAvg, FedProx, etc.). Therefore, to truly benefit from the deployment of federated learning in IoT networks, it is indispensable to design federated learning algorithms that consider these parameters. For instance, some of the end-devices might have poor performance due to limited computation power, noisy dataset, or limited communication resources. Such a type of end-device will have less influence on the global model than the device with better performance. Coping with these kind of issues due to the heterogeneity of devices, we must propose new federated learning algorithms.

		\item Considering the aforementioned software-aware design and hardware design schemes, it is necessary to consider hardware-software co-design for end-devices involved in federated learning. The hardware-software co-design allows the concurrent design of both hardware and software. Numerous ways that can be used for hardware-software co-design are high-level synthesis, co-verification-based embedded systems, and virtual prototyping. The virtual prototyping-based design leverages computer-aided engineering, computer-aided design, and computing automated design, to validate the design prior to implementation. A high-level synthesis design enables automated digital hardware using an algorithmic description of the desired system. Multiplexer and wired signal delays are prominent challenges in high-level synthesis-based design. In co-verification-based design, embedded systems are designed using testing and debugging of both hardware and software at the same time. Such a design offers better performance.\par   
		
		\item A malicious end-device can send a learning model updates to an aggregation server which can result in prolonging the federated learning convergence time. Therefore, one can propose novel federated learning frameworks that enable trust verification of end-device before using its local learning model updates updates. One way is through blockchain-based authentication schemes. Another way is through light-weight authentication schemes.   
		
		\item As federated learning involves iterative exchange of learning model updates between the end-devices and aggregation, efficient communication resources allocation schemes will be required. The two kinds of communication resources used in federated learning are core network and radio access network. For optical fiber-based core networks, there are enough communication resources. On the other hand, the radio access network has communication resources constraints. Therefore, wireless resources optimization problem will need to be solved for federated learning. Depending on the nature of the formulated problem, we can use various approaches. These approaches include heuristic approaches, matching theory-based schemes, and game theory-based schemes.    
		
		\item FML learning was proposed in \cite{smith2017federated} to enable learning of multi-tasks whose data are distributed among multiple nodes with federated learning settings (i.e., data and system heterogeneity). To enable federated learning for a different tasks, efficient scheduling over resource-constrained radio access network is required for all tasks. Moreover, resources must be assigned as per the nature of the learning task. For instance, more resources should be assigned to the critical tasks (i.e., industrial control) than ordinary tasks (i.e., keyboard keyword suggestion). Assigning more resources lowers the convergence time for federated learning. Therefore, novel joint scheduling and resource allocation schemes for FML will need to be developed.   
		
		\item In dispersed federated learning, the communication of the sub-global aggregation server with end-devices is more frequent than with the centralized aggregation server. Therefore, we can use high-performance channel coding schemes (i.e., turbo codes) having high overhead for reliable communication between the sub-global aggregation server and the global aggregation server. It must be noted that the sub-global model results from frequent, iterative aggregation and exchange of various local learning models, therefore, it must be provided with more reliable communication. On the other hand, we can use low-overhead linear block codes or convolutional codes for reliable communication between end-devices and sub-global aggregation server due to their frequent communication.  
		
		\item There is a need to propose novel communication resource-efficient federated optimization schemes for dispersed federated learning. One can use D2D communication-assisted dispersed federated learning. In D2D communication assisted dispersed federated learning, a set of closely located devices form clusters which is followed by cluster head selection using some attractive criteria (e.g., socially-aware interaction). The devices within a cluster compute their local learning model and exchange it with a cluster head in an iterative manner to yield a sub-global model. The advantage of sub-global model computation lies in the reuse of the already occupied spectrum by the cluster devices. The sub-global model updates from all the cluster heads are sent to the BS where the global model aggregation takes place. Finally, the global model updates are sent back to the cluster heads which disseminate the global model to the devices.     
		
		\item Mobility management for groups used to compute sub-global models in a dispersed federated learning is of significant importance. A set of devices within a sub-group used for sub-global model computation is desirable to remain within the group until the iterative computation of the sub-global model. Therefore, we must handle the issue of mobility.  
		
	\end{itemize}

	\section{Open Research Challenges}
	\label{sec:challenges}
	This section presents various open challenges that exist in enabling federated learning over IoT networks. The challenges presented in our paper are different than those of existing surveys and tutorials related to federated learning for wireless networks, as given in Table~\ref{tab:challengessummary} \cite{Dusit_FL_survey, li2019federated, li2019federated2, edgeAI}. We present causes and possible solutions to these challenges. Summary of the open research challenges with their possible solutions for federated learning are given in Table~\ref{tab:challenges}. 
	
	\begin{table}[]
		%\rowcolors{2}{gray!25}{white}
		\caption {Summary of existing surveys and tutorials open research challenges.} \label{tab:challengessummary} 
		\begin{center}
			\begin{tabular}{p{2.2 cm}p{5.7cm}}
				\toprule 
				\textbf{Reference}   & \textbf{Challenges} \\ \midrule
				Lim \textit{et al.}, \cite{Dusit_FL_survey} & Dropped participants, privacy concerns, unlabeled data,	interference among mobile devices, communication security, asynchronous FL, comparisons with other distributed learning methods, further studies on learning convergence, usage of tools to quantify statistical heterogeneity, combined algorithms for communication reduction, cooperative mobile crowd ML, applications of FL. \\ \midrule
				Li \textit{et al.}, \cite{li2019federated} & Extreme communication schemes, communication reduction and pareto frontier, novel models of asynchrony, heterogeneity diagnostics, granular privacy constraints, beyond supervised learning, productionizing federated learning, benchmarks. \\ \midrule
				Wang \textit{et al.}, \cite{edgeAI} & Accelerating AI tasks by
				edge communication and computing systems, efficiency of In-Edge AI for real-time
				mobile communication system toward 5G, incentive and business model of In-Edge AI.  \\ \midrule
				Our Tutorial & Sparsification-enabled federated learning, data-heterogeneity-aware-clustering-enabled federated learning, mobility-aware federated learning, homomorphic encryption-enabled federated learning, secure and trustful aggregation-enabled federated learning, interference-aware resource efficient federated learning, interference-aware association for federated learning, quantization-enabled federated learning, adaptive-resource-allocation-enabled federated learning.   \\ 
				\bottomrule 
			\end{tabular}
		\end{center}
	\end{table}

	%\subsection{Learning Time Optimization} 

	\subsection{Sparsification-Enabled Federated Learning}
	\setlength{\parindent}{0.7cm}How do we enable federated learning for a massive number of heterogeneous devices under strict wireless resource constraints? Optimal use of limited wireless resources in federated learning for a massive number of heterogeneous devices can be enabled via sparsification, which allows only a small set of devices to send their local learning model parameters to edge/cloud server. The criterion for choosing the set of devices to send their data to the edge/cloud server is a challenging task. One way is to choose the devices with gradients of magnitude greater than a certain threshold. However, computing the optimal value of the threshold for a massive number of heterogeneous devices is difficult. Another criterion for the selection of end-devices can be their local learning model accuracy which is affected by many factors such as clean datasets, local computational power, and backup power. Furthermore, devices with degraded performance over wireless channels result in prolonging the federated learning convergence time due to high packet error rates. To address the above challenges, there is a need for novel effective sparsification enabled federated learning protocols.\par

	\begin{table*}
		% \makegapedcells
		\caption {Summary of the research challenges and their guidelines.} \label{tab:challenges} 
		\centering
		\resizebox{\textwidth}{!}{
			\begin{tabular}{p{3cm}p{3.5cm}p{5cm}p{5cm}}
				\toprule 
				
				\textbf{Challenges} & \textbf{Taxonomy relevancy} &  \textbf{Causes} & \textbf{Guidelines} \\
				\midrule
				\textbf{Sparsification-Enabled Federated Learning} & Federated optimization schemes (client selection protocol)  &\begin{itemize} \item  Significant changes in the in end-devices local accuracy performance due to noisy datasets, local computational power, and local energy. \item Wireless resources constraints.  \end{itemize} & \begin{itemize} \item  Gradient-based sparsification scheme. \item Selection of devices with sufficient computational, clean datasets, and communication resources. \end{itemize}\\
				\midrule
				
				\textbf{Mobility-Aware Federated Learning}  & Federated optimization schemes  & \begin{itemize} \item Connectivity loss due to mobility during training phase. \item Variations in SINR due to mobility. \end{itemize} & \begin{itemize} \item Priority in selection of static and low-mobility devices for training. \item Deep learning-based mobility prediction schemes.  \end{itemize} \\
				\midrule
				
				\textbf{Data-Heterogeneity-Aware-Clustering-Enabled Federated Learning} &   Federated optimization schemes & \begin{itemize} \item Existence of significant statistical heterogeneity among massive number of devices datasets. \item Statistical heterogeneity significantly degrades the federated learning convergence performance. \end{itemize} & \begin{itemize} \item Clustering based on statistical homogeneity. \item Selection of most trustful end-devices acting for aggregation within clusters. \end{itemize} \\
				%    \midrule

				% \textbf{Distributed-resource-fairness-enabled dispersed federated learning} & \begin{itemize} \item Variable local learning model accuracy due to end-devices systems heterogeneity. \item Random variations in wireless channel conditions results in performance degradation in terms of loss in global federated learning accuracy.  \end{itemize} & \begin{itemize} \item Adaptive weights assignment to end-devices local learning models at the aggregation server based on local model accuracy. \item Selection of end-devices with better performance in federated learning.  \end{itemize} \\
				\midrule
				\textbf{Homomorphic encryption-enabled federated learning} & Security and privacy & \begin{itemize} \item Ability of a malicious user to alter learning model parameters during wireless transfer between the aggregation server and end-devices. \item Ability of man-in-the-middle to infer end-devices sensitive information from learning model parameters. \end{itemize} & \begin{itemize} \item Partially homomorphic encryption. \item Somewhat homomorphic encryption. \item Fully homomorphic encryption.   \end{itemize} \\
				\midrule
				\textbf{Secure and trustful aggregation-enabled federated learning}  &  Security and privacy & \begin{itemize} \item Presence of malicious end-devices. \item Wrong local learning model parameters prolong federated learning convergence time. \end{itemize} & \begin{itemize} \item Secure aggregation-enabled federated learning scheme. \item Consortium blockchain-based trustful verification of end-devices updates. \end{itemize} \\
				\midrule
				\textbf{Interference-aware resource efficient federated learning} & Resource optimization & \begin{itemize} \item Communication resources constraints. \item Protection of cellular users while reusing frequency by end-devices used in federated learning. \end{itemize} & \begin{itemize} \item Game theory-based resource allocation. \item Matching game with externalities for resource allocation. \item Relaxation-enabled resource allocation scheme. \end{itemize} \\
				
				\midrule 
				\textbf{Interference-aware association for federated learning} & Resource optimization & \begin{itemize} \item Dependency of SINR on association. \item Proportional effect of association on packet error rate. \end{itemize} & \begin{itemize} \item Game theory-based association. \item Relaxation-enabled association scheme. \end{itemize} \\
				
				\midrule 
				\textbf{Quantization-enabled federated learning} & Resource optimization & \begin{itemize} \item Communication resource constraints. \item Massive number of devices are expected in future for federated learning training-enabled IoT applications. \end{itemize} & \begin{itemize} \item Gradient-based selective end-devices selection. \item Selection of end-devices with better performance. \end{itemize} \\
				
				\midrule 
				\textbf{Adaptive-resource-allocation-enabled federated learning} & Resource optimization & \begin{itemize} \item Heterogeneity of computational resources (CPU-cycles/sec), local device energy, and communication resources results in variable local accuracy. \end{itemize} & \begin{itemize} \item Adaptive resource allocation (i.e., more wireless resources to devices with high local model computational time) to minimize the overall federated learning convergence time. \end{itemize} \\
				
				\bottomrule 
			\end{tabular}
		}
	\end{table*}

	\subsection{Data-Heterogeneity-Aware-Clustering-Enabled Federated Learning }
	\setlength{\parindent}{0.7cm}How does one enable federated learning for massively distributed devices having heterogeneous features? A massive number of IoT devices exhibit a significant amount of statistical heterogeneity in their datasets. FedAvg was developed to yield a global federated learning model using the assumption of synchronous amount of work at end-devices during each global iteration  \cite {mcmahan2017communication}. However, IoT devices show significant heterogeneity in their datasets and system parameters. To address this issue, FedProx was developed that is based on the addition of a weighted proximal term to the global model of FedAvg \cite {li2018federated}. Choosing the weights for FedProx for different applications is challenging. Additionally, it is not clear whether FedProx can provably improve convergence rate \cite{kairouz2019advances}. Therefore, there is a need to devise a novel protocol for federated learning to cope with data heterogeneity. One possible solution can be the formation of clusters of end-devices with statistical homogeneity. Within each cluster, a cluster head is selected which acts for aggregation of local learning models. The selection of cluster heads must follow attractive criteria. For instance, the criterion for cluster head selection can be an improvement in the overall throughput of the cluster. Another possible way can be socially-aware clustering. As cluster head has the capability to infer end-devices sensitive information from their local learning model \cite{Dusit_FL_survey}, it can be desirable to form cluster head with high social trust nature. Within each cluster, a sub-global model can be computed similarly to traditional federated learning. Next, cluster heads send their sub-global models to global aggregation server to yield a global federated learning model, which is then disseminated to cluster heads. Finally, the cluster heads disseminate the global model updates to devices of their corresponding clusters.

	\subsection{Mobility-Aware Federated Learning} 
	How do we enable seamless communication of mobile devices with the centralized edge/cloud server during training phase of federated learning? The set of mobile devices involved in learning must be connected to the edge/cloud server during the training phase. However, mobility of the devices might cause loss in coverage and thus, causes performance degradation of federated learning. Several solutions can be proposed to tackle this issue. One way is modification in the federated learning protocols that should consider the mobility of users in addition to other factors during the devices selection phase. The devices with no mobility or with less mobility than other high mobility nodes must be preferred. This approach seems to be promising in the scenarios where the minimum required number of devices for federated learning have no or limited mobility. However, if the devices have high mobility then it is necessary to predict the mobility of devices during the device selection phase. For this, deep learning-based mobility prediction schemes can be used.\par 
	
	\subsection{Homomorphic Encryption-Enabled Federated Learning} 
	How does one enable secure exchange of federated learning parameters between devices and the cloud/edge server? Although federated learning has been proposed to enable privacy-aware on-device machine learning in a distributed manner, it suffers from security challenges and privacy challenges. A malicious user can access the learning model updates during their transmission between end-devices and the aggregation server. Then, the malicious user abnormally alters the learning parameters. Moreover, the learning parameters can be used by a malicious user to infer end-device sensitive information \cite{Dusit_FL_survey}. Therefore, it is indispensable to ensure secure federated learning over wireless channels. One can use homomorphic encryption for ensuring secure federated learning. In homomorphic encryption-based federated learning, a key-pair is synchronized among all end-devices via a secure channel \cite {zhang2020batchcrypt, hardy2017private, borrego2019privacy,li2015secure}. Initially, all the devices encrypts their local learning model updates and send the ciphertexts to a central server. Next, the server sends back the aggregated result to the end-devices. The advantage of using homomorphic encryption is non-requirement of decryption at the aggregation server. However, generally, homomorphic encryption results in the use of extra computation and communication resources. Homomorphic encryption can be divided into three main categories such as partially homomorphic encryption, somewhat homomorphic encryption, and fully homomorphic encryption \cite{shrestha2019integration}. These categories of homomorphic encryption offer freedom of performing mathematical operation depending on the type used. For instance, partially homomorphic encryption allows only one type of mathematical operation (i.e., addition or multiplication), whereas fully homomorphic encryption allows a large number of different mathematical operations. On the other hand, ciphertext in homomorphic encryption contains a noise which increases proportionally with mathematical operations \cite{lepoint2014comparison}. Therefore, there is a need to devise a homophobic encryption scheme that accounts for such type of noise.

	\subsection{Secure and Trustful Aggregation-Enabled Federated Learning}
	How do we enable aggregation of learning models at the aggregation server in a secure and trustful way? A malicious user can send the wrong local learning model parameters to the aggregation server to slow the federated learning convergence rate. In some cases, the global federated learning model might not converge due to the presence of a malicious user. On the other hand, it is necessary to verify the end-devices' updates before considering them for global aggregation. Several possible ways can be used to enable secure aggregation. In \cite{bonawitz2017practical}, the authors proposed a secure aggregation protocol for federated learning. The protocol can offer security against $1/3$ of the total devices but at the cost of extra communication resources and complexity. Although the protocol in \cite{bonawitz2017practical} offers reasonable performance, there is a need to propose novel protocols than can handle malicious users more than $1/3$ of the total devices. In another work, a reputation-based metric was introduced for the trustful participation of end-devices in the federated learning process \cite{kang2020reliable}. Consortium blockchain was then applied to enable efficient reputation management. Although the scheme presented in \cite{kang2020reliable} can offer resilience against adversaries, it has high-latency associated with blockchain. To address the above challenges, there is a need for federated learning protocols that can offer security and trustful verification using low complexity schemes.

	\subsection{Interference-Aware Resource Efficient Federated Learning }
	\setlength{\parindent}{0.7cm}How do we reuse the uplink communication resources already in use by other cellular users for federated learning while protecting them? To protect cellular users from interference due to end-devices involved in federated learning, we will require resource allocation schemes. There can be two different ways of allocating resource blocks to end-devices. The first one is to assign a single resource block to only one end-device while fulfilling the constraint the maximum allowed interference. Another approach is to assign multiple end-devices to a single resource block to more effectively reuse the resources. For the later case, we can use one-to-one matching, whereas for this case we can use one-to-many matching theory with externalities for resource allocation due to the fact that preference profiles depend both on resource blocks along with other end-devices assigned to the same resource block \cite{bairagi2019matching, pham2018decentralized, kazmi2017hierarchical, kazmi2017mode}.\par       
	\subsection{Interference-Aware Association for Federated Learning}
	\setlength{\parindent}{0.7cm}How does one associate end-devices to the edge servers-enabled small cell BSs (SBSs) to reuse the uplink cellular spectrum while protecting cellular users? For a fixed resource block allocation and fixed transmit power, the end-devices' association has no impact on interference control for cellular users while reusing their frequencies. However, joint transmit power allocation and device association for end-devices can be performed to simultaneously minimize federated learning cost and cellular user interference. Such kind of joint transmit power allocation and the device-association optimization problem is a mixed-integer non-linear programming problem that can be solved sub-optimally in a variety of ways.   \par

	\subsection{Quantization-Enabled Federated Learning}
	How do we enable federated learning for a massive number of devices under communication resource constraints? Although sparsification-enabled federated learning can reduce communication resource consumption, it might not be sufficient alone for a large number of devices and limited communication resources. To address this issue, we can use quantization-enabled federated learning. However, quantization induces errors in global federated learning due to the distortion of the signals. Such kind of errors in the global federated learning model might prolong its convergence time. Therefore, there is a need to make a trade-off between a federated learning convergence time and communication resources consumption during training. Several quantization techniques (i.e., universal vector quantization \cite{shlezinger2020uveqfed}, low-precision quantizer \cite{reisizadeh2020fedpaq}, and hyper-sphere quantization \cite{dai2019hyper}) can be used for federated learning over IoT networks.

	\subsection{Adaptive-Resource-Allocation-Enabled Federated Learning} 
	How does one perform training of the federated learning model with low convergence time for end-devices with heterogeneous resources? The devices involved in federated learning are heterogeneous in terms of computational resources (CPU-cycles/sec), local device energy, and communication resources. These heterogeneous parameters have a significant effect on the performance of federated learning. For instance, computation time of local device learning models strictly depends on local model accuracy, local device operating frequency, and dataset size for the local learning model algorithm. On the other hand, synchronous federated learning is based on the computation of all the devices' local models within fixed time. However, the local computation time for devices with heterogeneous features shows significant variations in local learning model accuracy for fixed local model computation time. Therefore, these challenges of heterogeneity in the devices must be resolved. One way to reduce the impact of heterogeneity on the performance of federated learning is to allocate more computational resources to users with larger local training dataset sizes for achieving certain accuracy within the allowed time and vice versa. Another approach is to assign more communication resources to the end-devices with poor performance due to a lack of computational resources. Allocating resources adaptively can reduce the global federated learning time, and thus offer fast convergence.  
	
	%\subsection{Fairness-Enabled Multi-Task Federated Learning}

	% How do we enable federated learning more user and space specific?

	\section{Conclusions and Future Prospects}
	\label{conclusions}
	%\subsection{Conclusions}
	In this paper, we have discussed federated learning for IoT networks. Specifically, we have presented state-of-art-advances of federated learning towards enabling smart IoT applications. We have provided a taxonomy using various parameters. Furthermore, we have identified several important issues (i.e., robustness, privacy, and high-communication resources consumption) with the existing federated learning schemes based on a centralized aggregation server and presented guidelines to solve them. Finally, we have presented several open research challenges along with their causes and guidelines. We have concluded that dispersed federated learning will serve as a key federated learning scheme for future IoT applications using a massive number of end-devices. \par         
	%\subsection{Future Prospects}
	%\label{sec:future prospects}
	Although IoT networks enabled by federated learning and fifth-generation ($5$G) of cellular networks, can offer a wide variety of smart applications such as intelligent transportation, smart industry, smart health-care, among others. However, there are several Internet of Everything (IoE) applications that seem difficult to be fulfilled by $5$G. These IoE applications are haptics, flying vehicles, extended reality, brain-computer interfaces, and autonomous connected vehicles. The requirements of these applications seem difficult to be fulfilled by $5$G \cite{6Gvisionlatif}. Therefore, there is a need to propose a new generation of wireless systems, dubbed as sixth-generation ($6$G) wireless systems. Machine learning will be considered to be an integral part of $6$G \cite{saad2019vision, 9061001, akyildiz20206g, 6Gvisionlatif}. Additionally, privacy preservation will be given primary importance in $6$G \cite{wang2020security}. Therefore, dispersed federated learning with homomorphic encryption can be a promising candidate for the future $6$G-enabled IoE applications. In dispersed federated learning, homomorphic encryption will give privacy preservation against a malicious sub-global aggregation server, whereas sub-global aggregations will give privacy preservation against malicious global aggregation server.

	%we presented the recent advances of machine learning in making cities smarter. Then, we discussed the striking design aspects required for enabling smart cities by federated learning. Furthermore, SDN-based high-level framework for federated enables smart cities is proposed. A Stackelberg game-based solution is proposed to incentive based federated learning system. Finally, few indispensable open research challenges are outlined along with their causes and possible solutions. We concluded that the use of federated learning is a promising solution to enable various intelligent smart city applications. However, various challenges exist that must be addressed in the foreseeable future to turn the vision of federated learning enabled smart cities into reality. \par

	\bibliographystyle{IEEEtran}
	\bibliography{Database}

	\begin{IEEEbiography}[{\includegraphics[width=1in,height=1.25in,clip,keepaspectratio]{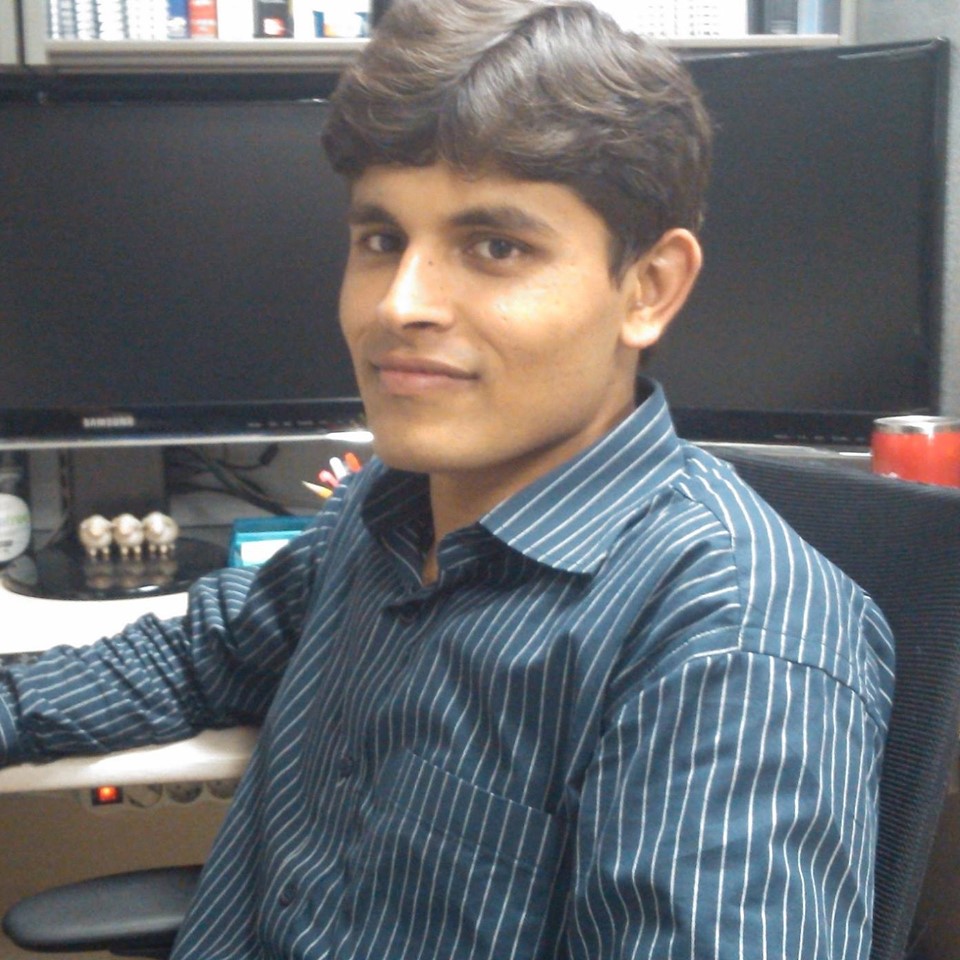}}]{Latif U. Khan} is currently pursuing his Ph.D. degree in Computer Science and Engineering at Kyung Hee University (KHU), South Korea. He is working as a leading researcher in the intelligent Networking Laboratory under a project jointly funded by the prestigious Brain Korea 21st Century Plus and Ministry of Science and ICT, South Korea. He received his MS (Electrical Engineering) degree with distinction from University of Engineering and Technology (UET), Peshawar, Pakistan in 2017. Prior to joining the KHU, he has served as a faculty member and research associate in the UET, Peshawar, Pakistan. He has published his works in highly reputable conferences and journals. He is the author/co-author of two conference best paper awards. He is also author of two books, such as "Network Slicing for $5$G and Beyond Networks" and "Federated Learning for Wireless Networks". His research interests include analytical techniques of optimization and game theory to edge computing, end-to-end network slicing, and federated learning for wireless networks. 
	\end{IEEEbiography}

	\begin{IEEEbiography}[{\includegraphics[width=1in,height=1.25in,clip,keepaspectratio]{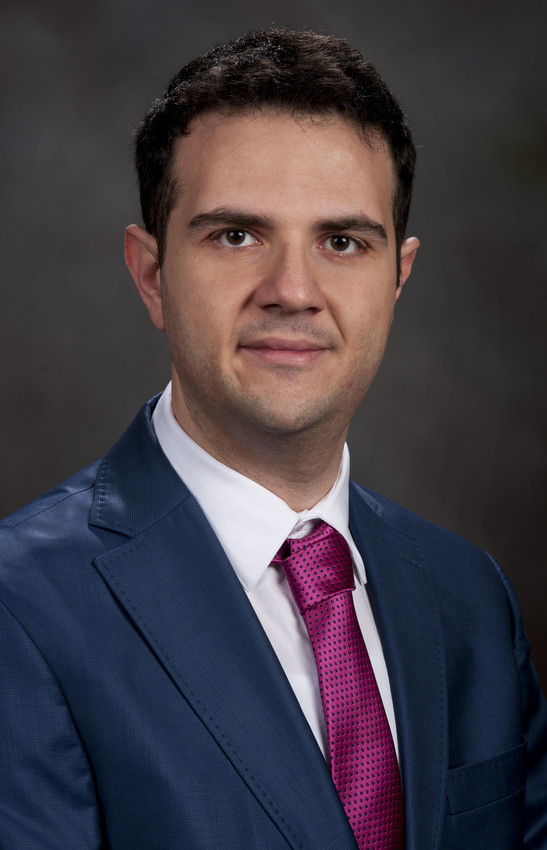}}]{Walid Saad } (S’07, M’10, SM’15, F’19) received the Ph.D. degree from the University of Oslo, Oslo, Norway, in 2010.,He is currently a Professor with the Department of Electrical and Computer Engineering, Virginia Tech, Blacksburg, VA, USA, where he leads the Network sciEnce, Wireless, and Security (NEWS) Laboratory. His research interests include wireless networks, machine learning, game theory, security, unmanned aerial vehicles (UAV), cyber-physical systems, and network science.,Dr. Saad was named the Stephen O. Lane Junior Faculty Fellow at Virginia Tech, from 2015 to 2017. In 2017, he was named as the College of Engineering Faculty Fellow. He was a recipient of the NSF CAREER Award in 2013, the Air Force Office of Scientific Research (AFOSR) Summer Faculty Fellowship in 2014, and the Young Investigator Award from the Office of Naval Research (ONR) in 2015. He was the author/coauthor of eight conference best paper awards at WiOpt in 2009, the International Conference on Internet Monitoring and Protection (ICIMP) in 2010, the IEEE Wireless Communications and Networking Conference (WCNC) in 2012, the IEEE International Symposium on Personal, Indoor and Mobile Radio Communications (PIMRC) in 2015, IEEE SmartGridComm in 2015, EuCNC in 2017, IEEE GLOBECOM in 2018, and the International Federation for Information Processing (IFIP) International Conference on New Technologies, Mobility and Security (NTMS) in 2019. He was also a recipient of the 2015 Fred W. Ellersick Prize from the IEEE Communications Society, the 2017 IEEE ComSoc Best Young Professional in Academia award, the 2018 IEEE ComSoc Radio Communications Committee Early Achievement Award, and the 2019 IEEE ComSoc Communication Theory Technical Committee. He received the Dean’s Award for Research Excellence from Virginia Tech in 2019. He also serves as an Editor for the IEEE Transactions on Wireless Communications, the IEEE Transactions on Mobile Computing, and the IEEE Transactions on Cognitive Communications and Networking. He is also an Editor-at-Large for the IEEE Transactions on Communications. He is also an IEEE Distinguished Lecturer. (Based on document published on 18 May 2020).
	\end{IEEEbiography}

	\begin{IEEEbiography}[{\includegraphics[width=1in,height=1.25in,clip,keepaspectratio]{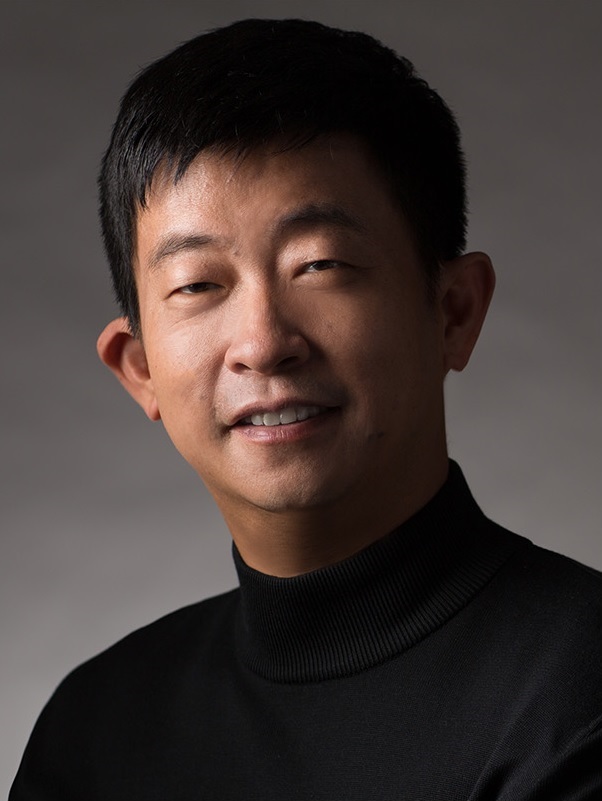}}]{Zhu Han}(S’01, M’04, SM’09, F’14) received the B.S. degree in electronic engineering from Tsinghua University, in 1997, and the M.S. and Ph.D. degrees in electrical and computer engineering from the University of Maryland, College Park, in 1999 and 2003, respectively. 
		
		From 2000 to 2002, he was an R\&D Engineer of JDSU, Germantown, Maryland. From 2003 to 2006, he was a Research Associate at the University of Maryland. From 2006 to 2008, he was an assistant professor at Boise State University, Idaho. Currently, he is a John and Rebecca Moores Professor in the Electrical and Computer Engineering Department as well as in the Computer Science Department at the University of Houston, Texas. His research interests include wireless resource allocation and management, wireless communications and networking, game theory, big data analysis, security, and smart grid.  Dr. Han received an NSF Career Award in 2010, the Fred W. Ellersick Prize of the IEEE Communication Society in 2011, the EURASIP Best Paper Award for the Journal on Advances in Signal Processing in 2015, IEEE Leonard G. Abraham Prize in the field of Communications Systems (best paper award in IEEE JSAC) in 2016, and several best paper awards in IEEE conferences. Dr. Han was an IEEE Communications Society Distinguished Lecturer from 2015-2018, AAAS fellow since 2019 and ACM distinguished Member since 2019. Dr. Han is 1\% highly cited researcher since 2017 according to Web of Science. Dr. Han is also the winner of 2021 IEEE Kiyo Tomiyasu Award, for outstanding early to mid-career contributions to technologies holding the promise of innovative applications, with the following citation: "for contributions to game theory and distributed management of autonomous communication networks."

	\end{IEEEbiography}

	\begin{IEEEbiography}[{\includegraphics[width=1.5in,height=1.25in,clip,keepaspectratio]{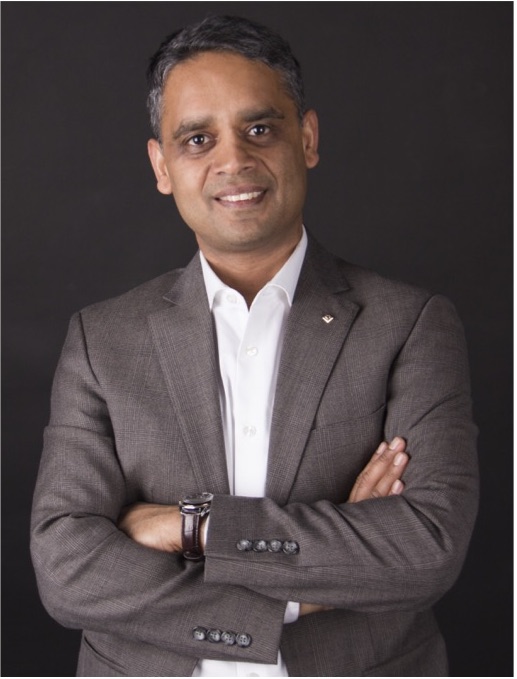}}]{Ekram Hossain}(F'15) is a Professor in the Department of Electrical and Computer Engineering at University of Manitoba, Canada (\url{https://home.cc.umanitoba.ca/~hossaina/}). He is a Member (Class of 2016) of the College of the Royal Society of Canada, a Fellow of the Canadian Academy of Engineering, and a Fellow of the Engineering Institute of Canada.  He was elevated to an IEEE Fellow ``for contributions to spectrum management and resource allocation in cognitive and cellular radio networks''.  He was listed as a Clarivate Analytics Highly Cited Researcher in Computer Science in 2017, 2018, 2019, and 2020. Currently he serves as the Editor-in-Chief of IEEE Press. Previously he served as the Editor-in-Chief for the IEEE Communications Surveys and Tutorials (2012--2016).
		
	\end{IEEEbiography}

	\begin{IEEEbiography}[{\includegraphics[width=1in,height=1.25in,clip,keepaspectratio]{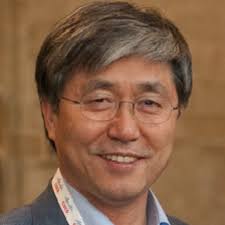}}]{Choong Seon Hong} (S’95-M’97-SM’11)  received the B.S. and M.S. degrees in electronic engineering from Kyung Hee University, Seoul, South Korea, in 1983 and 1985, respectively, and the Ph.D. degree from Keio University, Tokyo, Japan, in 1997. In 1988, he joined KT, Gyeonggi-do, South Korea, where he was involved in broadband networks as a member of the Technical Staff. Since 1993, he has been with Keio University. He was with the Telecommunications Network Laboratory, KT, as a Senior Member of Technical Staff and as the Director of the Networking Research Team until 1999. Since 1999, he has been a Professor with the Department of Computer Science and Engineering, Kyung Hee University. His research interests include future Internet, intelligent edge computing, network management, and network security. 
		Dr. Hong is a member of the Association for Computing Machinery (ACM), the Institute of Electronics, Information and Communication Engineers (IEICE), the Information Processing Society of Japan (IPSJ), the Korean Institute of Information Scientists and Engineers (KIISE), the Korean Institute of Communications and Information Sciences (KICS), the Korean Information Processing Society (KIPS), and the Open Standards and ICT Association (OSIA). He has served as the General Chair, the TPC Chair/Member, or an Organizing Committee Member of international
		conferences, such as the Network Operations and Management Symposium (NOMS), International Symposium on Integrated Network Management (IM), Asia-Pacific Network Operations and Management Symposium (APNOMS), End-to-End Monitoring Techniques and Services (E2EMON), IEEE Consumer Communications and Networking Conference (CCNC), Assurance in Distributed Systems and Networks (ADSN), International Conference on Parallel Processing (ICPP), Data Integration and Mining (DIM), World Conference on Information Security Applications (WISA), Broadband Convergence Network (BcN), Telecommunication Information Networking Architecture (TINA), International Symposium on Applications and the Internet (SAINT), and International Conference on Information Networking (ICOIN). He was an Associate Editor of the IEEE TRANSACTIONS ON NETWORK AND SERVICE MANAGEMENT and the IEEE JOURNAL OF COMMUNICATIONS AND NETWORKS. He currently serves as an Associate Editor for the International Journal of Network Management.
	\end{IEEEbiography}

	\balance
\end{document}